\documentclass[a4paper,20pt]{article}
\usepackage{multirow}
\usepackage{amsmath}
\usepackage{tgtermes}
\usepackage{xcolor}
\usepackage{amsfonts}
\usepackage{amssymb}
\usepackage{amstext}
\usepackage{import,breqn}
\usepackage{graphicx}
\usepackage{caption,subcaption}
\usepackage{float}
\usepackage{bm}
\usepackage{makecell}
\usepackage{authblk}
\usepackage{hyperref}
\usepackage{authblk}
\usepackage{orcidlink}
\usepackage{cite}
\usepackage{epsfig}%
\usepackage{pdfrender,xcolor}
\usepackage{cancel}
\usepackage{tabularx}
\usepackage{adjustbox}
\usepackage{makecell}

\hypersetup{
	colorlinks=true,
	linkcolor=blue,
	urlcolor=blue,
	citecolor=red
}
\title{Theory space and stability analysis of General Relativistic cosmological solutions in modified gravity}

\author[1]{Saikat Chakraborty\footnote{Saikat.C@chula.ac.th}}
\author[1]{Piyabut Burikham\footnote{Piyabut.B@chula.ac.th ; \, Corresponding author}}

\affil[1]{High Energy Physics Theory Group, Department of Physics, Faculty of Science, Chulalongkorn University, Bangkok 10330, Thailand}

\date{}

\begin{document}

\maketitle

\begin{abstract}
Some aspects of two General Relativistic cosmological solutions, an exact $\Lambda$CDM-like cosmological solution $j=1$ ($j$ is cosmographic jerk parameter), and a specifically designed toy cosmological solution $j=1+3\varepsilon(q-1/2)$ ($q$ is cosmographic deceleration parameter, $0<|\varepsilon|<1$) that is capable of accommodating a phantom crossing scenario as suggested by DESI DR2, are studied within the context of $f(R)$ gravity, by portraying them as a \emph{flow} in the 2-dimensional \emph{theory space} spanned by the quantities $r=\frac{R f'}{f}, m=\frac{R f''}{f'}$. For the $f(R)$ theories exactly reproducing a background $\Lambda$CDM-like expansion history $j=1$, it is shown by means of a \emph{cosmographic} reconstruction approach that the curvature degree of freedom need not necessarily behave like an effective cosmological constant, and that cosmologies under different possible such theories lead to different possible values of $\Omega_{m0}$. With the theory space analysis, it is also shown that $\Lambda$CDM-mimicking $f(R)$ cosmologies that asymptote to General Relativistic $\Lambda$CDM in the limit $q\to1/2$, are prone to instability under small homogeneous and isotropic perturbation, casting a doubt on achieving an exact $\Lambda$CDM-like cosmological solution $j=1$ within $f(R)$ gravity. Regarding the toy cosmological solution $j=1+3\varepsilon(q-1/2)$ that is capable of accommodating a phantom crossing scenario, it is shown that possible underlying $f(R)$ theories that admit it as a solution are inevitably plagued by tachyonic instability ($f''(R)<0$). All the above physically interesting conclusions are derived without explicitly reconstructing, even numerically, the functional form of the underlying $f(R)$, which demonstrates the edge of the $r$-$m$ theory space analysis over the traditional explicit reconstruction approach. The methodology is general and can be extended straightforwardly to modified teleparallel or modified symmetric teleparallel gravity. 
\end{abstract}

\tableofcontents

\section{Introduction}

Modified gravity research has seen a surge of interest following the discovery of the universe's accelerated expansion, and there is a plethora of modified gravity models that have been introduced over the past decades \cite{Papantonopoulos2015,CANTATA:2021asi}. It is crucial to recover General Relativistic (GR) solutions, such as Friedmann or Schwarzschild, within a modified gravity framework. The phenomenological motivation behind it is the empirical success of GR to account for astrophysical data (see, e.g. \cite{Creminelli:2017sry,Ezquiaga:2017ekz,Baker:2017hug,Copeland:2018yuh}). From the theoretical point of view, this is required for a well-defined GR limit of the modified gravity theory. For a solution to be physically viable, it must also be natural, i.e., not arising from only a fine-tuned set of initial conditions.

The subject of this work is to study some aspects of a given General Relativistic homogeneous and isotropic cosmological solution in modified gravity. In particular, we try to present a sound mathematical scheme to answer the following questions:
\begin{itemize}
    \item Can a given General Relativistic cosmological solution $a(t)$ be realized within the framework of a given modified gravity theory without incurring any known theoretical pathology?
    \item In the space of all such theories that admit the given cosmological solution $a(t)$, does GR behave like a cosmological past or future attractor?
    \item Even if the solution can be realized within the framework of a given modified gravity theory, is it stable under small homogeneous and isotropic perturbations?\footnote{Homogeneous and isotropic perturbations in the solution space of a given theory bear a slightly different significance than usual inhomogeneous cosmological perturbations. It is related to the naturalness of a solution. Homogeneous and isotropic perturbation of a solution trajectory becomes important in the study of inflationary attractors and tracker quintessence models, which were introduced to provide natural solutions to the fine-tuning problems in the early universe and the coincidence problem in the late universe.}
\end{itemize}
In this work, we illustrate our approach in the context of $f(R)$ gravity, which is one of the most well-known forms of modified gravity. However, the applicability of this approach extends beyond $f(R)$ to any $f$-class of modified gravity theories, e.g. $f(T),\,f(Q),\,f(G)$ (modified teleparallel, symmetric teleparallel and general teleparallel gravity).

Barrow and Ottewill were the first ones to systematically study the existence and stability of General Relativistic cosmological solutions in $f(R)$ gravity \cite{Barrow:1983rx}. In later studies, the existence and stability of different General Relativistic cosmological solutions of interest has been the subject of study in the context of $f(R)$ gravity \cite{Boehmer:2007tr,Barrow:2006xb,Toporensky:2006kc,Toporensky:2016kss}, $f(R,\phi)$ gravity \cite{Faraoni:2004dn,Faraoni:2005vk,Faraoni:2005ie}, Gauss-Bonnet and modified Gauss-Bonnet gravity \cite{Boehmer:2009fey,delaCruz-Dombriz:2011oii,Pozdeeva:2019agu,delaCruz-Dombriz:2011oii}. Whereas all the above-mentioned works are important to assess the theoretical viability of the respective modified gravity theory, it is to be appreciated that the cosmological solutions considered in those works are rather simple, like de-Sitter, Einstein static solutions, power-law Friedmann solutions, etc. In many of the above works, the authors could actually reconstruct in a simple functional form the underlying gravity Lagrangian admitting the given solution. From the dynamical system point of view \cite{Bahamonde:2017ize}, these solutions are different cosmological \emph{epochs} that can be expressed as a fixed point in the phase space of the underlying theory. More generic cosmological solutions are not fixed points but rather phase trajectories, connecting different epochs. For example, in the phase space of the General Relativistic $\Lambda$CDM model, the current state of our observable universe is a point on a heteroclinic trajectory connecting a matter-dominated saddle and a De-Sitter attractor. Therefore, it is more physically interesting to extend the above analysis to the case of realistic cosmological solutions that are trajectories. 

The question of \emph{existence} can in principle be addressed via the reconstruction method. For the $f$-class of modified gravity theories, one can, in principle, reconstruct the underlying form of $f$ starting from any given cosmological solution \cite{Nojiri:2009kx,Nojiri:2010wj,Nojiri:2010oco}. In particular, there have been works on realizing a $\Lambda$CDM-like cosmological solution within the framework of $f(R)$ gravity \cite{Dunsby:2010wg,He:2012rf,Choudhury:2019zod}, $f(\mathcal{G})$ gravity \cite{Myrzakulov:2010gt}, $f(R,\mathcal{G})$ and modified gauss-Bonnet gravity \cite{Elizalde:2010jx}, $f(Q)$ gravity \cite{Chakraborty:2025qlv} as well as in theories with nonminimal coupling between matter and geometry \cite{Ortiz-Banos:2021jgg,Kavya:2024bpj}. The caveat here is that, unless the underlying gravity Lagrangian can be reconstructed in a compact and simple functional form, it is hard to examine whether the reconstructed theory actually remains, throughout the course of cosmic evolution, free from various theoretical pathologies that plague modified gravity theories. For example, the conditions $f'(R)>0$ and $f''(R)>0$ are imposed for the absence of ghost and tachyonic instabilities in $f(R)$ gravity \cite{Sotiriou:2008rp,DeFelice:2010aj}). The generic $f(R)$ theory that reproduces a $\Lambda$CDM-like cosmological evolution can only be expressed in terms of complicated hypergeometric functions \cite{Dunsby:2010wg,He:2012rf,Choudhury:2019zod}, and it is extremely difficult to assess whether they actually satisfy those conditions.

Before the discoverty of the late-time acceleration, works by Damour and Nordvedt \cite{Damour:1992kf,Damour:1993id} and Mimoso \cite{Mimoso:1998dn,Mimoso:1999ai} showed that scalar-tensor theories contain an attractor mechanism towards GR, with the redshift at the onset of the matter-dominated epoch providing a measure for the present level of deviation from GR, providing a natural explanation for the observed tight constraints on the variation of the effective gravitational coupling. After the discovery of the late-time acceleration in the universe, this question was further explored in a series of papers by Jarv et.al.  in the context of scalar-tensor gravity \cite{Kuusk:2008ak,Saal:2012zb,Jarv:2008eb,Jarv:2010xm,Jarv:2011sm}, scalar-torsion gravity \cite{Jarv:2015odu}, and $f(Q)$ gravity \cite{Guzman:2024cwa}. A noteworthy characteristic in all the above-mentioned works is that the authors study the approach to GR in the vicinity of a fixed point, i.e., a cosmological epoch. As we have mentioned above, the present state of the Universe is more accurately described as a point on a trajectory connecting two epochs, rather than an epoch itself. From the mathematical language of dynamical systems, GR being an attractor around a fixed point does not necessarily imply GR will be an attractor along a heteroclinic trajectory.

Motivated by the current status of relevant research as outlined in the above paragraphs, we attempt to provide in this work a sound mathematical framework to answer the specific questions posed towards the beginning of this section, in a more general set-up. In particular, we aim to address those questions in the context of a cosmological solution that is a trajectory in the solution space of the underlying theory. We confine ourselves to $f(R)$ gravity in this work, motivated by its long history as a standard modified gravity framework, and the recently claimed strong statistical evidence of $f(R)$ over the standard General Relativistic $\Lambda$CDM model, taking into account DESI DR2+Pantheon+SH0ES datasets \cite{Plaza:2025gcv,Odintsov:2024woi,Odintsov:2025jfq}. The key idea that is implemented in this work, which enables us to address the above questions, is that one can represent an $f(R)$ cosmological solution as a \emph{flow} in the 2-dimensional space spanned by the dimensionless variables $\{r,m\}=\lbrace\frac{Rf'}{f},\frac{Rf''}{f'}\rbrace$, something that we call as the \emph{theory space}. Since we will start with a given cosmological solution $a(t)$ and not a given theory $f(R)$, the question of reconstruction automatically arises. As we will show by explicit examples later on in the text, the theory space analysis succeeds even when the explicit reconstruction method fails, in the sense that even if the theory cannot be reconstructed explicitly as $f=f(R)$, it can be reconstructed implicitly as $m=m(r)$. The quantity $m$ is of particular interest as it is related to the scalaron mass in $f(R)$ gravity, with $m\to0$ signifying the GR limit.

Let us clarify a point at the onset, which could be a source of confusion. Throughout the text, what we specify as a cosmological \emph{solution} is a particular kind of cosmological evolution $a(t)$, or equivalently $H(z)$ ($z=\frac{1-a}{a}$ being cosmological redshift). This is to be differentiated from a cosmological \emph{model}, which needs the specification of the underlying gravitational framework (e.g. GR or $f(R)$) and the matter components in the picture. For example, in our terminology, a cosmological evolution of the form 
\begin{equation}\label{eq:LCDM_rewrite}
    h^2(z) = \frac{2}{3}(1+q_0)(1+z)^3 + \frac{1}{3}(1-2q_0)
\end{equation}
is a cosmological solution. It is possible to realize such a cosmological solution within the framework of GR in the presence of a nonrelativistic fluid plus a cosmological constant, giving rise to the $\Lambda$CDM model\footnote{Throughout the paper, we will ignore the contribution from radiation.}. However, the same cosmological solution can be realized by other types of cosmological fluids within GR (see Subsection \ref{subsec:LCDM} below), or in the framework of other gravity theories e.g. $f(R)$. The latter situations give rise to the same evolution but different \emph{models}. We will adopt a way of specifying a spatially flat cosmological solution as a cosmographic condition $j=j(q)$ ($q,\,j$ being cosmographic deceleration and jerk parameter), inspired by the statefinder diagnostic approach \cite{Sahni:2002fz,Alam:2003sc}. 

The outline of the paper is as follows. In section \ref{sec:cosmography} we introduce the cosmographic parameters, the statefinder parameters and the cosmographic way of specifying a late-time cosmological solution as adopted in the statefinder diagnostic. In the subsections \ref{subsec:LCDM} and \ref{subsec:phantom} we introduce the particular cosmological solutions that we will be working with. Section \ref{sec:f(R)_cosmology} outlines the cosmological field equations and the effective dark energy fluid in $f(R)$ gravity. In section \ref{sec:recon_cosmo} we present a \emph{cosmographic} approach to the reconstruction method in $f(R)$ gravity, emphasizing the difference with earlier reconstruction methods. The notion of the \emph{theory space} is introduced in section \ref{sec:theory_space}. The routemap of investigating the stability of a given cosmological solution in the reconstructed $f(R)$ theory is outlined in section \ref{sec:stability}. In sections \ref{sec:LCDM_f(R)} and \ref{sec:phantom_crossing_f(R)}, we implement in detail our mathematical schemes to two General relativistic cosmological solutions; an exact $\Lambda$CDM-like cosmological solution and an almost $\Lambda$CDM-like phantom crossing cosmological solution. We summarize our framework, the results of applying it to the two given cosmological solutions, and discuss the future potential of this framework in section \ref{sec:concl}. Some more details about the theory space, which is not strictly required to understand the core part of the paper, are presented in the Appendix \ref{app:theory_space}.

\section{The cosmographic approach to specify a General Relativistic solution}\label{sec:cosmography}

The cosmographic parameters appear as the coefficients in the Taylor expansion of the scale factor $a(t)$ \cite{Dunsby:2015ers, Bolotin:2018xtq}. They are purely kinematic parameters characterizing \emph{how} the universe evolves, irrespective of whatever is the underlying dynamics. The $0$-th to $5$-th order cosmographic parameters are as follows:
\begin{subequations}\label{eq:CP_def}
    \begin{eqnarray}
        H &\equiv& \frac{\dot a}{a}\,,
        \\
        q &\equiv& -\frac{1}{aH^2}{\ddot a} = -1-\frac{\dot H}{H^2} \,,
        \\
        j &\equiv& \frac{1}{aH^3}\frac{d^3 a}{dt^3} = \frac{\ddot H}{H^3} - 3q - 2 \,,
        \\
        s &\equiv& \frac{1}{aH^4}\frac{d^4 a}{dt^4} = \frac{\dddot H}{H^4} + 4j + 3q(q+4) + 6 \,,
        \\
        l &\equiv& \frac{1}{aH^5}\frac{d^5 a}{dt^5} = \frac{H^{(4)}}{H^5} + 5s - 10(j+3q)(q+2) - 24 \,,
    \end{eqnarray}
\end{subequations}
and are called the Hubble, deceleration, jerk, snap, and lerk parameter, respectively. Of course, one could go on to construct even higher-order cosmographic parameters. There is an infinite hierarchy of them. The Taylor expansion of the Luminosity distance $d_L(z)$ (and other distance measures that are sometimes used in cosmology like the angular diameter distance $d_A(z)$, photon count distance $d_F(z)$ etc.) comprises solely of the present-day values of the cosmographic parameters \cite{Visser:2003vq,Visser:2004bf}. Therefore, these are precisely the parameters that we can directly measure from data, unlike, say, the density abundance parameters, whose measurements are always model-dependent. 

The following relations between the cosmographic parameters are helpful \cite{Dunsby:2015ers,Bolotin:2018xtq}
\begin{subequations}\label{eq:CP_rel}
    \begin{eqnarray}
        j &=& 2q^{2} + q - \frac{1}{H}\frac{dq}{dt}\,,
        \\
        s &=& \frac{1}{H}\frac{dj}{dt} - j(2 + 3q)\,,\label{s}
        \\
        l &=& \frac{1}{H}\frac{ds}{dt} - s(3+4q)\,.
    \end{eqnarray}
\end{subequations}

It is also helpful to write down the following expressions relating the Ricci scalar and its derivative to the cosmographic parameters
\begin{subequations}\label{eq:Ricci_CP}
    \begin{eqnarray}
        R &=& 6H^2(1-q) \quad \Rightarrow \quad \frac{R}{6H_0^2} = h^2(1-q)\,,
        \\
        \dot{R} &=& 6H^3(j-q-2) \quad \Rightarrow \quad \frac{\dot{R}}{6H_0^3} = h^3(j-q-2)\,,
    \end{eqnarray}
\end{subequations}
where $H_0$ is the present-day value of the Hubble parameter and we have defined the dimensionless quantity $h\equiv\frac{H}{H_0}$.

It is worth mentioning that the traditional cosmographic expansion with respect to $z$, around $z=0$, of a cosmological quantity faces a convergence issue, in the sense that the cosmographic series converges only within the range $0<z<1$. In practice, the cosmographic series of a quantity can be considered a good approximation only at low redshifts. To alleviate this problem, a cosmographic expansion with respect to a so-called \emph{$y$-redshift} was introduced in \cite{Cattoen:2007sk}, where $y=\frac{z}{1+z}$. This, in theory, raises the radius of convergence to $z\to\infty$ (since $y\to1$ as $z\to\infty$). In practice, it gives a better approximation than the traditional expansion with respect to $z$. More sophisticated and modern cosmographic methodologies involves use of Pad\'e polynomials \cite{Gruber:2013wua,Aviles:2014rma}.

When one tries to reconstruct the expansion history $H(z)$ (or $a(t)$) of the universe based on the cosmographic expansion with the present-day values of the cosmographic parameters as Taylor coefficients, one runs into the convergence issue as we have discussed above. However, that is not the path we take in this work. Rather, we will use an approach of kinematically specifying a given General Relativistic cosmological solution $a(t)$ or $H(z)$ as introduced in \cite{Dunajski:2008tg}, which does not involve a cosmographic series approximation, but merely utilizes the definition of the cosmographic quantities. The idea was to treat the parameters appearing in the Friedmann equations as integration constants and eliminate them by taking successive derivatives of the equation. In this way, any General Relativistic cosmological solution can be specified by an algebraic relation between a finite number of cosmographic parameters, some examples of which are presented in Ref.\cite{Dunajski:2008tg}. As an example, the standard $\Lambda$CDM model of cosmology\footnote{Throughout the paper, we will ignore contribution from radiation.} is a General Relativistic cosmological solution. The corresponding Friedmann equation and its two successive derivatives are 
\begin{subequations}\label{eq:lcdm}
    \begin{eqnarray}
      H^2 + \frac{k}{a^2} &=& \frac{\kappa\rho_{m0}}{3a^3} + \frac{\Lambda}{3}\,,
      \\
      \dot H - \frac{k}{a^2} &=& - \frac{\kappa\rho_{m0}}{2a^3}\,,
      \\
      \frac{\ddot H}{H} + \frac{2k}{a^2} &=& \frac{3}{2}\frac{\kappa\rho_{m0}}{a^3}\,.
    \end{eqnarray}
\end{subequations}
In the above equations, the CDM density $\kappa\rho_{m0}$ and the cosmological constant $\Lambda$ can be treated as parameters, which can be solved from the first two equations and substituted back into the third equation. The result is a cosmographic relation of the form 
\begin{equation}\label{eq:cosm_LCDM}
    j = \Omega_k + 1\,, 
\end{equation}
where $\Omega_k=\frac{k}{a^2H^2}$. Since $k=0,\pm1$, the above relation represents a purely kinematic constraint among the scale factor and the cosmographic parameters that specify the \emph{kinematics} of the General Relativistic $\Lambda$CDM solution\footnote{Even though Eq.\eqref{eq:cosm_LCDM} is already a relationship between kinematic quantities, one can still treat $k$ as an independent parameter and eliminate it by taking another derivative of the condition \eqref{eq:cosm_LCDM}. The result is a cosmographic condition involving the snap parameter: $s + 2(q+j) + qj = 1$ \cite{Dunajski:2008tg}.}. In general, during the late time, one can take $|\Omega_k|\ll1$ in the observable patch of the universe as a generic prediction of the inflationary paradigm, irrespective of its actual global topology. This brings us to the well-known cosmographic condition $j=1$ for specifying the $\Lambda$CDM cosmology, which is the basis of the statefinder diagnostic \cite{Sahni:2002fz,Alam:2003sc}.

We now claim the following: 
\begin{itemize}
    \item \emph{Any General Relativistic cosmological solution $a(t)$ with two non-interacting perfect barotropic fluids can be kinematically specified by a cosmographic condition involving only up to the jerk parameter.}
\end{itemize}
The proof is as follows. Suppose we have a General Relativistic cosmological solution $a(t)$ involving two perfect barotropic fluids with equation of state of the form $P=P(\rho)=w(\rho)\rho$, each separately conserved. This includes many of the dark energy models within the framework of GR, including Chaplygin gas, quintessence and, under some conditions, even non-canonical fields \cite{Arroja:2010wy}. The Friedmann equation and its first two derivatives are
\begin{subequations}\label{eqs:2_baro_fluid}
    \begin{eqnarray}
        H^2 + \frac{k}{a^2} &=& \frac{\kappa}{3}\rho_1 + \frac{\kappa}{3}\rho_2\,,
      \\
      \dot H - \frac{k}{a^2} &=& - \frac{\kappa}{2}\left[(1+w_1(\rho_1))\rho_1 + (1+w_2(\rho_2))\rho_2\right]\,,
      \\
      \frac{\ddot H}{H} + \frac{2k}{a^2} &=& \frac{3}{2}\kappa\left[(1+w_1(\rho_1))^2 \rho_1 + (1+w_2(\rho_2))^2 \rho_2 + (1+w_1(\rho_1))w_1'(\rho_1)\rho_1^2 + (1+w_2(\rho_2))w_2'(\rho_2)\rho_2^2\right]\,. \nonumber
      \\
      &&
    \end{eqnarray}
\end{subequations}
Once the equation of state parameters $w_1(\rho_1)$ and $w_2(\rho_2)$ are known, one can solve for $\rho_1,\rho_2$ from the first two equations and substitute back into the third equation. Using the definitions of the cosmographic parameters \eqref{eq:CP_def}, one can then express the result as an algebraic relation involving $q,j$ and $\Omega_k=\frac{k}{a^2 H^2}$. In fact, if one takes into consideration the generic inflationary prediction $|\Omega_k|\ll1$, then the $a(t),H(t)$ do not appear explicitly in the result, and the ultimate cosmographic condition is an algebraic condition involving $j$ and $q$. This is the basis of the statefinder diagnostic \cite{Sahni:2002fz,Alam:2003sc}, which is a kinematic diagnostic for dark energy models based on the three statefinder parameters
\begin{subequations}\label{eq:statefinder}
\begin{eqnarray}
    q &=& \frac{1}{2} + \frac{3}{2}w_{\rm DE}\Omega_{\rm DE}\,,
    \\
    j &=& 1 + \frac{9}{2}\Omega_{\rm DE}w_{\rm DE}(1+w_{\rm DE}) - \frac{3}{2}\Omega_{\rm DE}\frac{\dot{w_{DE}}}{H}\,,
    \\
    \tilde{s} &=& \frac{j-1}{3(q-1/2)} = 1 + w_{\rm DE} - \frac{1}{3}\frac{\dot{w_{\rm DE}}}{H w_{\rm DE}}\,\label{eq:statefinder_s}.
\end{eqnarray}
\end{subequations}
\footnote{The standard notation for the third statefinder parameters is $s$. To avoid confusion with the cosmographic snap parameter $s$, we denote the it here by $\tilde{s}$.} In the above $\Omega_{\rm DE}=\frac{\kappa\rho_{\rm DE}}{3H^2}$ is the density abundance parameter for the (effective) dark energy fluid, and $w_{\rm DE}$ is its equation of state parameter. The parameter $j$ traditionally defined in the original statefinder papers \cite{Sahni:2002fz,Alam:2003sc} is the same as the cosmographic jerk parameter $j$. In general, for a late-time cosmological model with nonrelativistic baryonic matter and some kind of an additional ``dark fluid'', one can take $w_1=0$ and $w_2=w_{\rm DE}$ ($w_{\rm DE}$ not necessarily constant) in Eq.\eqref{eqs:2_baro_fluid}. Then the statefinder equations \eqref{eq:statefinder} can be calculated straightforwardly from Eqs.\eqref{eqs:2_baro_fluid}.

In this work, we will consider two such General Relativistic cosmological solutions within the framework of $f(R)$ gravity, which we specify below.

\subsection{$\Lambda$CDM-like cosmological solution}\label{subsec:LCDM}

The General Relativistic $\Lambda$CDM \emph{model} has served as the standard cosmological model for a long time now. The corresponding cosmological solution $a(t)$, under the assumption of negligible spatial curvature and negligible contribution from radiation, can be conveniently specified by the cosmographic condition $j=1$ \cite{Sahni:2002fz,Alam:2003sc}. In the statefinder plane $\{j,\tilde{s}\}$, this is defined by the point $\{j,\tilde{s}\}=\lbrace j,\frac{j-1}{3(q-1/2)} \rbrace = \{1,0\}$. Setting $j=1$ in the definition of $j$ in Eq.\eqref{eq:CP_def}, and integrating backwards twice, one gets
\begin{equation}\label{eq:LCDM}
    h^2(z) = \frac{2}{3}(1+q_0)(1+z)^3 + \frac{1}{3}(1-2q_0)\,,
\end{equation}
$z$ being the cosmological redshift, and the subscript `$0$' denotes present day values. Henceforth, we will refer to the cosmological evolution given by Eq.\eqref{eq:LCDM} as the \emph{$\Lambda${\rm CDM}-like cosmic evolution/solution}. The two integration constants are fixed by setting $q(z=0)=q_0$ and $h(z=0)=1$. 

Note that although $j=1$ is usually taken to specify the $\Lambda$CDM model in the statefinder diagnostic, such a \emph{kinematic} specification does \emph{not} automatically guarantee the $\Lambda$CDM \emph{model} itself. This is best understood by substituting $\tilde{s}=0$ in Eq.\eqref{eq:statefinder_s}, which gives 
\begin{equation}\label{wDE_for_s0}
    \frac{dw_{\rm DE}}{d\ln a} = 3w_{\rm DE}(1+w_{\rm DE})\,.
\end{equation}
Since $w_{\rm DE}\neq0$, one of the solutions of the above is $w_{\rm DE}=-1$, which corresponds to the $\Lambda$CDM model. However, this is not the only solution. In general, the above equation admits a 1-parameter family of solutions
\begin{equation}\label{eq:DDE}
    w_{\rm df}(a) = -\frac{w_0 a^{3}}{w_0 a^{3} - (1+w_0)} \Leftrightarrow w_{\rm df}(z) = -\frac{w_0}{w_0 - (1+w_0)(1+z)^3}\,, \qquad\qquad (w_0=w_{\rm df}(z=0))\,.
\end{equation}
This corresponds to the so-called \emph{unified dark fluid model} with vanishing sound speed \cite{Luongo:2014nld}, where the equation of state parameter of the additional fluid varies smoothly from the value $0$ at the limit $z\to\infty$ to the value $-1$ at the limit $z\to-1$, thus explaining, in a unified manner, the dark matter and an \emph{emergent} dark energy. Note that \eqref{eq:DDE} does NOT allow phantom crossing. To make it explicit that the above equation of state parameter corresponds not only to the dark energy but to a unified dark fluid as a whole, we have changed from the notation $w_{\rm DE}$ to $w_{\rm df}$. All the unified dark fluid models freeze into the $\Lambda$CDM model (which corresponds to the particular case $w_0=-1$ in Eq.\eqref{eq:DDE}) in the asymptotic future, $z\to-1$.

Substituting the above expression for $w_{\rm df}(a)$ into the continuity equation
\begin{equation}
    \frac{d\ln\rho_{\rm df}}{d\ln a} + 3(1 + w_{\rm df}) = 0\,,
\end{equation}
it can be calculated that $w_{\rm df}\propto\frac{1}{\rho_{\rm df}}$, so that $P_{\rm df}$ is constant. If one chooses the dark fluid to satisfy the Weak Energy Condition (WEC), then the proportionality constant is negative, since $w_{\rm df}$ varies monotonically from $0$ to $-1$ as $z$ goes from $\infty$ to $-1$. Then, one can also write 
\begin{align}
        \rho_{\rm df} &\propto - \frac{1}{w_{\rm df}} \nonumber
        \\
        &\propto \left[\left(\frac{1+w_0}{-w_0}\right)\frac{1}{a^3} + 1\right]\,.\label{rho_df}
\end{align}
In Eq.\eqref{rho_df}, the proportionality constant is now positive. From \eqref{rho_df}, one can see that the unified dark fluid scales partially as a pressureless fluid and partially as a cosmological constant, with the energy density corresponding to both the sectors being positive.

Therefore, even though the line $\{j,q\}=\{1,q\}$ or the point $\{j,\tilde{s}\}=\{1,0\}$ is commonly associated with the $\Lambda$CDM model in the statefinder diagnostic, one must always keep in mind that the statefinder diagnostic is oblivious to this degeneracy. This degeneracy has been discussed earlier in \cite{Chakraborty:2022evc}. Very recently, the equivalence between the kinematic and the dynamic specification of the $\Lambda$CDM model has been strongly questioned in \cite{Chakraborty:2025rvc}.

\subsection{A phantom crossing cosmological solution}\label{subsec:phantom}

The cosmographic condition $j=1$, which generically gives rise to a dynamical dark energy with the equation of state parameter \eqref{eq:DDE}, cannot accommodate a phantom crossing. In light of the second data release of the DESI collaboration (henceforth DESI DR2; see, e.g. \cite{DESI:2025zgx}), which strongly hints towards a scenario with a phantom crossing from the phantom region $w_{\rm DE}<-1$ to the non-phantom region $w_{\rm DE}>-1$, we will also consider a somewhat more realistic cosmological solution that can accommodate a phantom crossing. To specify a simple such solution cosmographically, we notice that it is the statefinder parameter $\tilde{s}$ that determines whether there exists a phantom crossing, and if there is, in which sense. Eq.\eqref{eq:statefinder_s} can be written as
\begin{eqnarray}\label{wDE_for_snon0}
    \frac{dw_{\rm DE}}{d\ln a} = 3w_{\rm DE}(1+w_{\rm DE}-\tilde{s})\,.    
\end{eqnarray}
For a phantom crossing to occur, one needs $\displaystyle{\frac{dw_{\rm DE}}{d\ln a}\bigg\vert_{w_{\rm DE}=-1}=3\tilde{s}\big\vert_{w_{\rm DE}=-1}\neq0}$. For the phantom crossing to occur from the phantom to the non-phantom regime, one needs $\tilde{s}\big\vert_{w_{\rm DE}\neq-1}>0$. In general, the statefinder parameter $\tilde{s}$ is a function of time $\tilde{s}=\tilde{s}(a)$, but as a toy model incorporating a phantom crossing, we consider a simple toy model of cosmological evolution with a constant positive statefinder parameter $\tilde{s}$. The Eq.\eqref{eq:statefinder_s} gives
    \begin{equation}\label{cosm_cond_DESI}
        \tilde{s} = \frac{j-1}{3\left(q-\frac{1}{2}\right)} = \varepsilon ({\rm constant}) >0 \quad\Rightarrow\quad j = 1 + 3\varepsilon\left(q - \frac{1}{2}\right)\,.
    \end{equation}
Substituting $\tilde{s}=\varepsilon$ ($\varepsilon=$constant) in Eq.\eqref{eq:statefinder_s} and integrating, one obtains the dark energy equation of state
    \begin{equation}\label{eq:DDE_phantom}
        w_{\rm DE}(a) = \frac{-w_0(1-\varepsilon)a^{3(1-\varepsilon)}}{w_0 a^{3(1-\varepsilon)} - (1-\varepsilon+w_0)} \Leftrightarrow w_{\rm DE}(z) = \frac{-w_0(1-\varepsilon)}{w_0 - (1-\varepsilon+w_0)(1+z)^{3(1-\varepsilon)}}\,, \qquad (w_0=w_{\rm DE}(z=0))\,.
    \end{equation}
For $\tilde{s}=\varepsilon>0$, Eq.\eqref{wDE_for_snon0} can be expressed as
\begin{equation}
    -(1+z)\frac{dw_{\rm DE}}{dz} = 3w_{\rm DE}(1 + w_{\rm DE} - \varepsilon)\,,
\end{equation}
which, at $z=0$, gives explicitly
\begin{equation}
    w'_0 = -3w_0(1 + w_0 - \varepsilon)\,, \qquad (w'_0=w_{\rm DE}'(z)\vert_{z=0})\,.
\end{equation}
DESI DR2 hints at $-1<w_0<0$ and $w'_0<0$ \cite{DESI:2025zgx}, which signifies a phantom crossing at some early time. This translates into a lower bound on the parameter $\varepsilon$;
\begin{equation}\label{eps_bound}
    \varepsilon>1+w_0>0\,.
\end{equation}
As long as the parameter $\varepsilon$ satisfies the bound \eqref{eps_bound}, the cosmological evolution corresponding to the cosmographic condition $j=1+3\varepsilon\left(q - \frac{1}{2}\right)$ accommodates a phantom crossing scenario at some early time, just like what is hinted by DESI DR2. The exact redshift at which the phantom crossing occurs depends on the values of $\varepsilon$. 

Again, a word of caution is in order. We have noticed earlier that the cosmographic condition $j=1$ can imply either a model with a constant dark energy equation of state parameter $w_{\rm DE}=-1$ (the $\Lambda$CDM model) \emph{or}, more generically, a dynamical dark energy model (the unified dark fluid model). Similarly, the cosmographic condition $j = 1 + 3\varepsilon\left(q - \frac{1}{2}\right)$ can imply either a model with a constant dark energy equation of state parameter $w_{\rm DE}=-1+\varepsilon$ (a $w$CDM model) \emph{or}, more generically, a dynamical dark energy model \eqref{eq:DDE_phantom}. The former possibility is understood clearly by looking at Eq.\eqref{wDE_for_snon0}, which, in principle, also provides a solution $w_{\rm DE}=-1+\varepsilon$ for a constant statefinder parameter $\tilde{s}=\varepsilon$. All the models of Eq.\eqref{eq:DDE_phantom}, whether or not accommodating a phantom crossing at some early time, freeze into this particular $w$CDM model in the asymptotic future $z\to-1$.

In a way, the toy model \eqref{eq:DDE_phantom} that we have presented in this subsection also acts as a cousin of the already known unified dark fluid model in the literature. For $\tilde{s}=\epsilon$(constant), Eq.\eqref{wDE_for_snon0} gives
\begin{equation}
    3(1+w_{\rm DE}) = \frac{1}{w_{\rm DE}}\frac{dw_{\rm DE}}{d\ln a} + 3\varepsilon = \frac{d\ln w_{\rm DE}}{d\ln a} + 3\varepsilon\,.
\end{equation}
Substituting it into the continuity equation
\begin{equation}
    \frac{d\ln\rho_{\rm DE}}{d\ln a} + 3(1 + w_{\rm DE}) = 0\,,
\end{equation}
one gets\begin{equation}
    \frac{d\ln\rho_{\rm DE}}{d\ln a} + \frac{d\ln w_{\rm DE}}{d\ln a} + 3\varepsilon = 0\,,
\end{equation}
which can be written as
\begin{equation}
    \frac{d}{d\ln a}[\ln(\rho w_{\rm DE})]+\frac{d}{d\ln a}[3\varepsilon\ln a] = 0\,.
\end{equation}
The last step is possible because we have assumed a simple toy model of phantom crossing cosmology with the statefinder $\tilde{s}=\varepsilon$ a constant. The above equation can now be solved straightforwardly to obtain $\rho_{\rm DE}\propto\frac{1}{w_{\rm DE}a^{3\varepsilon}}$. Like in the case of the unified dark fluid model, if one again demands the dark energy fluid to satisfy the WEC, then the proportionality constant in the relation $\rho_{\rm DE}\propto\frac{1}{w_{\rm DE}a^{3\varepsilon}}$ is negative. One can then write
\begin{align}
        \rho_{\rm DE} &\propto  - \frac{1}{w_{\rm DE}a^{3\varepsilon}}\nonumber\\
    &\propto \left[\left(\frac{1+w_0-\varepsilon}{-w_0(1-\varepsilon)}\right)\frac{1}{a^3} + \frac{1}{a^{3\varepsilon}(1-\varepsilon)}\right]\,,\label{eq:rho_almost_LCDM}
\end{align}
In eq.\eqref{eq:rho_almost_LCDM}, the proportionality constant is now positive. From Eq.\eqref{eq:rho_almost_LCDM}, one can see that the fluid scales partially as a pressureless fluid with negative energy density~(or {\it phantom}) and partially as a fluid with equation of state parameter $-1+\varepsilon$~(or dark energy for $\epsilon < 2/3$). Note the crucial difference with the energy density of the unified dark fluid model in eq..\eqref{rho_df}. In both Eq.\eqref{rho_df} and eq.\eqref{eq:rho_almost_LCDM}, the total energy density of the fluid is positive since the fluid is demanded to satisfy the WEC. However, in Eq.\eqref{eq:rho_almost_LCDM}, the part of the fluid that scales as a pressureless fluid actually has a negative energy density, as enforced by the condition \eqref{eps_bound}. Existence of both phantom and dark energy in a single model is a distinctive aspect of the quintom model studied in e.g. Ref.~\cite{Panpanich:2019fxq}, where the phantom crossing is also present. 

One can substitute the cosmographic condition $j=1+3\varepsilon(q-1/2)$ and $\frac{d}{dt}=-H(1+z)\frac{d}{dz}$ into the first relation of \eqref{eq:CP_rel} 
\begin{equation}
    (1+z)q'(z) = - 2q^2 - q + j(q) = -2q^2 - q + 1 + 3\varepsilon\left(q-\frac{1}{2}\right)\,,
\end{equation}
integrating which one gets
\begin{equation}\label{eq:q_almost_LCDM}
    q(z) = \frac{1}{2}\frac{c_1(1+z)^3 - (1-c_1)(2-3\epsilon)(1+z)^{3\varepsilon}}{c_1(1+z)^3 + (1-c_1)(1+z)^{3\epsilon}} \,, \qquad c_1=\frac{\left(\frac{\frac{1}{3}(1-2q_0)}{\frac{2}{3}(1+q_0)-\varepsilon}\right)^{\frac{\varepsilon}{1-\varepsilon}}}{\left(\frac{\frac{1}{3}(1-2q_0)}{\frac{2}{3}(1+q_0)-\varepsilon}\right)^{\frac{1}{1-\varepsilon}}+\left(\frac{\frac{1}{3}(1-2q_0)}{\frac{2}{3}(1+q_0)-\varepsilon}\right)^{\frac{\varepsilon}{1-\varepsilon}}}\,,
\end{equation}
where the constant of integration is set such that $q(z=0)=q_0$. Using $\frac{d}{dt}=-H(1+z)\frac{d}{dz}$ and $h\equiv H/H_0$, the definition of the deceleration parameter $q$ (the second relation in \eqref{eq:CP_def}) can be rewritten as
\begin{equation}
    (1+z)h'(z)=(1+q)h\,.
\end{equation}
Substituting into the above equation the expression of $q$ from \eqref{eq:q_almost_LCDM} and integrating, one arrives at the form of the Hubble parameter evolution.
\begin{equation}\label{eq:almost_LCDM}
    h^2(z) = c_1(1+z)^3 + (1-c_1)(1+z)^{3\varepsilon}\,,
\end{equation}
 The cosmological solution \eqref{eq:almost_LCDM} is a General Relativistic cosmological solution that can be achieved within GR in the presence of a nonrelativistic fluid and another fluid with equation of state parameter given by Eq.\eqref{eq:DDE_phantom}. As a consistency check, one can verify that in the limit $\varepsilon\to0$, the constant $c_1$ reduces to $\frac{2}{3}(1+q_0)$, so that the cosmological solution \eqref{eq:almost_LCDM} smoothly reduces to the $\Lambda$CDM-like solution of the form \eqref{eq:LCDM}. The solution with small positive $\tilde{s}$ can thus be considered as an \emph{almost $\Lambda$CDM-like cosmic evolution/solution} specifically designed to accommodate a phantom crossing scenario. Dynamical dark energy in the context of almost $\Lambda$CDM-like cosmic evolutions, which are cosmographically very close to $\Lambda$CDM, has very recently been considered in \cite{Chakraborty:2025rvc}. 

\section{Cosmological field equations in $f(R)$ gravity}\label{sec:f(R)_cosmology}

$f(R)$ gravity can be derived from the following action \cite{Sotiriou:2008rp,DeFelice:2010aj}:
\begin{equation}\label{action}
    S = \frac{1}{2\kappa} \int \sqrt{-g} d^{4}x [f(R) + 2\mathcal{L}_{m}]\,,
\end{equation}
 where $\kappa = 8\pi G$, $G$ being Newton's gravitational constant, $g$ is the metric determinant, $f(R)$ is a function of the Ricci scalar $R$, $\mathcal{L}_m$ is the matter action. Varying the action with respect to the metric yields the field equations 
 \begin{equation}\label{field_eq}
G_{\mu\nu} \equiv R_{\mu\nu}-\frac{1}{2}g_{\mu\nu}R = \frac{\kappa T_{\mu\nu}}
{F(R)} + g_{\mu\nu}\frac{\left[f(R)-RF(R)\right]}{2F(R)} + \frac{\nabla_{\mu}\nabla_{\nu}F(R)-g_{\mu\nu}\square F(R)}
{F(R)}\,,
\end{equation}
where the prime denotes a derivative with respect to $R$, $T_{\mu\nu}$ is the energy momentum tensor and $F(R)=f'(R)$. In writing the field equation in the above manner, we have, naturally, assumed $F(R)\neq0$, otherwise, there would be no gravity. $f(R)$ theories of gravity propagate an extra scalar degree of freedom in the gravity sector, sometimes dubbed as a `curvaton', as made clear from the trace field equation
 \begin{equation}\label{trace_field_eq}
    R F(R) - 2f(R) + 3\square F(R) = \kappa T \,,   
\end{equation}
$f(R)=R$ is the trivial case for which the action reduces to GR, in which case the above equation becomes algebraic. Thus, there is no extra propagating scalar degree of freedom. The absence of ghost and tachyonic instability requires $f'(R)=F(R)>0$ and $f''(R)=F'(R)\geq0$ respectively \cite{Sotiriou:2008rp,DeFelice:2010aj}.
 
 For a Friedmann-Lema\^itre-Robertson Walker (FLRW) metric
 \begin{equation}\label{metric}
     ds^{2}=-dt^{2}+a^{2}(t)\left[\frac{dr^{2}}{\left(1-kr^{2}\right)}+r^{2}d\theta^{2}+r^{2}\sin^{2}\theta d\phi^{2}\right]\,,
 \end{equation}
 and in the presence of a perfect fluid $T^{\mu}_{\nu} = (-\rho, P, P, P)$, the modified Friedmann and Raychaudhuri equations can be neatly expressed as
\begin{subequations}\label{eq:f(R)_fieldeqs}
    \begin{eqnarray}
        3F\left(H^{2} + \frac{k}{a^{2}}\right) = \kappa \rho_{\rm eff} = \kappa(\rho + \rho_{\rm curv})\,,\label{eq:fried}\\
        -F\left(2\dot{H} + 3H^{2} + \frac{k}{a^{2}}\right) = \kappa P_{\rm eff} = \kappa(P + P_{\rm curv})\,,\label{eq:Raychoudhuri}
    \end{eqnarray}     
\end{subequations}
where $\rho,\,P$ are the fluid energy density and pressure, and the curvaton energy density and pressure are
\begin{subequations}
    \begin{eqnarray}\label{eq:curv}
        \kappa \rho_{\rm curv}&=&\frac{1}{2}(RF-f)-3H\dot{F}\label{eq:rho_curv}\\
        \kappa P_{\rm curv}&=&\ddot{F}+2H\dot{F}-\frac{1}{2}(RF-f)\label{eq:P_curv}
    \end{eqnarray}
\end{subequations}
Over dot represents the differentiation with respect to cosmic time throughout this work. We will confine our attention to the case of a constant equation of state parameter $w$ so that $P=w\rho$ and the fluid follows the continuity equation
\begin{equation}
    \dot{\rho} = - 3H(1+w)\rho \,.   
\end{equation}

One needs to be careful to formally define a dark energy equation of state parameter in modified gravity theories like $f(R)$. The dark energy equation of state has to be defined in such a way that the effective dark energy fluid, arising from the curvature degree of freedom, is separately conserved in the absence of any dark sector interaction \cite{Nesseris:2022hhc}. This is required for the equation of state parameter to make any physical sense. If one ignores $|\Omega_k|$ for the observable patch of the universe due to the inflation in the early universe, the cosmological field equations \eqref{eq:f(R)_fieldeqs} can be rearranged in the following form
\begin{subequations}
    \begin{eqnarray}\label{eq:f(R)_fieldeqs_new}
        3H^{2} &=& \kappa\rho_{\rm tot} = \kappa\rho + \kappa\rho_{\rm DE}\,,\label{eq:fried_new}
        \\
        -\left(2\dot{H} + 3H^{2}\right) &=& \kappa P_{\rm tot} = \kappa P_{\rm DE}\,,\label{eq:Raychoudhuri_new}
    \end{eqnarray}     
\end{subequations}
where the energy density and pressure of the \emph{effective curvature fluid}, which, now, is interpreted as the dark energy fluid, are defined as 
\begin{subequations}\label{eq:curv_fluid}
    \begin{align}
         \kappa\rho_{\rm DE}&=\frac{1}{2}(Rf'-f)-3H\dot{F}+3H^{2}(1-f')\,,\label{eq:DEed}\\
         \kappa P_{\rm DE}&=\ddot{f'}+2H\dot{f'}-\frac{1}{2}(RF-f)-(2\dot{H}+3H^2)(1-f')\,. \label{eq:DEp}
    \end{align}
\end{subequations}
The dark energy equation of state $w_{\rm DE} = \frac{P_{\rm DE}}{\rho_{\rm DE}}$ is 
\begin{equation}\label{eq:eos_curv}
    w_{\rm DE} = \frac{\ddot{f'}+2H\dot{f'}-\frac{1}{2}(Rf'-f)-(2\dot{H}+3H^2)(1-f')}{\frac{1}{2}(Rf'-f)-3H\dot{f'}+3H^{2}(1-f')}\,.
\end{equation}
In this form, the non-relativistic matter and the dark energy fluid are separately conserved
\begin{subequations}\label{eq:cons}
    \begin{eqnarray}
        && \dot\rho + 3H\rho = 0\,.\\
        && \dot\rho_{\rm DE} + 3H\rho_{\rm DE}(1+w_{\rm DE}) = 0\,.
    \end{eqnarray}
\end{subequations}
Using the field equations rewritten as in \eqref{eq:f(R)_fieldeqs_new}, the dark energy equation of state parameter can also be written as
\begin{equation}\label{eq:DE_eos}
    w_{\rm DE} = - \frac{2\dot{H}+3H^{2}}{3H^{2}-\kappa\rho} = \frac{H^{2}-\frac{R}{3}}{3H^{2}-\kappa\rho}\,.
\end{equation}
Utilizing the relation between the Ricci scalar and the deceleration parameter in Eq.\eqref{eq:Ricci_CP}, $w_{\rm DE}$ can be conveniently expressed as
\begin{equation}\label{eq:wDE}
 w_{\rm DE} = \frac{2q-1}{3-3\Omega_m} = \frac{2q-1}{3\Omega_{\rm DE}}\,,
\end{equation}
$\Omega_m\equiv\frac{\kappa\rho}{3H^2},\,\Omega_{\rm DE}\equiv\frac{\kappa\rho_{\rm DE}}{3H^2}$ being the standard density abundance parameter for the nonrelativistic fluid and the dark energy fluid (which, in this case, is the effective curvature fluid). The above expression will be used in Sec.\ref{sec:LCDM_f(R)} to plot the dark energy equation of state for different possible $\Lambda$CDM-mimicking $f(R)$ theories.

In rewriting the $f(R)$ cosmological field equations \eqref{eq:f(R)_fieldeqs} in the form \eqref{eq:f(R)_fieldeqs_new}, we have effectively reduced them to a form that is equivalent effectively to the two fluid scenario of General Relativistic cosmology as in Eq.\eqref{eqs:2_baro_fluid}, each of which is separately conserved. What this reveals is that, if an $f(R)$ cosmology in presence of a nonrelativistic fluid is to cosmographically mimic a General Relativistic cosmological solution, e.g. a $\Lambda$CDM-like cosmological solution \eqref{eq:LCDM} or an almost $\Lambda$CDM-like phantom crossing cosmological solution \eqref{eq:almost_LCDM}, then, physically speaking, the equation of state parameter of the effective curvature fluid \eqref{eq:eos_curv} has to behave as \eqref{eq:DDE} or \eqref{eq:DDE_phantom} respectively. This physical insight will prove useful in making a comparison of our cosmographic reconstruction approach with the earlier works, which we will explain in detail in Sec.\ref{subsec:comparison}. 

\section{A cosmographic approach to the reconstruction method in $f(R)$ gravity}\label{sec:recon_cosmo}
 
For the purpose of this work, which will follow a cosmographic approach, we strategically express the modified Friedmann and Raychaudhuri equation above as algebraic equations involving the cosmographic parameters, the matter density abundance parameter, as well as $f$ and its derivatives (scaled with respect to $H_0^2$ to make them dimensionless)
\begin{subequations}\label{eq:field_eqs}
    \begin{eqnarray}
    \Omega_m &=& -6h^2(-j+q+2\Omega_k+2)(f''H_0^2) + q f' + \frac{1}{6h^2}\left(\frac{f}{H_0^2}\right)\,,\label{eq:energy_eq}
    \\
    w \Omega_m &=& -12h^4(-j+q+2 \Omega_k+2)^2(f^{(3)}H_0^4) - 2h^2[2j + q(q+2\Omega_k+6)+s+2 \Omega_k+2](f'' H_0^2) \nonumber
    \\
    && - \frac{1}{3}(q-2(\Omega_k+1))f' - \frac{1}{6h^2}\left(\frac{f}{H_0^2}\right)\,.
    \end{eqnarray}
\end{subequations}
In the above $\{\Omega_m,\Omega_k\}=\lbrace\frac{\kappa\rho_m}{3H^2},\frac{k}{a^2 H^2}\rbrace$ are the standard matter density abundance parameters and curvature abundance parameters. 

Note that, had we been using a cosmographic series expansion in terms of the \emph{present-day values} of the cosmographic parameters e.g. $q_0,j_0,s_0$ etc, then the statement that the field equations become algebraic would only hold strictly at $z=0$. However, we remind the reader again that we are not using a cosmographic series expansion here, but merely the definitions of the standard cosmographic parameters\footnote{Replacing any appearance of an $\ddot{a}(t)$ by $-a(t)H_0^2 h^2(z)q(z)$, replacing any appearance of a $\frac{d^3a(t)}{dt^3}$ with $a(t)H_0^3 h^3(z)j(z)$, and so on. The differential nature of the field equations can be recovered back by explicitly substituting the definitions of the cosmographic parameters from \eqref{eq:CP_def} into the Eq.\eqref{eq:field_eqs}.}. Therefore, the field equations \eqref{eq:field_eqs}, as we have written above, are valid globally for all $z$.

The two field equations in Eq.\eqref{eq:field_eqs} can be combined to remove $\Omega_m$, resulting in a third-order homogeneous differential equation of $f(R)$ with respect to $R$ 
\begin{equation}\label{eq:master_eq_1}
    \mathcal{A}\,R^3 f^{(3)}(R) + \mathcal{B}\,R^2 f''(R) + \mathcal{C}\,R f'(R) + \mathcal{D}\,f(R) = 0\,,
\end{equation}
where
\begin{subequations}
   \begin{eqnarray}
       \mathcal{A} &=& (-j+q+2\Omega_k+2)^2 \,,
   \\
   \mathcal{B} &=& (1-q+\Omega_k)\left[j(3w+2)+q^2+q(-3 w+2 \Omega_k+6)+s-2(3w-1)(\Omega_k+1)\right]\,,
   \\
   \mathcal{C} &=& (1-q+\Omega_k)^2(3qw+q-2\Omega_k-2)\,,
   \\
   \mathcal{D} &=& 3 (w+1)(1-q+\Omega_k)^3\,.
   \end{eqnarray}
\end{subequations}
The equation \eqref{eq:master_eq_1} provides a new, \emph{cosmographic} approach to reconstruct the $f(R)$ gravity theory that admits any given General Relativistic cosmological solution $a(t)$. Recall that, as we detailed in Sec.\eqref{sec:cosmography}, for any General Relativistic cosmological solution involving two perfect fluids, which encapsulates most dark energy models where the dark energy component can be modelled as a perfect fluid, can be expressed by an algebraic relation $j=j(q)$. Hence, the snap parameter $s$ is not an independent cosmographic parameter but expressible in terms of $a,h,q,j$. The success of this new reconstruction method, therefore, relies on the condition that $R=R(h,q)$ must be invertible to ultimately obtain the expressions $a(R),h(R),q(R),j(R)$. Indeed, such invertibility conditions are a common aspect in all the reconstruction methods for $f(R)$ gravity that have been proposed so far (see, e.g. \cite{Dunsby:2010wg,He:2012rf,Choudhury:2019zod,Carloni:2010ph,Nojiri:2010oco,Nojiri:2009kx}).

There are two crucial differences between the existing reconstruction methods in $f(R)$ gravity and our so-called \emph{cosmographic} approach, as given by the reconstruction differential equation \eqref{eq:master_eq_1}, which must be mentioned. 

Firstly, unlike in the other approaches, our reconstruction differential equation is a homogeneous differential equation. This is because we have chosen to eliminate $\Omega_m$ to derive the reconstruction differential equation. The cost to pay is that our reconstruction equation is a third-order differential equation in $f(R)$, whereas in most of the other approaches (except in \cite{He:2012rf}) the reconstruction equation is a second-order differential equation in $f(R)$. A homogeneous differential equation can offer certain simplifications over an inhomogeneous differential equation while trying to find a solution, because one does not need to bother about the particular integral.

Secondly, all the coefficients arising in our reconstruction differential equation \eqref{eq:master_eq_1} are kinematic cosmographic quantities and the matter density $\rho_m$, or equivalently the matter density abundance parameter $\Omega_m$ does not explicitly enter into the reconstruction differential equation. In view of the cosmographic way of specifying a General Relativistic cosmological solution as detailed in Sec.\eqref{sec:cosmography}, Eq.\eqref{eq:master_eq_1} is specially equipped to reconstruct $f(R)$ gravity theories admitting such a given solution. Moreover, it can be recalled that $\Omega_{m0}$ cannot be measured from data in a model-independent manner. One can constrain $\Omega_{m0}$ using data only after assuming a model (e.g. a dark energy with equation of state $w_{\rm DE}=w_0+w_a(1-a)$; see the DESI DR2 paper \cite{DESI:2025zgx}). On the other hand, the present-day values of the cosmographic parameters can be constrained in a model-independent way straight from various data sets \cite{Rapetti:2006fv,Dialektopoulos:2023dhb,Mukherjee:2024wix,Rodrigues:2025tfg}. Therefore, if one wants to proceed in a model-independent way to reconstruct the underlying $f(R)$ gravity that is most suited to the data, then the reconstruction differential equation Eq.\eqref{eq:master_eq_1} is the way to go.

\section{The \emph{theory space}}\label{sec:theory_space}

Even though the analytical cosmographic reconstruction approach we have outlined in the previous section offers certain advantages over the earlier reconstruction frameworks provided in the literature, let us confess at this point that, just like with the other reconstruction methods, our reconstruction procedure is also prone to failing in uncovering the underlying gravity theory whenever the given cosmological solution gets a little complicated. The failure manifests as an inability to obtain a compact functional form $f(R)$, or even to derive the reconstruction differential equation. We will come across both these instances in this work.

The present section introduces the idea of portraying a given cosmological solution in a so-called \emph{theory space}, which is one of the main ideas of our paper. The underlying motivation is that, even if one cannot reconstruct the underlying $f(R)$ theory admitting a given cosmological solution $a(t)$, neither in the compact form nor even numerically, one can still get a glimpse of the nature of the underlying theory and draw some important qualitative conclusions. In a way, the theory space analysis is a systematic procedure to bypass the explicit reconstruction of $f(R)$ when the latter fails. This analysis sets apart the present work from earlier works dealing with the reconstruction of $f(R)$ gravity (or any modified gravity for that matter). 

Let us introduce the following dimensionless quantities 
\begin{equation}\label{eq:define_th_space}
    r \equiv \frac{R f'(R)}{f(R)} = \frac{d\ln f}{d\ln R} \,, \qquad\qquad m = \frac{R f''(R)}{f'(R)} = \frac{d\ln f'}{d\ln R} \,.
\end{equation}
The 2-dimensional space defined by these two quantities is what we call the \emph{theory space}, since the quantities depend only on the functional form of the $f(R)$ theory and nothing else. A given $f(R)$ theory corresponds to a curve or a family of curves in the theory space\footnote{Unless one considers a monomial theory $f(R)\propto R^n$ ($n=$constant), in which case it is a point in the theory space}. Conversely, any curve $m(r)$ in the theory space corresponds to a particular 2-parameter $f(R)$ form\footnote{The reason we say it will be a 2-parameter $f(R)$ is because any relation $m=m(r)$ gives a second order differential equation on $f(R)$; hence solving it will introduce two integration constants.}. Physically, the quantity $m$ is related to the mass of the scalaron, the scalar gravitational degree of freedom, and hence is related to observables like the post parametrized Newtonian parameters, effective gravitational coupling $\mu(a,k)$ characterizing the structure growth and the lensing parameter $\Sigma(a,k)$ affecting the gravitational lensing. These observables serve as model-independent characterizations of deviations from GR. The quantity $r$, on the other hand, can be interpreted as characterizing the appearance of an effective cosmological constant in the GR limit of an $f(R)$ (i.e. $f(R)$ linear in $R$), with $r$ being frozen to the value $r=1$ in the absence of any effective cosmological constant. Pure GR ($f(R)=R$) is given by the single point $\{r,m\}=\{1,0\}$, whereas GR with a cosmological constant ($f(R)=-2\Lambda+R$) is given by the line $m=0$. The notion of the theory space has been utilized to differentiate between classes of late time $f(R)$ models \cite{Amendola:2006we}\footnote{Although the authors did not term it as the `theory space'. Also, notice the different signature in the definition of $r$.}.

Some more elaborate details and explanations about the theory space, which may not be strictly necessary for understanding the core results of the paper, but can nonetheless be important for rigorous understanding of the concept of the theory space, are presented in appendix \ref{app:theory_space}.

\subsection{A cosmological solution in the theory space}

To investigate a given cosmic evolution/solution $a(t)$ as some kind of a \emph{flow} in the theory space, we proceed as follows. Firstly, with some manipulation, the main reconstruction equation \eqref{eq:master_eq_1} can be expressed in terms of $r,m(r),m'(r)$
\begin{equation}\label{eq:master_eq_2}
    \mathbb{A}\,r^2\,m(r)\,m'(r) + \mathbb{B}\,r^3\,m'(r) + \mathbb{C}\,r^2\,m'(r) + \mathbb{D}\,r\,m^2(r) + \mathbb{E}\,r\,m(r) + \mathbb{F}\,r + \mathbb{G} = 0\,,
\end{equation}
where
\begin{subequations}
    \begin{eqnarray}
        \mathbb{A} &=& (-j+q+2\Omega_k+2)^2\,,
        \\
        \mathbb{B} &=& - (-j+q+2\Omega_k+2)^2\,,
        \\
        \mathbb{C} &=& (-j+q+2\Omega_k+2)^2\,,
        \\
        \mathbb{D} &=& (-j+q+2\Omega_k+2)^2\,,
        \\
        \mathbb{E} &=& -j^2+3 j (-q w+w \Omega_k+w+2 \Omega_k+2)-q^3+q^2 (3 w-\Omega_k-6)\\\nonumber
        &&+q ((\Omega_k+1) (3 w+2 \Omega_k)-s)+(\Omega_k+1) (s-2 (3 w+1) (\Omega_k+1))
        \\
        \mathbb{F} &=& (-q+\Omega_k+1)^2 (3 q w+q-2 (\Omega_k+1))\,,
        \\
        \mathbb{G} &=& 3 (w+1) (-q+\Omega_k+1)^3\,.
    \end{eqnarray}
\end{subequations}
From Eq.\eqref{eq:master_eq_2}, one can determine the slope of the tangent to the theory curve $m(r)$ in the theory space
\begin{equation}\label{eq:m-r_slope}
    m'(r) = -\frac{\mathbb{D}\,r\,m^2(r) + \mathbb{E}\,r\,m(r) + \mathbb{F}\,r + \mathbb{G}}{\mathbb{A}\,r^2\,m(r) + \mathbb{B}\,r^3 + \mathbb{C}\,r^2}
\end{equation}
Now, notice that one can write
\begin{equation}
    \frac{dr}{dN} = r(m-r+1)\frac{\dot{R}}{HR} = r(m-r+1)\left(\frac{j-q-2}{1-q}\right)\,,
\end{equation}
where $N=\ln(a)$. Coupled with Eq.\eqref{eq:m-r_slope}, one can write
\begin{equation}
    \frac{dm}{dN} = m'(r)\frac{dr}{dN} = -r(m-r+1)\left(\frac{j-q-2}{1-q}\right)\left(\frac{\mathbb{D}\,r\,m^2(r) + \mathbb{E}\,r\,m(r) + \mathbb{F}\,r + \mathbb{G}}{\mathbb{A}\,r^2\,m(r) + \mathbb{B}\,r^3 + \mathbb{C}\,r^2}\right)\,.
\end{equation}
One then arrives at the following coupled system
\begin{subequations}\label{eq:autonomous}
    \begin{eqnarray}
       \frac{dr}{dN} &=& r(m-r+1)\left(\frac{j-q-2}{1-q}\right)\,,
       \\
       \frac{dm}{dN} &=& -r(m-r+1)\left(\frac{j-q-2}{1-q}\right)\left(\frac{\mathbb{D}\,r\,m^2 + \mathbb{E}\,r\,m + \mathbb{F}\,r + \mathbb{G}}{\mathbb{A}\,r^2\,m + \mathbb{B}\,r^3 + \mathbb{C}\,r^2}\right)\,,
       \\
       \frac{dq}{dN} &=& 2q^2 + q - j\,.
    \end{eqnarray}
\end{subequations}

The modified Friedmann equation \eqref{eq:energy_eq} determines a \emph{normalized} density abundance parameter
\begin{equation}\label{eq:energy_eq_mr}
    \frac{\Omega_m}{f/(6H^2)} = 1 + \frac{q\,r-2\,r\,m}{1-q+\Omega_k} - \frac{(3q-j)r\,m}{(1-q+\Omega_k)^2}\,.
\end{equation}

In general, the system \eqref{eq:autonomous} is not autonomous, because of the explicit appearance of $\Omega_k,\,j$ and $s$. However, in the late-time universe, one can take $|\Omega_k|\ll1$ in the observable patch of the universe as a generic prediction of inflation. All spatially flat cosmological solutions with two perfect fluids, which include cosmological solutions under a large number of General Relativistic dark energy models, can be expressed as a cosmographic condition $j=j(q)$, i.e. a curve in the statefinder plane; see the discussion below Eq.\eqref{eqs:2_baro_fluid}. In particular, this includes the $\Lambda$CDM-like cosmological solution $j=1$ and the simple toy model of the phantom crossing cosmological solution \eqref{cosm_cond_DESI} $j=1+3\epsilon(q-\frac{1}{2})$.  When $j=j(q)$ is known, one can use Eq.\eqref{s} to obtain $s=s(q)$. In this case, the system \eqref{eq:autonomous} becomes autonomous with the dimensionless dynamical variables $r,\,m,\,q$. 

The autonomous system is regular in the domain
\begin{equation}
    \mathcal{D} = \{(r,m,q)\in\mathbb{R}^3: q\neq1 \land \mathbb{A}(r,m,q)r^2m(r)+\mathbb{B}(r,m,q)r^3+\mathbb{C}(r,m,q)r^2\neq0\}
\end{equation}
In the domain where the autonomous system is regular, it defines a 3-dimensional phase flow. Physically speaking, the phase trajectories represent all possible cosmological solutions of the underlying $f(R)$ theory, which could have otherwise been reconstructed starting from the given cosmological solution $j=j(q)$. 

It is interesting that the dynamical system \eqref{eq:autonomous} is spanned by a set of quantities such that the \emph{kinematics} of the universe as given by the cosmographic parameter $q$, and the \emph{dynamics} of the underlying theory as given by the completely theory-dependent parameters $\{r,m\}$, arise as completely disjoint directions in the phase space $r$-$m$-$q$. Such a structure of the dynamical system is different from traditional dynamical system formulation for $f(R)$ gravity in terms of the Hubble-normalized dynamical variables, e.g. in \cite{Amendola:2006we,Carloni:2007br}. In the traditional approach, the quantities $r$ and $m$ are taken to be just auxiliary variables, whereas the Hubble normalized dynamical variables are usually a mixture of kinematics and dynamics by definition (i.e., can be expressed as a combination of $\{r,m,H,q,j\}$). In particular, one can compare with the work \cite{Chakraborty:2021jku}, which analyzed the phase space of $\Lambda$CDM-mimicking $f(R)$ models in terms of the phase space variables $\{x,\Omega,q\}=\lbrace\frac{f''\dot{R}}{f'H},\frac{\rho}{3f'H^2},q\rbrace$. We believe that the set $\{r,m,q\}$ allows for a better understanding of the underlying theory, in particular to assess its extent of deviation from GR. This is the basis of our subsequent theory space analysis.

From the autonomous system \eqref{eq:autonomous}, one may be tempted to perform a fixed point analysis. However, we will not pursue in that direction in this work. When one does not know beforehand the compact functional form of the underlying gravity theory\footnote{One actually \emph{never} know beforehand the underlying gravity theory exactly, but can only build different models and test its viability.}, typical questions of interest are how close or distant from GR are the underlying $f(R)$ theories admitting a given cosmological solution, and whether there is an attractor behaviour of GR in the past or future in the space of all such theories. To address such qualitative questions, we feel, rather than a fixed point analysis of the autonomous system \eqref{eq:autonomous}, analyzing the 2-dimensional \emph{non-autonomous} system spanned solely by the theory space variables $\{r,m\}$ is more useful.

Note that the $q$-equation decouples in the autonomous system for constant $j$, e.g. as in \eqref{eq:autonomous_LCDM}. For cosmological solutions corresponding to realistic late-time cosmological models, the deceleration parameter $q$ decreases monotonically along the course of cosmic evolution, starting from a value close to $\frac{1}{2}$ near matter domination and passing through a value $q_0\approx-0.55$ at $z\approx0$. This suggests that one can define a `flow' in the theory space $r-m$ by a \emph{non-autonomous} system, taking $-q$ as a proxy time variable:
\begin{subequations}\label{eq:nonautonomous}
    \begin{eqnarray}
        && \frac{dr}{d(-q)} = - \frac{dr/dN}{dq/dN} = r(m-r+1)\left[\frac{j(q)-q-2}{(1-q)(2q^2+q-j(q))}\right]\,,
        \\
        && \frac{dm}{d(-q)} = - \frac{dm/dN}{dq/dN} = r(m-r+1)\left[\frac{j(q)-q-2}{(1-q)(2q^2+q-j(q))}\right]\left(\frac{\mathbb{D}\,r\,m^2 + \mathbb{E}\,r\,m + \mathbb{F}\,r + \mathbb{G}}{\mathbb{A}\,r^2\,m + \mathbb{B}\,r^3 + \mathbb{C}\,r^2}\right)\,.\nonumber
        \\
        &&
    \end{eqnarray}
\end{subequations}
Below Eq.\eqref{eq:energy_eq_mr} we have pointed out that for a given spatially flat cosmological solution $j=j(q)$, the system \eqref{eq:autonomous} becomes autonomous. In this case, the corresponding non-autonomous system \eqref{eq:nonautonomous} can be solved by setting some initial conditions $\{r_{\rm in},m_{\rm in}\}$ at some value of $q$, which typically produces a curve in the theory space $r$-$m$. Since any curve in the theory space implies an $f(R)$ theory, the resulting curve, in a way, can be considered as a visual representation of the underlying $f(R)$ theory in the theory space. The cosmological evolution along decreasing $q$ induces a sense of directionality on this curve, which allows us to understand how the deviation of the underlying $f(R)$ from GR changes with time.

Since ``$-q$'' serves as a proxy time variable for the non-autonomous system \eqref{eq:nonautonomous}, and what we do using the non-autonomous system \eqref{eq:nonautonomous} is essentially to obtain numerical solutions $\{r(q),m(q)\}$, it is important to specify the domain of $q$ within which this can be done meaningfully:
\begin{align}
\mathcal{D} =& \{q\in\mathbb{R}: q\neq1 \land 2q^2-j(q)\neq0    \nonumber
\\
& \land \mathbb{A}(r(q),m(q),q)r^2(q)m(q)+\mathbb{B}(r(q),m(q),q)r^3(q)+\mathbb{C}(r(q),m(q),q)r^2(q)\neq0\}\,.
\end{align}
Any meaningful result from the numerical treatment of the system \eqref{eq:nonautonomous} can only be obtained within this domain of validity. It would be easier to express the domain of validity while we consider the specific cosmological solutions later in this paper. 

Notice that, in the theory space $m$-$r$, the region $m<0$ is a region of instability for the theory. Since $R=6H^2(1-q)>0$ for a late-time cosmic evolution, $m<0$ implies either $f'<0$ or $f''<0$, implying either a ghost instability or a tachyonic instability. Therefore, any solution curve lying on the region $m<0$ of the theory space necessarily implies a cosmological solution under a physically non-viable theory. Note, however, that the converse is not necessarily true; it is possible for a physically non-viable theory to lie on the $m>0$ region of the theory space, if it is plagued by both ghost and tachyonic instability simultaneously ($f'<0,\,f''<0$). 

The theory space analysis, as we have constructed above, is particularly suited to investigate the very important question of whether GR acts as a cosmological past or future attractor in a broader space of theories admitting a given cosmological solution. The latter question is an interesting one and has been occasionally addressed in relation to scalar-tensor theories \cite{Damour:1992kf,Damour:1993id,Mimoso:1998dn,Mimoso:1999ai,Saal:2012zb}, scalar-torsion theories \cite{Jarv:2015odu}, $f(Q)$ theories \cite{Guzman:2024cwa}.

Eq.\eqref{eq:master_eq_2}, in a way, is a modified reconstruction equation, which reconstructs the underlying theory, at least numerically, in the theory space as $m=m(r)$, rather than explicitly as $f=f(R)$. The subsequent theory space analysis, in principle, should succeed even when the explicit reconstruction of $f(R)$ fails. However, there is a slight limitation of the theory space analysis as compared to explicit reconstruction. Reconstructing the underlying theory as $f=f(R)$ allows one to obtain $\Omega_m(z)$ from the modified Friedmann equation \eqref{eq:energy_eq}, and hence $w_{\rm DE}(z)$ from Eq.\eqref{eq:wDE}, corresponding to a given cosmological solution $j=j(q)$. This is not possible if one rather reconstructs the theory as $m=m(r)$, because the corresponding Friedmann constraint \eqref{eq:energy_eq_mr} allows one to obtain only the quantity $\frac{\Omega_m}{f}$ as a whole, and not $\Omega_m$ separately. The knowledge of $\Omega_m(z)$ and $w_{\rm DE}(z)$ is sometimes of particular physical interest.

\section{Stability of the reconstructed solution}\label{sec:stability}

Reconstruction of a healthy theory free of ghost or tachyonic instability, starting from a given cosmological solution, ensures the \emph{existence} of the given solution within the solution space of the generic class of theories under consideration, e.g. $f(R)$ in our case. But the mere existence of a solution does not ensure that it can actually arise as a stable solution within the solution space. An unbounded growth of perturbation in the geometry of a solution usually indicates that the solution can be achieved with very fine-tuned initial conditions. Such a solution is usually deemed unnatural. When a cosmological solution in a theory is unnatural, even though it may exist for fine-tuned initial conditions, it does not present enough motivation to do a detailed study of inhomogeneous perturbation (e.g. structure formation etc) on this solution.

Stability with respect to small homogeneous and isotropic perturbation of the form $a(t)\to a(t)[1+\varepsilon(t)]$ serves as a first criterion for a physically acceptable cosmological solution. The history of such studies dates back to the era of Einstein, when Einstein's static solution in GR was deemed unphysical because it was found to be unstable by Eddington under simple homogeneous and isotropic small perturbations of the form $a_0\to a_0[1+\varepsilon(t)]$. Later, it was demonstrated that such a solution can be stable within the framework of $f(R)$ gravity \cite{Boehmer:2007tr}\footnote{At this point, it is worth mentioning that the study of homogeneous and isotropic perturbation serves only as a \emph{first check, and not a guarantee} for the stability of a solution. For example, although it was shown in \cite{Boehmer:2007tr} that the Einstein static solution can be stable in some $f(R)$ gravity under homogeneous and isotropic perturbation, later it was shown that it is unstable under more general inhomogeneous perturbation \cite{Seahra:2009ft}.}. Homogeneous and isotropic perturbations have been investigated for interesting cosmological solutions in the framework of various modified gravity theories, e.g. $f(R)$ gravity \cite{Barrow:1983rx}, $f(R,G)$ gravity \cite{delaCruz-Dombriz:2011oii}, $f(R,\phi)$ gravity \cite{Faraoni:2005vk} and $f(Q)$ gravity \cite{Guzman:2024cwa}. In all these works, the cosmological solution that was considered can be expressed as a fixed point in the phase space, e.g. de-Sitter solutions or the FLRW solution with a single fluid. Such studies have not been performed on a more realistic cosmological solution that cannot be expressed as a fixed \emph{point}, but rather is a phase \emph{trajectory} in the phase space. 

In this section, following the idea employed in \cite{Bamba:2013fha}, we attempt to provide a framework for investigating the stability of a given cosmological solution in the generic solution space of the underlying $f(R)$ theory with respect to homogeneous and isotropic perturbations. Unlike \cite{Bamba:2013fha}, which could reconstruct the compact functional form of $f(R)$ corresponding to simple nonsingular bouncing solutions, in our case, we cannot find compact functional forms. The underlying $f(R)$ theories are implicitly specified as the solutions of the corresponding reconstruction differential equations.

Since we are focusing on a spatially flat FLRW solution, it suffices for our purpose to consider a small homogeneous and isotropic perturbation in the form $h(N) \to h(N) + \delta h(N)$. We substitute this into the master equation Eq.\eqref{eq:master_eq_1}. Written this way, $h(N)$ represents the given cosmological solution, and $\delta h(N)$ represents a time-dependent homogeneous and isotropic perturbation on it. After some straightforward but rather tedious manipulations, one arrives at a perturbation evolution equation of the form
\begin{equation}\label{eq:ptbn_f(R)}
 a_3(r,m,q)\frac{d^3 \delta h}{dN^3} + a_2(r,m,q)\frac{d^2 \delta h}{dN^2} + a_1(r,m,q)\frac{d\delta h}{dN} + a_0(r,m,q)\delta h = 0\,.  
\end{equation}

The completely generic forms of the coefficients $a_0,\,a_1,\,a_2,\,a_3$ are very complicated, and we do not give them here. However, once a certain background cosmological solution $j=j(q)$ is specified, the forms become somewhat more presentable, and in the two particular examples that will be considered below, namely the $\Lambda$CDM-mimicking cosmological solution $j=1$ and the phantom crossing cosmological solution $j=1+3\epsilon\left(q-\frac{1}{2}\right)$, the explicit forms of the coefficients will be presented.

Note that the $f(R)$ field equations are fourth order in metric, i.e., third order in the Hubble parameter, which is consistent with the third order perturbation equation \eqref{eq:ptbn_f(R)} in terms of $\delta h$. Physically, one can make sense of the perturbation equation \eqref{eq:ptbn_f(R)} in the following way. For any given $f(R)$, one can imagine the solution space to be a phase space spanned by the quantities $\{h(N), h'(N), h''(N)\}$\footnote{In essense this is equivalent to the phase space formulation of $f(R)$ gravity with respect to the quantities $\{H,R,\dot{R}\}$ \cite{deSouza:2007zpn}}. Within this phase space, if one sets some initial conditions at $N=N_0$, say $\{h(N_0), h'(N_0), h''(N_0)\}=\{h_0,h'_0,h''_0\}$ and solves the system, one gets a particular cosmological solution. Whether solutions arrived at by setting slightly different initial conditions produce a trajectory that converges towards the original trajectory is addressed by the perturbation equation \eqref{eq:ptbn_f(R)}. In other words, the stability of the solution vis-a-vis Eq.\eqref{eq:ptbn_f(R)} determines whether the effect of small deviations in the initial conditions is washed away, or becomes more significant, along the course of the evolution. The latter situation usually hints towards a fine-tuning issue.

For a given background solution $\{r(N),m(N),q(N)\}$, the coefficients of the homogeneous and isotropic perturbation equation \eqref{eq:ptbn_f(R)} generically has time-dependent coefficients. It is not easy to infer statements about the global stability of a solution trajectory, and one typically needs sophisticated mathematical tools such as Lyapunov functions or Jacobi stability analysis \cite{Boehmer:2010jqg,Harko_2016}. Such an analysis is beyond the scope of this paper, and we do not pursue them here. Instead, we perform a crude stability analysis as follows. 

Note that, to obtain a numerical solution of the system \eqref{eq:autonomous} mathematically, one needs to set an initial condition $\{r_{\rm in},m_{\rm in},q_{\rm in}\}$ at some cosmic moment $N_0$. At any given cosmic moment $N_0$, the perturbation equation \eqref{eq:ptbn_f(R)} can be considered an equation with constant coefficients. The condition for stability of the corresponding solution requires that all the roots of the corresponding characteristic equation $a_3x^3+a_2x^2+a_1x+a_0=0$ must have a negative real part. This is ensured by the Routh-Hurwitz stability criteria:
\begin{equation}\label{coeff_conds_N0}
    \frac{a_0}{a_3}\bigg\vert_{N0}>0, \quad \frac{a_1}{a_3}\bigg\vert_{N0}>0, \quad \frac{a_2}{a_3}\bigg\vert_{N0}>0 \quad \frac{a_1 a_2}{a_0 a_3}\bigg\vert_{N0} > 1 \,,
\end{equation}
which strictly holds for an equation with constant coefficients. Since $a_i$s are functions of $r,m,q$, the condition \eqref{coeff_conds_N0} provides some restrictions on the values of $\{r(N_0),m(N_0),q(N_0)\}$. If the choice of the initial conditions $\{r_{\rm in},m_{\rm in},q_{\rm in}\}$ obeys those restrictions, then the resulting solution can be inferred to be stable with respect to small homogeneous and isotropic perturbation $\delta h(N_0)$ at $N=N_0$.

If we now demand a particular trajectory (i.e., a given solution) to be stable with respect to small perturbation $\delta h$, irrespective of at what $N$-value the initial condition was set, then a condition like \eqref{coeff_conds_N0} can be demanded to be held at \emph{all} values of $N$:
\begin{equation}\label{coeff_conds}
    \frac{a_0}{a_3}>0, \quad \frac{a_1}{a_3}>0, \quad \frac{a_2}{a_3}>0 \quad \frac{a_1 a_2}{a_0 a_3} > 1 \,,
\end{equation}
The above condition singles out a 3-dimensional volume within the 3-dimensional phase space $r$-$m$-$q$. For the trajectories that cross outside the region specified by the condition \eqref{coeff_conds} at some point during their course of evolution, the corresponding cosmological solutions become unstable with respect to a small homogeneous and isotropic perturbation $\delta h(N)$. Such a trajectory cannot be considered a physically stable cosmological solution. Ideally, a physically acceptable stable cosmological solution must be a trajectory that remains \emph{always} within the 3-dimensional volume defined by the condition \eqref{coeff_conds}, so that it reduces the effect of the variation in initial conditions and draws nearby trajectories towards it.\footnote{The idea is essentially the same as in the scenario of inflationary slow-roll attractor solutions or tracker quintessence solutions.}

Notice that the stability analysis in this case becomes more straightforward as compared to the one, say, in \cite{Bamba:2013fha}, because we have devised our reconstruction method in such a way that the reconstruction differential equation is homogeneous. This is precisely the reason why one can analyze the stability conditions here from the signs of the coefficients of the characteristic equation. Also, note that utilizing the set $\{r,m\}$ instead of $\{f,f',f'',f'''\}$ allows for investigating the stability in the reduced 3-dimensional phase space spanned by $r,m,q$, or, as we will see, even in the 2-dimensional theory space $\{r,m\}$.

\section{$\Lambda$CDM-mimicking $f(R)$ theories}\label{sec:LCDM_f(R)}

Let us first establish the applicability of our framework by applying it to $\Lambda$CDM-mimicking $f(R)$ models. At this point, let us clarify that what we are after is the family of $f(R)$ theories that give rise to a cosmological evolution of the form \eqref{eq:LCDM}. As we will explicitly show in this section, the particular identification $\{\Omega_{m0},\Omega_{\rm{DE}0}\}=\lbrace\frac{2}{3}(1+q_0),\frac{1}{3}(1-2q_0)\rbrace$ holds only for the General Relativistic $\Lambda$CDM model, but \emph{not} for the $\Lambda$CDM-mimicking $f(R)$ models. 

For $j=1$, the second relation in \eqref{eq:CP_rel} gives $s=-(2+3q)$. Substituting $j=1,\,s=-(2+3q),\,w=0,\,\Omega_k=0$ in Eq.\eqref{eq:master_eq_1} gives 
\begin{equation}\label{eq:master_eq_1_LCDM}
    -(q+1)^2 R^3 f^{(3)}(R)+(q+1) (q+2) (q-1) R^2 f''(R)+(2-q) (q-1)^2 R f'(R)+3 (q-1)^3 f(R)=0\,.
\end{equation}
The next step is to express $q$ as a function of $R$. For the $\Lambda$CDM-like cosmic evolution \eqref{eq:LCDM}, one can calculate that
\begin{equation}\label{eq:dec_LCDM}
    q(z) = -1 + (1+z)\frac{h'(z)}{h(z)} = \frac{1}{2} - \frac{1}{2}\left(\frac{1-2q_0}{h^2}\right)\,,
\end{equation}
inverting which, one can get
\begin{equation}
    h^2 = \frac{1-2q_0}{1-2q}\,.
\end{equation}
Then we have 
\begin{equation}\label{eq:R_LCDM}
    R=6(\dot{H} + H^2) = 6H_0^2 h^2 (1-q) = 6H_0^2(1-2q_0)\left(\frac{1-q}{1-2q}\right)\,.
\end{equation}
For the sake of compactness, let us denote the constant $H_0^2(1-2q_0)$ by a constant $\Lambda$. Then the above expression of $R(q)$ can be easily inverted to obtain
\begin{equation}
    q(R) = \left(\frac{R-6\Lambda}{2R-6\Lambda}\right)\,.
\end{equation}
Substituting this back into Eq.\eqref{eq:master_eq_1_LCDM}, we get the reconstruction differential equation for $\Lambda$CDM-mimicking $f(R)$
\begin{equation}\label{eq:recon_LCDM_1}
    6 (R-3 \Lambda ) (R-4 \Lambda )^2 f^{(3)}(R) + (5 R-18 \Lambda ) (R-4 \Lambda ) f''(R) - (R-2 \Lambda ) f'(R) + f(R) = 0\,.
\end{equation}
The General Relativistic $\Lambda$CDM solution corresponds to an $f(R)$ where the correction term over the Einstein-Hilbert term is just a cosmological constant, i.e. $f(R)=-2\Lambda+R$. One can check by direct substitution that this is a trivial solution of the above equation, thus identifying the constant $\Lambda=H_0^2(1-2q_0)$ that we defined before with the cosmological constant. However, this is not the only possible $\Lambda$CDM-mimicking $f(R)$.

To find the generic solution to the reconstruction differential equation \eqref{eq:recon_LCDM_1}, it proves easier to define the variables
\begin{equation}\label{def:x_g}
    x = -3 + \frac{R}{\Lambda}\,, \qquad \qquad g(x) = \frac{f(R)}{\Lambda}\,.
\end{equation}
Note that, for the $\Lambda$CDM-like cosmic evolution \eqref{eq:LCDM}, $q$ varies monotonically from $\frac{1}{2}$ to $-1$. From Eq.\eqref{eq:R_LCDM}, therefore, one always has $R>3\Lambda$, or equivalently $x>0$. In terms of $g$ and $x$, the reconstruction equation for $\Lambda$CDM-mimicking $f(R)$ can be written as
\begin{equation}\label{eq:recon_LCDM_2}
    6 x (x-1)^2 g^{(3)}(x)+(5 x-3) (x-1) g''(x)-(x+1) g'(x)+g(x)=0\,.
\end{equation}
The General Relativistic cosmological solution $f(R)=-2\Lambda+R$, is given by $g(x)=1+x$, and one can check by direct substitution that it trivially satisfies the above equation. In fact, this is the only possible polynomial solution of Eq.\eqref{eq:recon_LCDM_2}. The generic solution of Eq.\eqref{eq:recon_LCDM_2} comes in terms of hypergeometric functions. The exact form is too complicated and not suitable for any practical purposes ; so it is not presented here.

\subsection{Comparison with earlier reconstruction methods}\label{subsec:comparison}

There have been a few works before on reproducing the $\Lambda$CDM model within the framework of $f(R)$ gravity \cite{Dunsby:2010wg,He:2012rf,Choudhury:2019zod}. There is a subtle difference between the earlier approaches and our \emph{cosmographic approach}; so a comparative discussion is in order. As we have seen in Eq.\eqref{eq:LCDM}, for a $\Lambda$CDM-like cosmic evolution, kinematically, the coefficient of $(1+z)^3$ in the expression of $h^2(z)$ comes out to be $\frac{2}{3}(1+q_0)^3$ once the initial conditions $\{h(z=0),q(z=0)\}=\{1,q_0\}$ is imposed. In \cite{Dunsby:2010wg,He:2012rf,Choudhury:2019zod}, the authors reconstruct the $f(R)$ based on a cosmological evolution of the form
\begin{equation}\label{eq:LCDM_GR}
    h^2 = \Omega_{m0}(1+z)^3 + \Omega_{\Lambda0}\,.
\end{equation}
The above way of specifying the $\Lambda$CDM evolution fixes $\frac{2}{3}(1+q_0)$ to be precisely equal to $\Omega_{m0}$. In other words, this implies that it is only the nonrelativistic matter that scales as $\sim(1+z)^3$. Recall that, the $f(R)$ field equations can be written in the form of the standard General Relativistic Friedmann equation with the nonrelativistic fluid and the effective curvature fluid. The approach starting from Eq.\eqref{eq:LCDM_GR} enforces that, whatever is the underlying form of $f(R)$, the entirety of the effective curvature fluid must behave as an effective cosmological constant and thus must scale totally as $\sim(1+z)^0$. In effect, this approach particularly rejects the possibility that some part of the curvature fluid can also scale as $\sim(1+z)^3$, while still reproducing a $\Lambda$CDM-mimicking cosmology. 

On the contrary, our approach is based on the cosmographic condition $j=1$ (or equivalently the cosmic evolution of the form \eqref{eq:LCDM}) itself, and \emph{not} \eqref{eq:LCDM_GR}. That way, it does not impose the restriction $\frac{2}{3}(1+q_0)=\Omega_{m0}$ a-priori. Therefore, our approach admits the possibility that, in general, the curvature fluid can scale partially as $\sim(1+z)^0$ and partially as $\sim(1+z)^3$. At the same time, the cosmic evolution remains still indistinguishable from the General Relativistic $\Lambda$CDM model at the background level. In other words, our approach acknowledges the possibility that the effective curvature fluid can behave more like the unified dark fluid \eqref{eq:DDE}, while still giving rise to the $\Lambda$CDM-like cosmic evolution \eqref{eq:LCDM}. Mathematically, however, our cosmographic approach also yields a hypergeometric form for the reconstructed $f(R)$ just like the earlier approaches, ensuring consistency.

The fact that the curvature degree of freedom can scale partially as $\sim(1+z)^3$ and partially as $\sim(1+z)^0$ for different solutions to the reconstruction differential equation \eqref{eq:recon_LCDM_1} can be made apparent by calculating $\Omega_{m}$ explicitly. Imposing $j=1$, $\Omega_k=0$ and using the expressions of $h(z),q(z)$ from Eqs.\eqref{eq:LCDM},\eqref{eq:dec_LCDM}, one can compute the expression for $\Omega_{m}$ explicitly from the first of the field equations \eqref{eq:field_eqs}:
\begin{equation}\label{eq:Omega_m}
    \Omega_m = -6H_0^2 (1+q_0)(1+z)^3 f'' + \left(\frac{(1+q_0)(1+z)^3}{h^2}-1\right) f' + \frac{1}{6H^2}f\,.
\end{equation}
which, in general, is \emph{not} equal to $2(1+q_0)(1+z)^3/3h^2$, as one would expect if it was only the nonrelativistic matter fluid contributing to the $\sim(1+z)^3$ part of the scaling. Therefore, in general, a fraction of the effective curvature fluid in a $\Lambda$CDM-mimicking $f(R)$ model also scales as $\sim(1+z)^3$, with the fraction varying for different $\Lambda$CDM-mimicking $f(R)$ models. As a consistency check, one can verify that, if one imposes $f(R)=-2\Lambda+R$, $R=6H^2(1-q)$ and $\Lambda=H_0^2(1-2q_0)$ (see Eq.\eqref{eq:R_LCDM} and the sentence below it), one gets $\Omega_m = \frac{\frac{2}{3}(1+q_0)}{h^2}(1+z)^3$, in which case one can say that the fraction of the effective curvature fluid that scales as $\sim(1+z)^3$ is zero, and the entirety of the effective curvature fluid is effectively reduced to a cosmological constant. 

In particular, the present-day value of the density abundance parameter for the nonrelativistic fluid for a $\Lambda$CDM-mimicking $f(R)$ model is given by
\begin{equation}
    \Omega_{m0} = -6H_0^2 (1+q_0) f''_0 + q_0 f'_0 + \frac{1}{6H_0^2}f_0\,,
\end{equation}
which, in general, is \emph{not} equal to the $\frac{2}{3}(1+q_0)$ that one would expect for $\Omega_{m0}$ in the General Relativistic $\Lambda$CDM model (or for the particular $\Lambda$CDM-mimicking $f(R)$ for which the entirety of the curvature fluid is equivalent to a cosmological constant in an FLRW background). Again, only if one imposes $f(R)=-2\Lambda+R$, $R_0=6H_0^2(1-q_0)$ (from Eq.\eqref{eq:R_LCDM}) and $\Lambda=H_0^2(1-2q_0)$, one gets $\Omega_{m0}=\frac{2}{3}(1+q_0)$, which is simply the case of GR. 

Since $\Omega_m$ depends on $f,f',f''$ (Eq.\eqref{eq:Omega_m}), it is easy to see from Eq.\eqref{eq:wDE} that the effective curvature fluid will in general have a dynamical equation of state parameter for $\Lambda$CDM-mimicking $f(R)$ theories, with different $w_{\rm DE}(z)$ for different $\Lambda$CDM-mimicking $f(R)$s. This is in line with our earlier interpretation that our approach admits the possibility that the effective curvature fluid can behave like a unified dark fluid \eqref{eq:DDE} in general, while still being \emph{kinematically} $\Lambda$CDM-mimicking (i.e., reproducing $j(z)=1$).

\subsection{Numerical reconstruction}\label{subsec:num_rec}

\begin{figure}[H]
    \centering
    
    \begin{subfigure}[b]{0.45\linewidth}
    \includegraphics[width=\linewidth]{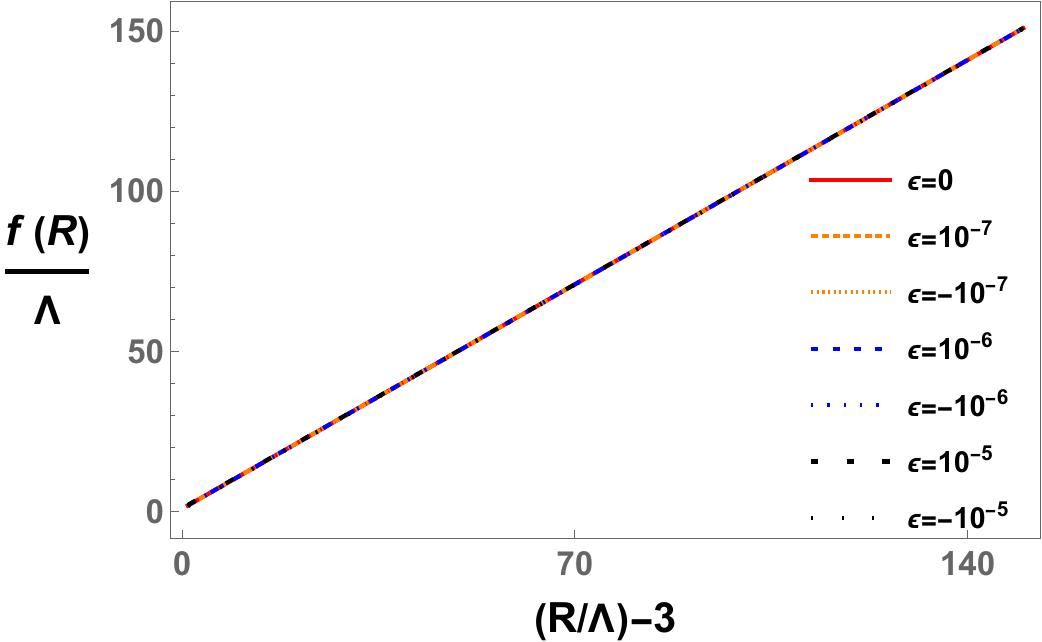}
     \caption{}
    \label{fig:f(R)_plot}
    \end{subfigure}
    \hspace{1.6mm}
    \vspace{0.5cm}
    \begin{subfigure}[b]{0.45\linewidth}
    \includegraphics[width=\linewidth]{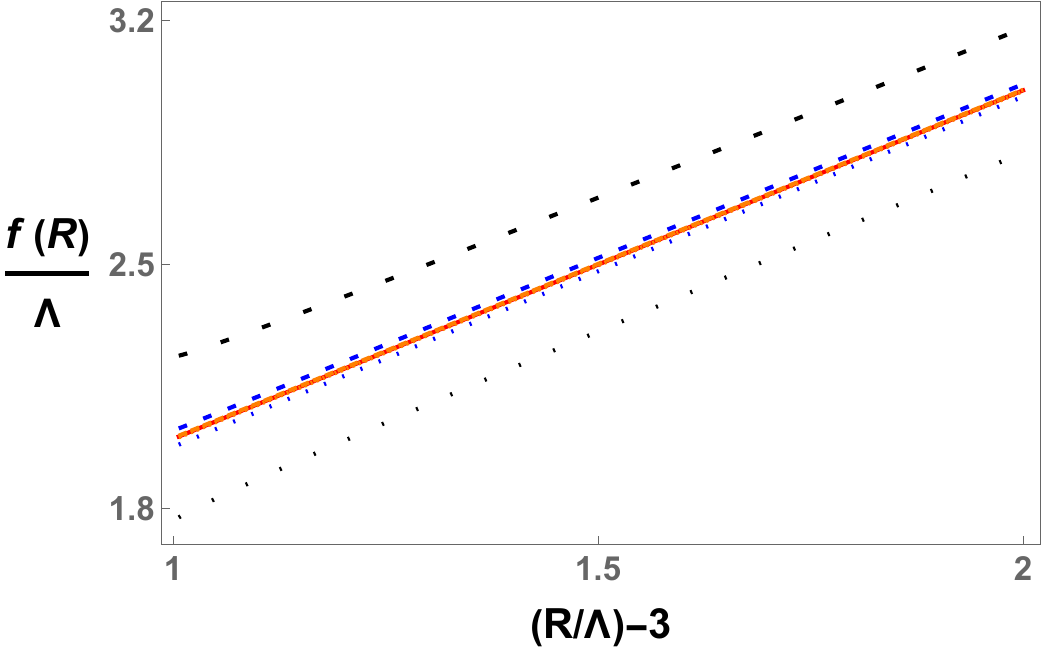}
     \caption{}
        \label{fig:f(R)_plot_zoomed_1}
    \end{subfigure}
    
    \begin{subfigure}[b]{0.45\linewidth}
    \includegraphics[width=\linewidth]{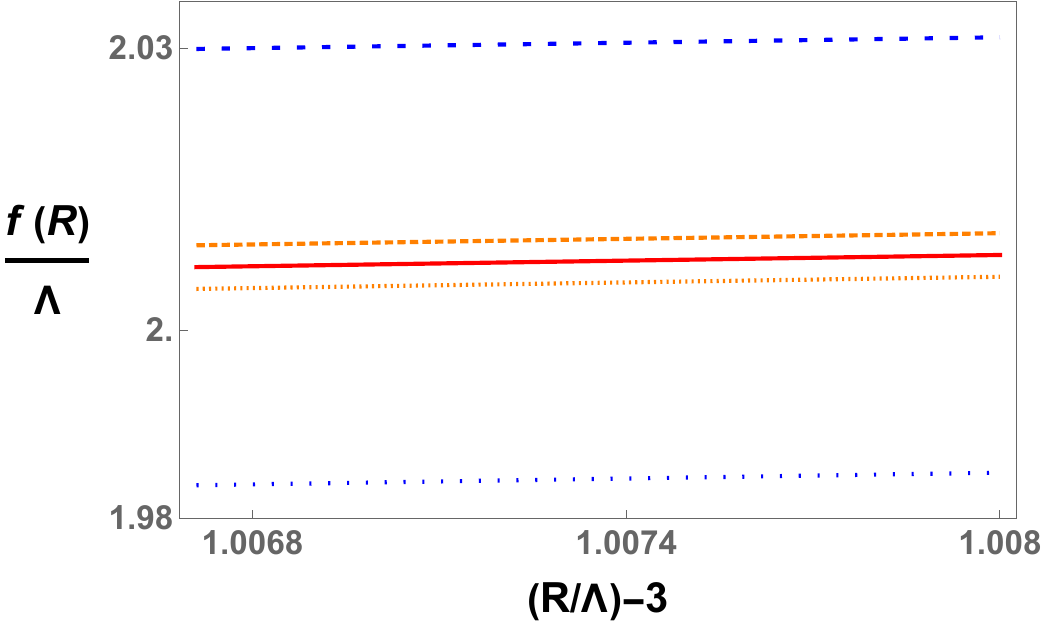}
     \caption{}
        \label{fig:f(R)_plot_zoomed_2}
    \end{subfigure}
    \hspace{1.6mm}
    \vspace{0.5cm}
    \begin{subfigure}[b]{0.45\linewidth}
    \includegraphics[width=\linewidth]{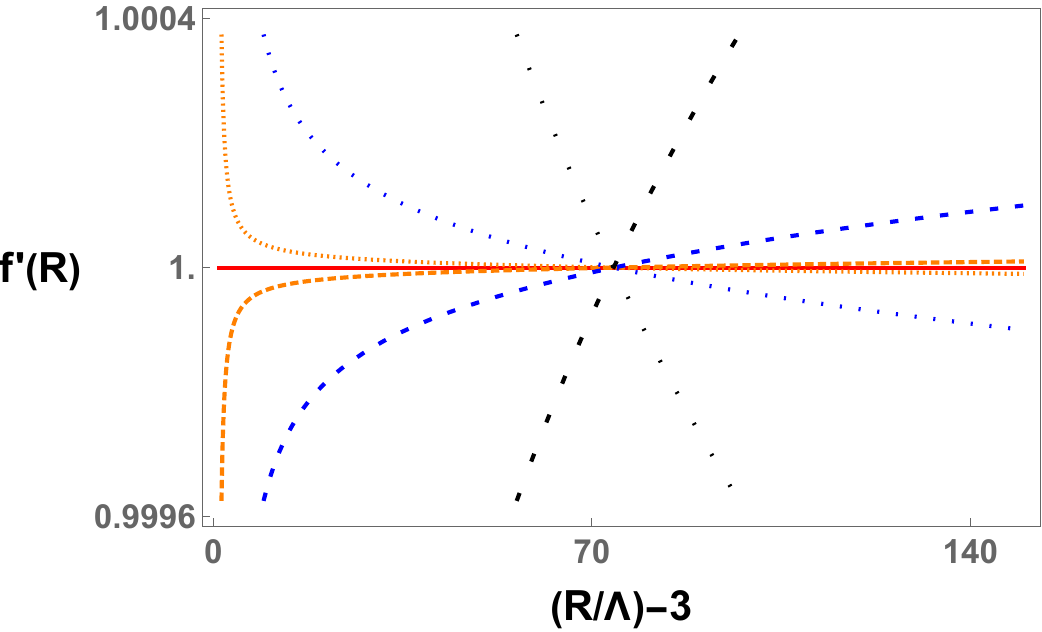}
     \caption{}
        \label{fig:f'(R)_plot}
    \end{subfigure}
    
    \begin{subfigure}[b]{0.45\linewidth}
    \includegraphics[width=\linewidth]{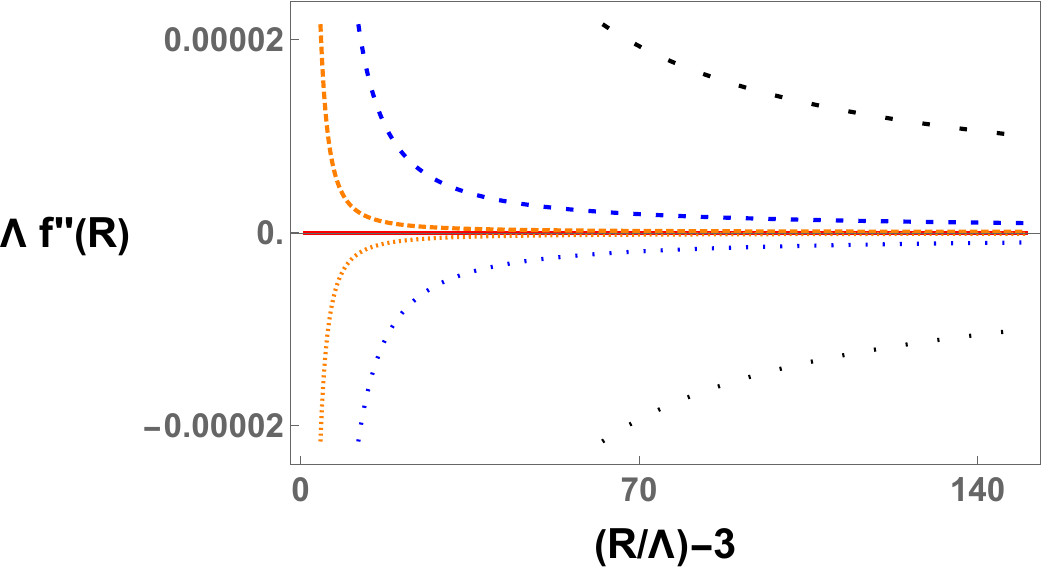}
     \caption{}
        \label{fig:f''(R)_plot}
    \end{subfigure}
    \hspace{1.6mm}
    \vspace{0.5cm}
    \begin{subfigure}[b]{0.45\linewidth}
    \includegraphics[width=\linewidth]{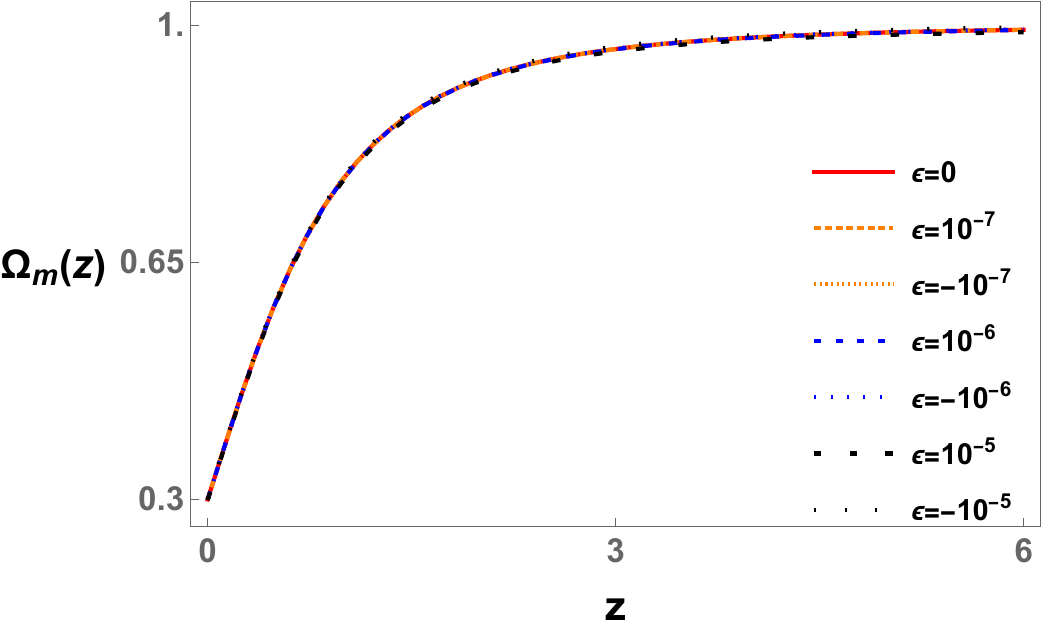}
     \caption{}
        \label{fig:Omega_m_plot}
    \end{subfigure}
    
    \begin{subfigure}[b]{0.45\linewidth}
    \includegraphics[width=\linewidth]{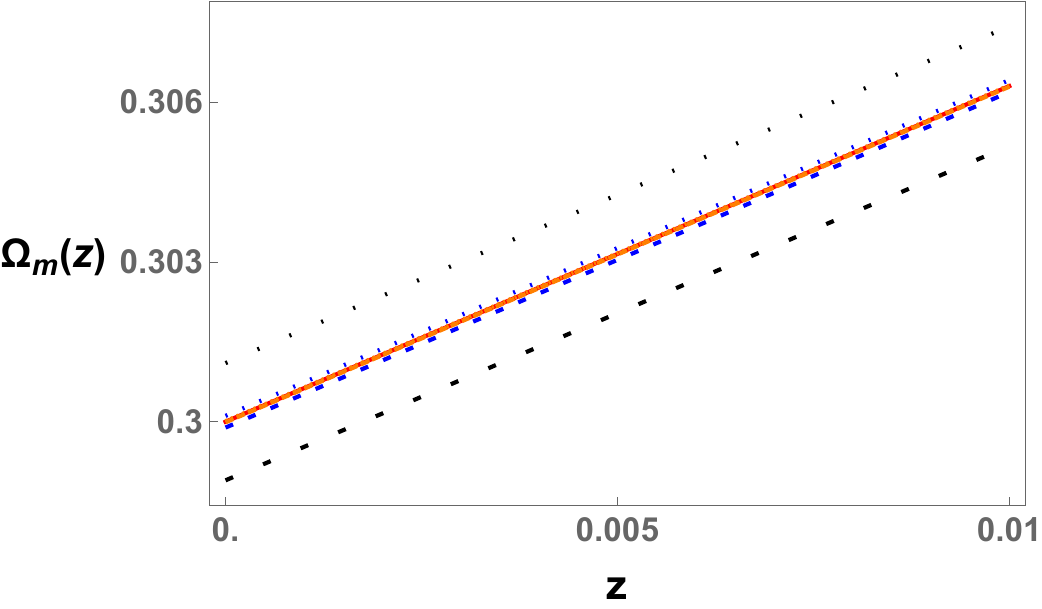}
     \caption{}
        \label{fig:Omega_m_plot_zoomed_1}
    \end{subfigure}
    \hspace{1.6mm}
    \begin{subfigure}[b]{0.45\linewidth}
    \includegraphics[width=\linewidth]{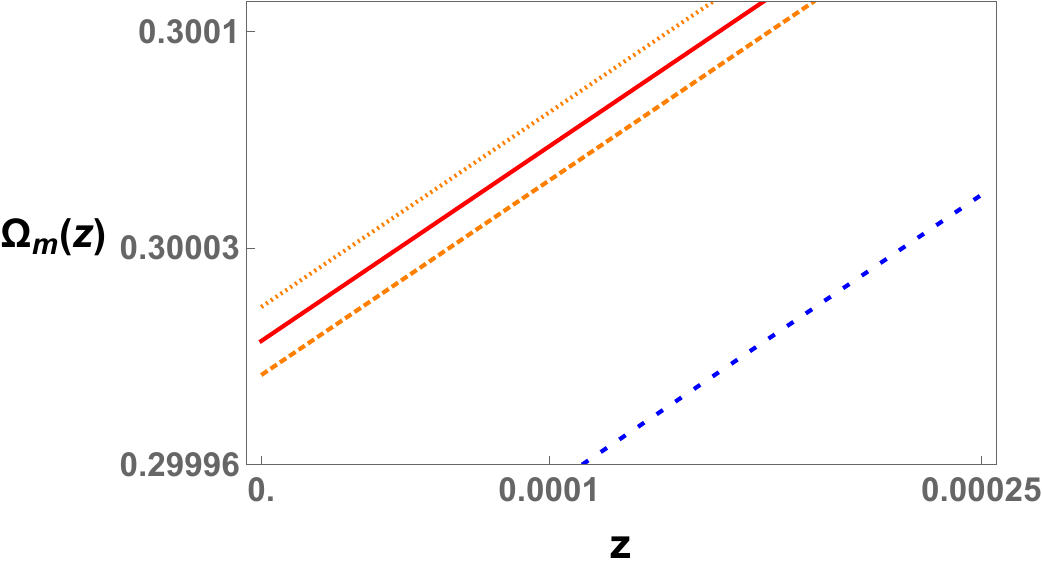}
     \caption{}
        \label{fig:Omega_m_plot_zoomed_2}
    \end{subfigure}
    
    \caption{(\ref{fig:f(R)_plot}),(\ref{fig:f(R)_plot_zoomed_1}),(\ref{fig:f(R)_plot_zoomed_2}) $f(R)$, (\ref{fig:f'(R)_plot}) $f'(R)$, (\ref{fig:f''(R)_plot}) $f''(R)$ and (\ref{fig:Omega_m_plot}),(\ref{fig:Omega_m_plot_zoomed_1}),(\ref{fig:Omega_m_plot_zoomed_2}) $\Omega_m$ corresponding to different $\Lambda$CDM-mimicking $f(R)$ models that are obtained from solving equation \eqref{eq:recon_LCDM_2}. The initial conditions are $f(R)=(-2+10^4\epsilon)\Lambda+R,\,f'(R)=1+10^2\epsilon,\,f''(R)=\frac{\epsilon}{\Lambda}$. $\epsilon$ is a dimensionless smallness parameter parametrizing small deviations from GR at a high redshift of $z_{\rm in}=6.0316$ (at which $q_{\Lambda\rm{CDM}}=0.49$). The red, dashed orange, dotted orange, dashed blue, dotted blue, dashed black, dotted black curves correspond to $\epsilon=0,10^{-7},-10^{-7},10^{-6},-10^{-6},10^{-5},-10^{-5}$ respectively. In the plots, $\Lambda$ denotes the constant quantity $H_0^2(1-2q_0)$, which is identified with the cosmological constant only for GR (the red curve). $\epsilon$ values are taken to be small enough that the $\Lambda$CDM-mimicking $f(R)$s remain very close to GR and the $\Omega_m$ evolution remains almost the same as the General Relativistic $\Lambda$CDM model, as is apparent from Figs.(\ref{fig:f(R)_plot}) and (\ref{fig:Omega_m_plot}). However, there are actually distinct $\Lambda$CDM-mimicking $f(R)$ solutions, giving rise to distinct evolution of $\Omega_m$. This can be made apparent by zooming more and more into the corresponding figures (Figs.(\ref{fig:f(R)_plot_zoomed_1}),(\ref{fig:f(R)_plot_zoomed_2}) and Figs.(\ref{fig:Omega_m_plot_zoomed_1}),(\ref{fig:Omega_m_plot_zoomed_2})). Differences in $f'(R)$ and $f''(R)$ are more prominent, as evident from Figs.(\ref{fig:f'(R)_plot}) and (\ref{fig:f''(R)_plot}).}
    \label{fig:LCDM_mimicking_f(R)s}
\end{figure}

The plots for several $\Lambda$CDM-mimicking $f(R)$s that can be numerically solved from the reconstruction differential equation \eqref{eq:recon_LCDM_2}, as well as $f'(R)$, $f''(R)$ and $\Omega_m$ are shown in Fig.\ref{fig:LCDM_mimicking_f(R)s}. 

The range for $x$ taken in Figs.\ref{fig:f(R)_plot},\ref{fig:f'(R)_plot},\ref{fig:f''(R)_plot} is motivated as follows. Recall that for the General Relativistic $\Lambda$CDM cosmology, the constant $\Lambda=H_0^2(1-2q_0)$ that we have defined can actually be identified with the cosmological constant, so that one can write (using \eqref{eq:lcdm})
\begin{equation}
    \Lambda_{\rm GR}=H_0^2(1-2q_0)=H^2(1-2q)\,,
\end{equation}
where the subscript `GR' signifies that the relations hold only for the General Relativistic $\Lambda$CDM model. Correspondingly one gets
\begin{equation}
    x_{\rm GR} = \frac{3}{1-2q}\,, \quad \text{with} \quad x_{\rm GR}(q\to0.49)=150 \quad \text{and} \quad x_{\rm GR}(q\to-0.99)=1.00671\,.
\end{equation}
We do not expect the $\Lambda$CDM-mimicking $f(R)$s to be too different from GR, as that would involve significant deviations in the perturbation signature and the variation of the effective gravitational coupling, both of which are highly constrained. Therefore, we take the range of $x$ to be $[1.00671,150]$ for the plots. For a $\Lambda$CDM-like evolution \eqref{eq:LCDM}, the $q=0.49$ occurs at a redshift value $z_{\rm in}=6.0316$. The plot for $\Omega_{m}$ is done in the range $z\in[0,6.0316]$.

The initial conditions used for the numerical solution are chosen such that at the redshift $z_{\rm{in}}=6.0316$, when $q=0.49$, the $\Lambda$CDM-mimicking $f(R)$ models deviate slightly from GR, with the small deviations parametrized by a dimensionless smallness parameter $\epsilon$. The precise initial conditions are mentioned in the caption.

A notable feature is revealed in the above figures. $f'(R)>0$ and $f''(R)>0$ are usually taken to be the conditions for the absence of the ghost and the tachyonic instability in $f(R)$ gravity. None of the $\Lambda$CDM-mimicking $f(R)$s remain free from this instability for all values of $R$. We have explored several different ways in which the deviation from GR can be parametrized using a dimensionless smallness parameter $\epsilon$, and this same feature seems to persist. Although we cannot draw a generic conclusion from here since there are infinite ways small deviations from GR can be parametrized, and of course, we have not explored them all. Nonetheless, it does cast a doubt on the feasibility of mimicking $\Lambda$CDM-like evolution \eqref{eq:LCDM} with $f(R)$ gravity. 

\begin{figure}[H]
    \centering
    \begin{subfigure}[b]{0.45\linewidth}
    \includegraphics[width=1.28\linewidth]{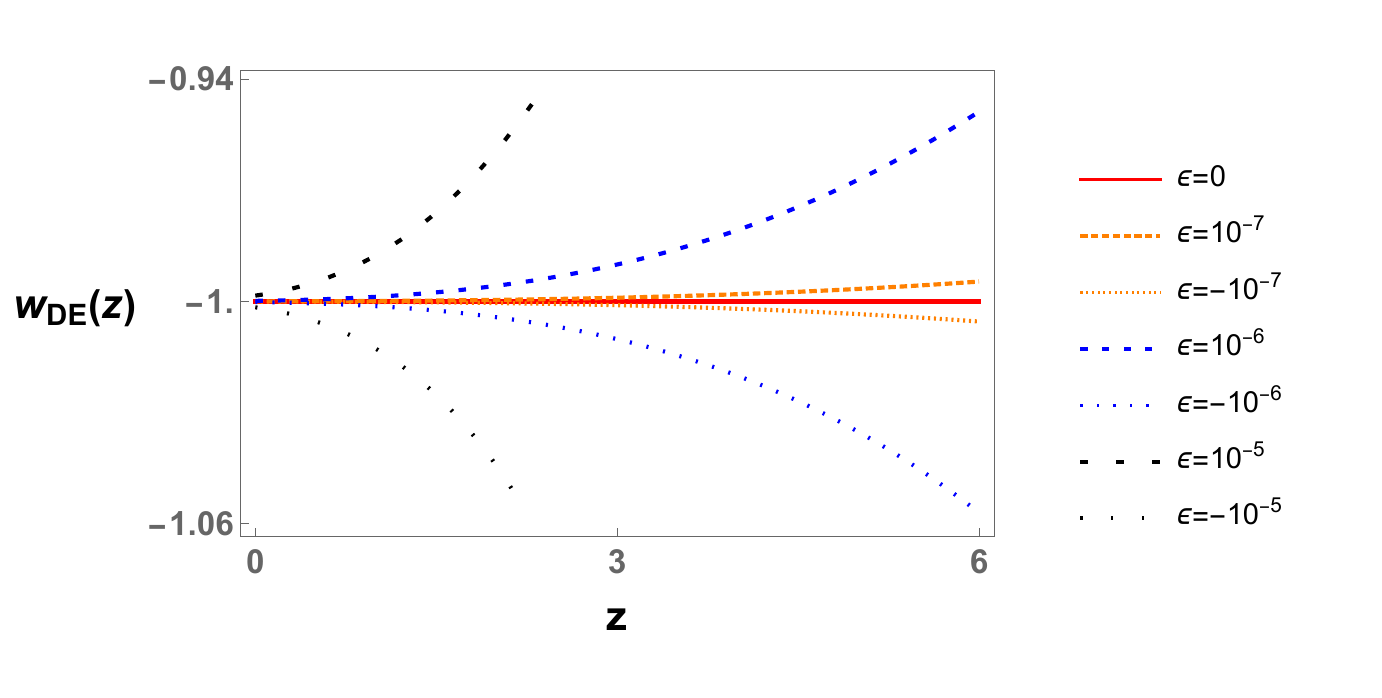}
     \caption{}
     \label{fig:wDE plot}
    \end{subfigure}
    \hfill 
    \begin{subfigure}[b]{0.45\linewidth}
    \includegraphics[width=0.95\linewidth]{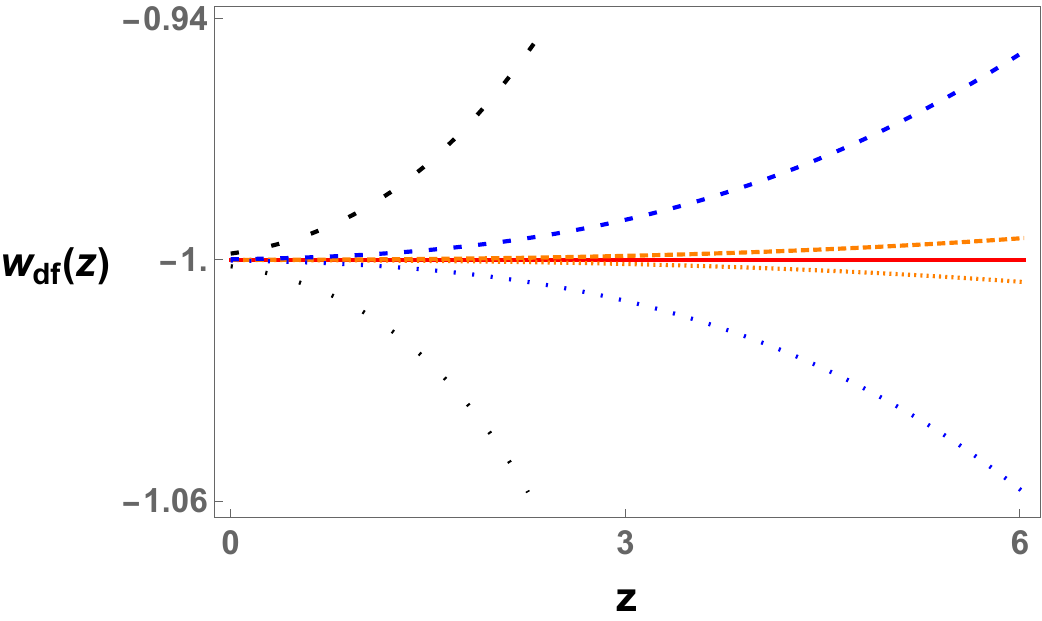}
     \caption{}
     \label{fig:wdf plot}
    \end{subfigure}
    \caption{Panel \ref{fig:wDE plot} shows $w_{\rm DE}(z)$ versus $z$ corresponding to different $\Lambda$CDM-mimicking $f(R)$ models that are obtained by numerically solving the reconstruction differential equation \eqref{eq:recon_LCDM_2}. The red, dashed orange, dotted orange, dashed blue, dotted blue, dashed black, dotted black curves correspond to $\epsilon=0,10^{-7},-10^{-7},10^{-6},-10^{-6},10^{-5},-10^{-5}$ respectively. Panel \ref{fig:wdf plot} shows the evolution of the unified dark fluid equation of state $w_{\rm df}(z)$ (\eqref{eq:DDE}) for different values of $w_0$, that are chosen to coincide with the values of $w_{\rm DE}(z=0)$ for the respective plots of panel \ref{fig:wDE plot}.}
    \label{fig:wDEvsz}
\end{figure}

The behaviours of $w_{\rm DE}(z)$ for different $f(R)$ cosmologies cosmographically equivalent to $\Lambda$CDM (i.e., reproducing the cosmographic constraint $j(z)=1$), which we are calling as \emph{$\Lambda$CDM-mimicking $f(R)$ theories} throughout this work, can be calculated using Eq.\eqref{eq:wDE}, and are shown in Fig.~\ref{fig:wDE plot}. Although $q(z)$ is unique (Eq.\eqref{eq:dec_LCDM}) for the $\Lambda$CDM-like evolution \eqref{eq:LCDM}, $\Omega_m(z)$ is different for different such models (see Eq.\eqref{eq:Omega_m}), which gives rise to different evolution for $w_{\rm DE}(z)$ according to Eq.\eqref{eq:wDE}. 

The interpretation of the $f(R)$ cosmological field equations as effective two fluid scenarios of general relativity leaves two physical possibilities compatible with the $f(R)$ cosmology cosmographically mimicking the $\Lambda$CDM; the effective curvature fluid can behave either as an effective cosmological constant or as an effective unified dark fluid. Different $\Lambda$CDM-mimicking models generally give rise to a unified dark fluid scenario as demonstrated in Fig.\eqref{fig:wdf plot}. Fig.\eqref{fig:wdf plot} is a plot of the unified dark fluid equation of state parameter $w_{\rm df}$ (Eq.\eqref{eq:DDE}) for six different values of $w_0$. The values of $w_0$ for the plots in Fig.\ref{fig:wdf plot} are chosen such that they coincide with values of $w_{\rm DE}(z=0)$ for the corresponding plots in Fig.\ref{fig:wDE plot}. 
There is only one possibility for the effective curvature fluid to behave as a cosmological constant, the horizontal red curve in the middle of each of the panels. In all other cases, the effective curvature fluid is physically similar to the corresponding unified dark fluid scenario, while still giving rise to a $\Lambda$CDM-mimicking cosmology. Even very small deviations in the initial conditions at $z=z_{in}=6.0316$ can lead to very different evolution for the corresponding equation of state parameter, as shown in Fig.\ref{fig:wDEvsz}. However, the differences in the density parameters $\Omega_{m}(z)$ and $\Omega_{\rm DE}(z)=1-\Omega_m(z)$ are hardly noticeable, and the redshift profiles of $H(z)$ are identical among every mimicking model. 

\subsection{Theory space analysis}

$\Lambda$CDM-mimicking $f(R)$ theories arise as solutions of the hypergeometric differential equation \eqref{eq:recon_LCDM_2}, and their generic form cannot be expressed in a compact functional form. Earlier works following different reconstruction methods \cite{Dunsby:2010wg,He:2012rf,Choudhury:2019zod} have arrived at the same conclusion, and focused mostly on trying to understand the behaviours of the $\Lambda$CDM-mimicking $f(R)$ theories employing numerical solutions and associated plots. We have also done the same in the subsection \ref{subsec:num_rec}. The analysis of the equation of state parameter in the subsection \ref{subsec:num_rec}, in particular, helped us understand how the effective curvature fluid behaves for the $\Lambda$CDM-mimicking $f(R)$ theories. The theory space analysis helps us visualize how different $\Lambda$CDM-mimicking $f(R)$ theories deviate from GR along the course of cosmic evolution. This, in turn, helps us answer the question whether GR behaves as a cosmological past or future attractor in the space of all such $\Lambda$CDM-mimicking theories. As we will show below, even though the $\Lambda$CDM-mimicking $f(R)$ theories have a rather complicated non-compact form, the above question can be addressed rather elegantly by considering their flow in the theory space.


For the $\Lambda$CDM-mimicking $f(R)$ theories, the autonomous system \eqref{eq:autonomous} can be simplified by substituting $\Omega_k=0,\,w=0,\,j=1,\,s=-(2+3q)$:
\begin{subequations}\label{eq:autonomous_LCDM}
    \begin{eqnarray}
       \frac{dr}{dN} &=& -r(m-r+1)\left(\frac{1+q}{1-q}\right)\,,
       \\
       \frac{dm}{dN} &=& \frac{r \left(m^2 (q+1)^2+m (q+1) \left(-q^2-2 q+1\right)-(2-q) (1-q)^2\right)+3 (1-q)^3}{(1+q) (1-q) r}\,,\label{eq:autonomous_LCDM_m}
       \\
       \frac{dq}{dN} &=& (2q-1)(q+1)\,.
    \end{eqnarray}
\end{subequations}

One can that $q=\frac{1}{2}$ acts as a repelling plane and $q=-1$ acts as an attracting plane in the 3-dimensional phase space $r$-$m$-$q$. Also, note that these two planes cannot be crossed by any trajectory, since $dq/dN=0$ at $q=1/2,-1$\footnote{Note that the dynamical system is formally ill-defined on the plane $q=-1$, due to the existence of a pole in \eqref{eq:autonomous_LCDM_m}. Mathematically, this can be regularized by redefining the phase space time variable as $N\to\tilde{N}:d\tilde{N}=\frac{dN}{1+q}$, and rewriting the dynamical system in terms of the new time variable $\tilde{N}$. This change of variable is justified because, during a $\Lambda$CDM-like cosmic evolution, $1+q>0$. Such a change of time does not alter the \emph{qualitative} dynamics of the system.}. The entire relevant cosmological dynamics occurs in between these two planes, with all the $\Lambda$CDM-mimicking trajectories originating from the plane $q=\frac{1}{2}$ in the asymptotic past and approaching the $q=-1$ plane in the asymptotic future. In other words, since the $\Lambda$CDM cosmographic condition $j(z)=1$ is already built within the autonomous system \eqref{eq:autonomous_LCDM}, all the phase trajectories connecting the two planes $q=\frac{1}{2}$ and $q=-1$ are possible $\Lambda$CDM-mimicking $f(R)$ cosmologies\footnote{The phase trajectories in the region $q>\frac{1}{2}$ or $q<-1$ of the phase space correspond to other possible cosmological solutions of the generic hypergeometric $f(R)$ theory that arise as the solution of the reconstruction equation. An $f(R)$ theory that satisfies the reconstruction equation \eqref{eq:recon_LCDM_2} must definitely admit $\Lambda$CDM-like cosmological solutions, but those need not necessarily be the only possible type of cosmological solutions that it admits.}.

In particular, for the General Relativistic $\Lambda$CDM model with the correction term over the Einstein-Hilbert term being constant, $f(R)=-2\Lambda+R$, one has
\begin{eqnarray}\label{eq:GR}
    r\vert_{\rm GR} = \frac{R}{-2\Lambda+R} = \frac{6H^2(1-q)}{-2H^2(1-2q) + 6H^2(1-q)} = \frac{3q-3}{q-2}\,, \quad m\vert_{\rm GR} = 0\,,
\end{eqnarray}
where at the second step of the first equality we have replaced $\Lambda=H^2(1-2q)$, a direct consequence of the field equations of the General relativistic $\Lambda$CDM model \eqref{eq:lcdm} (assuming $|\Omega_k|\ll1$). $r\vert_{\rm GR}$ is a monotonically increasing function of time, increasing monotonically from the value $r\vert_{\rm GR}(q\to1/2)=1$ to the value $r\vert_{\rm GR}(q\to-1)=2$ with the $\Lambda$CDM evolution \eqref{eq:LCDM}. The General Relativistic $\Lambda$CDM solution is, therefore, represented by the line $m=0$ in the theory space, with any deviation from this line signifying deviation from GR.

Consider, within the 3-dimensional phase space spanned by $r,m,q$, the curve given by
\begin{equation}\label{eq:GR_curve}
    \lbrace\mathcal{C}(r,q)\equiv r-\frac{3q-3}{q-2}=0,m=0\rbrace
\end{equation}
It can be checked using the dynamical equations \eqref{eq:autonomous} that 
\begin{equation}
    \frac{d\mathcal{C}}{dN}\bigg\vert_{\mathcal{C}=0,m=0} = \left[\frac{\partial\mathcal{C}}{\partial r}\frac{dr}{dN} + \frac{\partial\mathcal{C}}{\partial q}\frac{dq}{dN}\right]_{\mathcal{C}=0,m=0} = 0\,, \quad \frac{dm}{dN}\bigg\vert_{\mathcal{C}=0,m=0} = 0\,,
\end{equation}
which proves that the curve given by Eq.\eqref{eq:GR_curve} is an actual phase trajectory in the phase space. Given Eq.\eqref{eq:GR}, we can identify this curve as the cosmological solution associated with the General Relativistic $\Lambda$CDM model, with all other curves representing cosmological solutions associated with different $\Lambda$CDM-mimicking $f(R)$ models. 

\subsubsection{The non-autonomous system analysis}

For the $\Lambda$CDM-mimicking $f(R)$ theories, the non-autonomous system \eqref{eq:nonautonomous} can be simplified by substituting $\Omega_k=0,\,w=0,\,j=1,\,s=-(2+3q)$:
\begin{subequations}\label{eq:nonautonomous_LCDM}
    \begin{eqnarray}
        && \frac{dr}{d(-q)} = - \frac{dr/dN}{dq/dN} = \frac{r(m-r+1)}{(1-q)(2q-1)}\,,
        \\
        && \frac{dm}{d(-q)} = - \frac{dm/dN}{dq/dN} = -\frac{r \left(m^2 (q+1)^2 - m(q+1) \left(q^2+2q-1\right) - (2-q)(1-q)^2\right)+3 (1-q)^3}{(q+1)^2 (1-q) (2q-1) r}\,.\nonumber
        \\
        &&
    \end{eqnarray}
\end{subequations}
The domain of validity of $q$ in which the system \eqref{eq:nonautonomous_LCDM} is regular is 
\begin{equation}
    \mathcal{D} = \{q\in\mathbb{R}:q\neq1 \land q\neq\frac{1}{2} \land q\neq-1 \land r(q)\neq0\}\,.
\end{equation}
The system \eqref{eq:nonautonomous_LCDM} can be solved to find meaningful results within a range of $q$ where the above validity condition is always satisfied. For the $\Lambda$CDM-like cosmic evolution, the deceleration parameter $q$ varies monotonically from $1/2$ to $-1$, so that the non-autonomous system \eqref{eq:nonautonomous_LCDM} is valid within the range $-1<q<1/2$. The value $q=1$ is never reached in a $\Lambda$CDM-like cosmic evolution. Later on, we will solve the system within a range $-0.55\leq q\leq0.49$, and as we will see, in this range we will always have $r(q)>1$ for any solution; the singular surface $r(q)=0$ is never crossed. Therefore, we can safely say that the system \eqref{eq:nonautonomous_LCDM} remains regular in the domain of $q$ that we consider, and the subsequent results remain meaningful.

\begin{figure}[H]
    \centering
    \begin{subfigure}[b]{0.49\linewidth}
    \includegraphics[width=\linewidth]{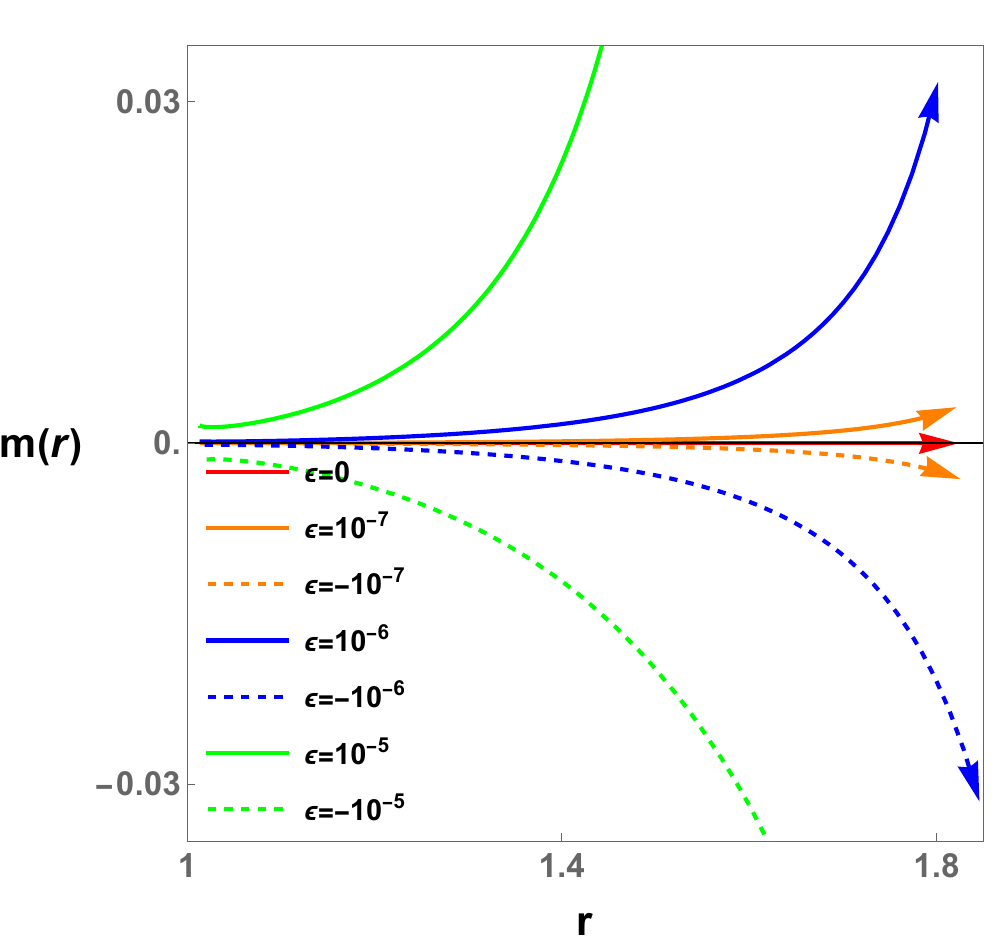}
     \caption{}
     \label{fig:m-r plot_zoomed}
    \end{subfigure}
    \hfill
    \begin{subfigure}[b]{0.49\linewidth}
    \includegraphics[width=\linewidth]{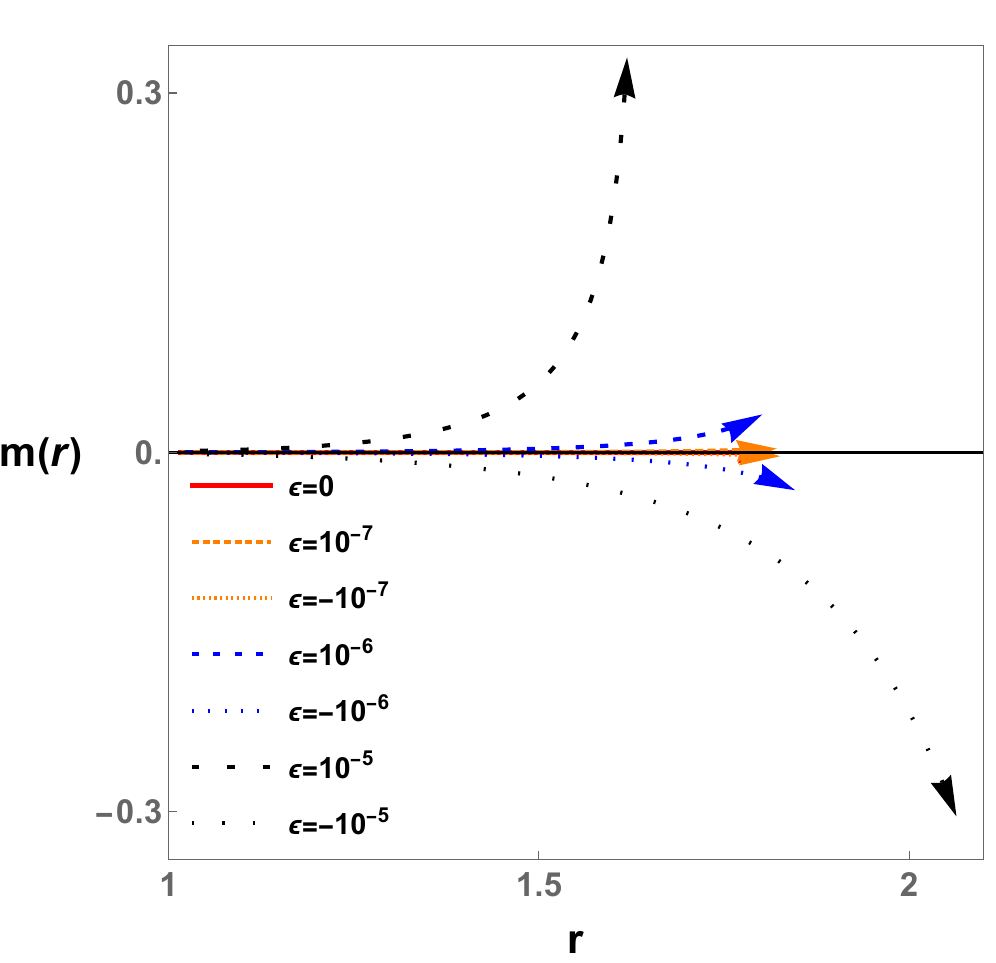}
     \caption{}
     \label{fig:m-r plot}
    \end{subfigure}
    \caption{Cosmological dynamics of $\Lambda$CDM-mimicking $f(R)$ theories in Figure~\ref{fig:LCDM_mimicking_f(R)s}, portrayed in the $\{r,m\}$ theory space, where $r=Rf'/f$ and $m=Rf''/f'$. The curves are numerical solutions of the reconstruction equation~\eqref{eq:recon_LCDM_2}, starting at $z_{\rm in}=6.0316$ with initial conditions $f(R)=(-2+10^4\epsilon)\Lambda+R$, $f'(R)=1+10^2\epsilon$, $f''(R)=\epsilon/\Lambda$, and evolving to $z=0$. The red, dashed orange, dotted orange, dashed blue, dotted blue, dashed black, dotted black curves correspond to $\epsilon=0,10^{-7},-10^{-7},10^{-6},-10^{-6},10^{-5},-10^{-5}$ respectively, with arrows indicating the direction of cosmic time. The $\Lambda$CDM model is the line $m(r)=0$. The present-day values ${r_0,m_0}$ are at the curve tips. Only positive $\epsilon$ yields non-pathological mimicking models. Panel \ref{fig:m-r plot} is a zoomed-out version of panel \ref{fig:m-r plot_zoomed}.}
    \label{fig:LCDM_mimicking_m(r)s}
\end{figure}

It is interesting to demonstrate the deviation from GR as a flow in the theory space for the particular $\Lambda$CDM-mimicking $f(R)$ theories depicted in Fig.\ref{fig:LCDM_mimicking_f(R)s}. Notice that $\{r,m\}$ can be related with $\{x,g\}$ defined in \eqref{def:x_g} as
\begin{equation}
    r = (3+x)\frac{g'(x)}{g(x)}\,, \qquad m = (3+x)\frac{g''(x)}{g'(x)}\,.
\end{equation}\label{eq:g(x) to m(r)}
One can then use the same numerical solutions $g(x)$ plotted in Fig.~\ref{fig:LCDM_mimicking_f(R)s} to make a parametric plot in the $m-r$ plane, which shows the corresponding solution curves $m(r)$ in the theory space. Fig.\ref{fig:LCDM_mimicking_m(r)s} shows how the corresponding $\Lambda$CDM-mimicking $f(R)$ theories deviate from GR during the period $q=0.49$ to today ($q=-0.55$). To produce the parametric plot, the following limits have been used\footnote{We remind the reader that the justification behind taking $x=x_{\rm GR}$ is that we do not want the underlying theory to deviate too much from GR.}
\begin{equation*}
    \{x_{\rm min},x_{\rm max}\} \approx \{x_{\rm GR}(q=0.49),x_{\rm GR}(q=-0.55)\}\,.
\end{equation*}

Notice that although Fig.\ref{fig:LCDM_mimicking_f(R)s} could already show the existence of different possible $\Lambda$CDM-mimicking theories other than GR, it could not really show how the deviations from GR of such theories evolve with time. The latter is nicely portrayed in the theory space of Fig.\ref{fig:LCDM_mimicking_m(r)s}. Fig.\ref{fig:LCDM_mimicking_m(r)s} reveals that the solution curves obtained corresponding to $\epsilon<0$ lie in the region $m<0$ in the theory space, signalling the existence of instability of the corresponding theories. On the contrary, the theory curves obtained corresponding to $\epsilon>0$ lie in the region $m>0$. Combined with Fig.\eqref{fig:f''(R)_plot}, which shows the corresponding theories have $f''>0$ for their entire range of $R$, one can conclude that the $\Lambda$CDM-mimicking $f(R)$ theories obtained corresponding to $\epsilon>0$ are indeed physically viable.

\begin{figure}[H]
    \centering
    \begin{subfigure}[b]{0.49\linewidth}
    \includegraphics[width=\linewidth]{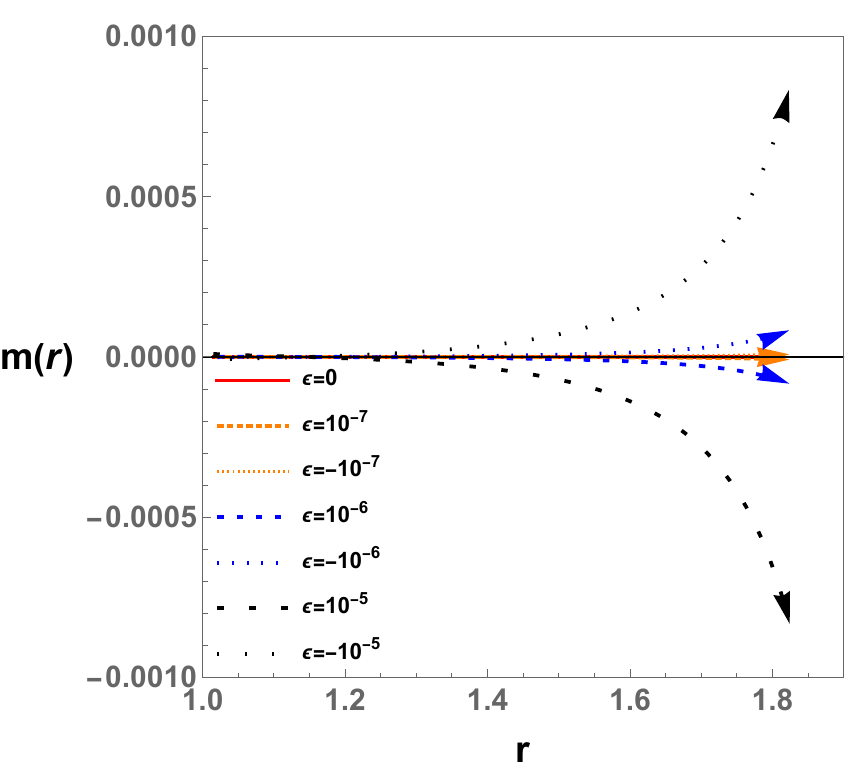}
     \caption{}
     \label{fig:m-r plot_1}
    \end{subfigure}
    \hspace{1.6mm}
    \begin{subfigure}[b]{0.49\linewidth}
    \includegraphics[width=\linewidth]{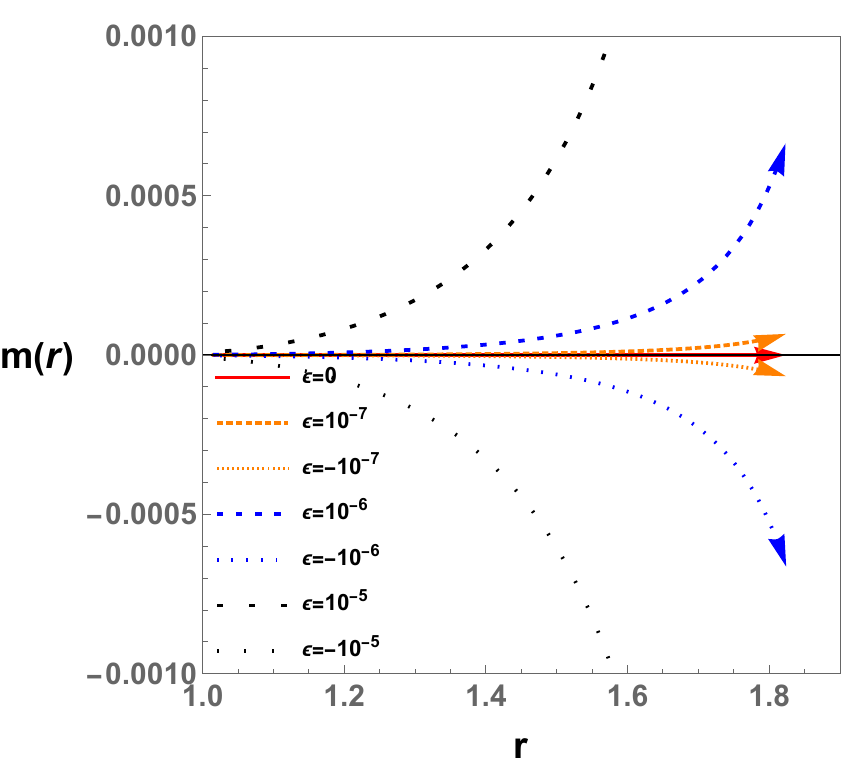}
     \caption{}
     \label{fig:m-r plot_2}
    \end{subfigure}
    \begin{subfigure}[b]{0.49\linewidth}
    \includegraphics[width=\linewidth]{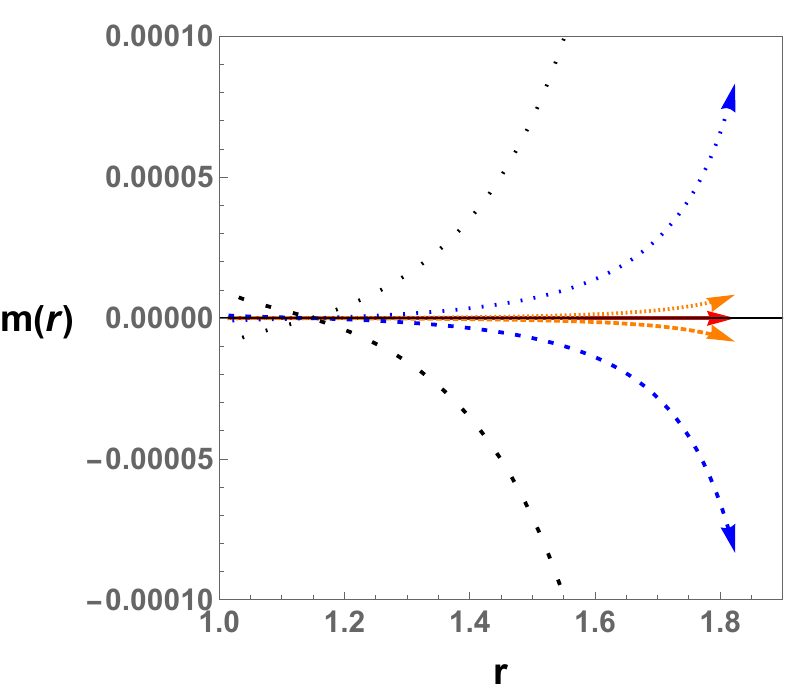}
     \caption{}
     \label{fig:m-r plot_1_zoomed}
    \end{subfigure}
    \hspace{1.6mm}
    \begin{subfigure}[b]{0.49\linewidth}
    \includegraphics[width=\linewidth]{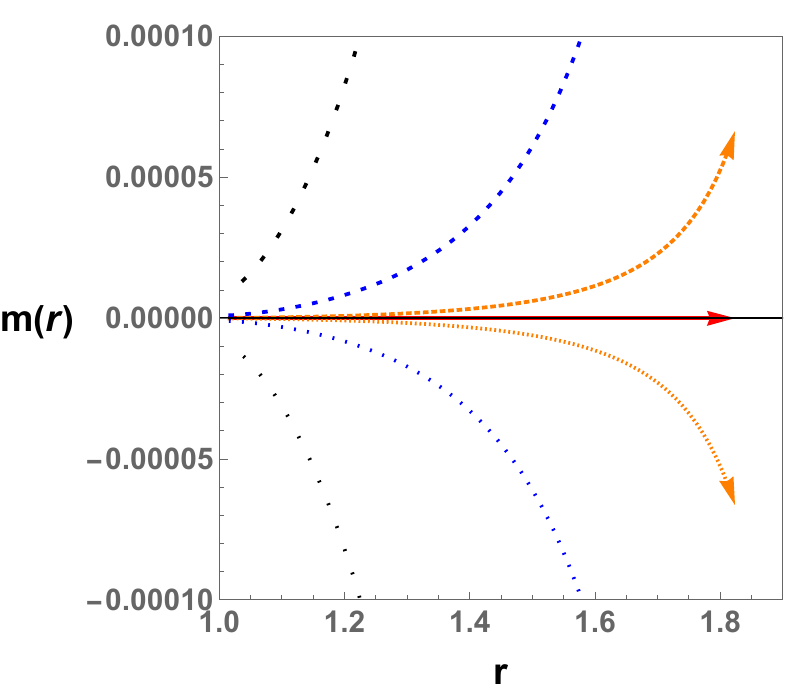}
     \caption{}
     \label{fig:m-r plot_2_zoomed}
    \end{subfigure}
    \caption{The dynamics of $\Lambda$CDM-mimicking $f(R)$ theories are shown as parametric curves ${r(q),m(q)}$ in the $m-r$ plane. The GR $\Lambda$CDM model ($m=0$, $r=(3q-3)/(q-2)$) is the central red line. Panels show numerical solutions of the nonautonomous system. In Panel (a), trajectories start from the GR line shifted by $+\epsilon$ in both $r$ and $m$; in Panel (b), from the GR line shifted by $-\epsilon$ in $r$ and $+\epsilon$ in $m$. The red, dashed orange, dotted orange, dashed blue, dotted blue, dashed black, dotted black curves correspond to $\epsilon=0,10^{-7},-10^{-7},10^{-6},-10^{-6},10^{-5},-10^{-5}$ respectively. The evolution runs from $q=0.49$ to $q_0 \approx -0.55$ (today). Panel (c) and (d) are zoomed-in version of the figure (a) and (b), respectively. Zoomed panels reveal that GR is not a generic past attractor. Trajectories can start in the physically viable region ($m>0$) and cross into the theoretically unstable region ($m<0$).}
    \label{fig:LCDM_mimicking_m(r)s_forward}
\end{figure}

The solution curves in Fig.\ref{fig:LCDM_mimicking_m(r)s} are obtained from the numerical solution of the reconstruction differential equation \eqref{eq:recon_LCDM_2} to show the behaviour of the corresponding solutions in the theory space. Instead, one can obtain $\Lambda$CDM-mimicking $f(R)$ solution curves directly by solving the nonautonomous system \eqref{eq:nonautonomous} numerically. We show some such solution curves in Figs.\ref{fig:m-r plot_1} and \ref{fig:m-r plot_2}, where we set the initial conditions such that at $q=0.49$ the underlying $f(R)$ theory is very close to GR, with small deviations from GR parametrized by a smallness parameter $\epsilon$. The exact initial conditions taken to produce these curves are specified in the caption.

The parametric plots in Figs.\ref{fig:m-r plot_zoomed}, \ref{fig:m-r plot}, \ref{fig:m-r plot_1}, \ref{fig:m-r plot_2} give an apparent impression that GR acts as the cosmological past attractor for the $\Lambda$CDM-mimicking $f(R)$ theories; that all such $f(R)$ theories asymptote to GR in the asymptotic past. However, we stress that there is no such generic tendency. This is made explicit in Figs.\ref{fig:m-r plot_1_zoomed} and \ref{fig:m-r plot_2_zoomed}, which are zoomed-in versions of the Figs.\ref{fig:m-r plot_1} and \ref{fig:m-r plot_2} respectively. Although the parametric plots for solutions corresponding to the initial condition $\lbrace r(0.49),m(0.49) \rbrace = \lbrace \frac{3q-3}{q-2}\vert_{q=0.49} - \epsilon,\epsilon\rbrace$ do seem to asymptote to GR in the far past (Fig.\ref{fig:m-r plot_2_zoomed}), this behaviour is absent for solutions corresponding to the initial condition $\lbrace r(0.49),m(0.49) \rbrace = \lbrace \frac{3q-3}{q-2}\vert_{q=0.49} + \epsilon,\epsilon\rbrace$ (Fig.\ref{fig:m-r plot_1_zoomed}). On the contrary, rather, for the solutions depicted in Fig.\ref{fig:m-r plot_1_zoomed}, it appears that the underlying theory actually starts deviating from GR also in the past. This latter class of solutions also show that a $\Lambda$CDM-mimicking $f(R)$ cosmology can start with a physically healthy theory near the matter-dominated epoch but end up as one plagued by either ghost or tachyonic instability (i.e., the solution curve switching from the $m>0$ to $m<0$ region).

\begin{figure}[H]
    \centering
    \begin{subfigure}[b]{0.49\linewidth}
    \includegraphics[width=\linewidth]{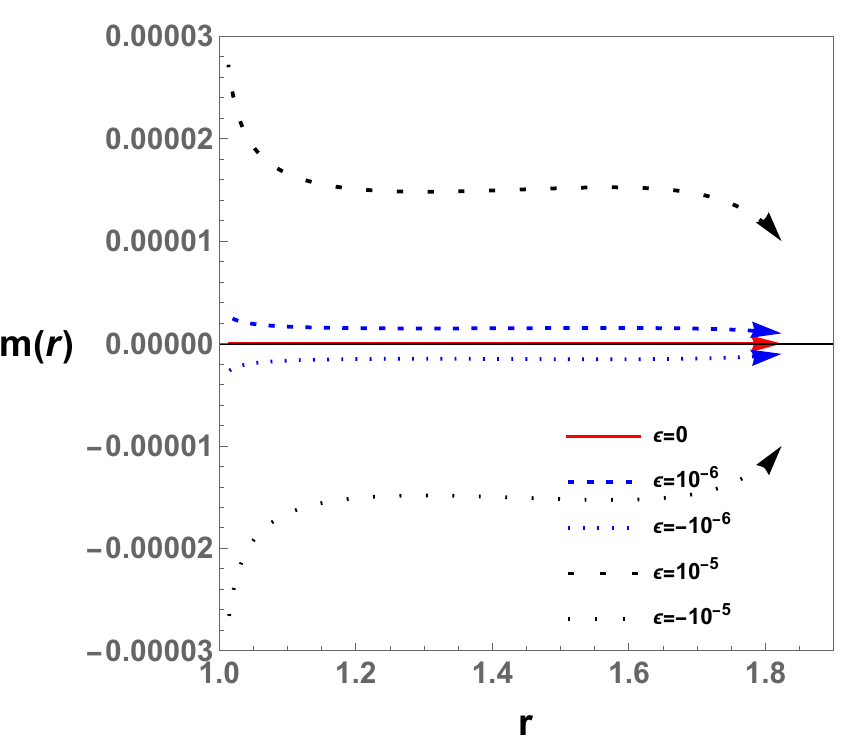}
     \caption{}
     \label{fig:m-r plot_3}
    \end{subfigure}
    \hspace{1.6mm}
    \begin{subfigure}[b]{0.49\linewidth}
    \includegraphics[width=\linewidth]{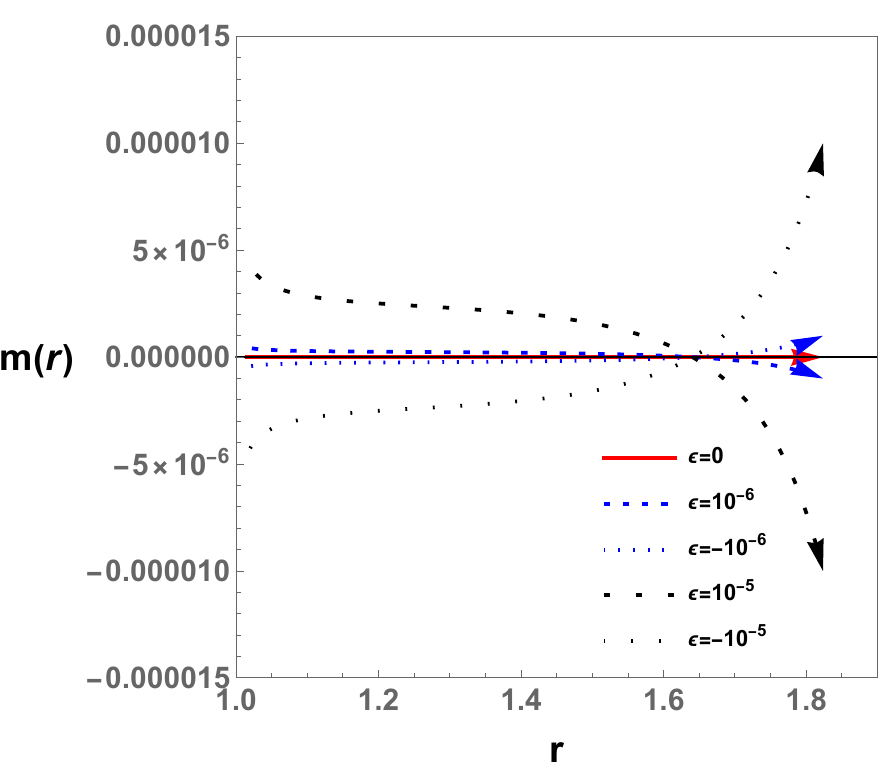}
     \caption{}
     \label{fig:m-r plot_4}
    \end{subfigure}
    \caption{The curves in the panel \eqref{fig:m-r plot_3} corresponds to solutions of the nonautonomous system \eqref{eq:nonautonomous_LCDM} with the initial condition $\lbrace r(-0.55),m(-0.55) \rbrace = \lbrace \frac{3q-3}{q-2}\vert_{q=-0.55} + \epsilon,\epsilon\rbrace$, whereas that in the panel \eqref{fig:m-r plot_4} corresponds to the initial conditions $\lbrace r(-0.55),m(-0.55) \rbrace = \lbrace \frac{3q-3}{q-2}\vert_{q=-0.55} + \epsilon,-\epsilon\rbrace$. The red, dashed blue, dotted blue, dashed black, dotted black curves correspond to $\epsilon=0,10^{-6},-10^{-6},10^{-5},-10^{-5}$ respectively. It appears that even though the underlying $\Lambda$CDM-mimicking $f(R)$ theories are very close to GR at the present epoch, it is possible for them to be different from GR near the matter-domination.}
    \label{fig:LCDM_mimicking_m(r)s_backward}
\end{figure}

All the theory space portraits presented so far in this section are obtained by assigning the condition that the underlying $f(R)$ theory is very close to GR during the matter-domination ($q=0.49\approx1/2$). A natural question to ask is what will happen if one assigns the condition that the underlying $f(R)$ theory is very close to GR at the present epoch ($q=-0.55$). Do such theories approach even closer to GR in the past? The answer to this question is, surprisingly, non-affirmative! This is demonstrated in the theory space portraits of Fig.\ref{fig:LCDM_mimicking_m(r)s_backward}. 

In summary, the theory space analysis of the $\Lambda$CDM-mimicking $f(R)$ theories, as demonstrated by the different parametric plots presented in this figure, reveals that there is no generic tendency of GR to behave as either a cosmological past or a future attractor. Even if the underlying theory was only slightly different from GR in the past, it deviates from GR pretty quickly. On the other hand, even if the underlying theory is only slightly different from GR \emph{now}, it can be quite different from GR in the past. 

A $\Lambda$CDM-like cosmological solution has served as a benchmark cosmological solution for a long time. Even after DESI's second data release, it has been argued that purely kinematic diagnostic agnostic of any \emph{a priori} chosen parametrization (e.g. CPL) may significantly reduce the tension between the DESI data and a $\Lambda$CDM-like cosmological solution \cite{Dinda:2025svh}, or outright show no phantom-crossing at all \cite{Roy:2025cxk}. The fact that GR doesn't have an attractive nature in either the past or the future direction within the theory space of $\Lambda$CDM-mimicking $f(R)$ theories has serious ramifications in cosmological model-building. One can never truly know the exact form of the underlying gravitational theory driving the cosmological dynamics at the largest observable scale. Assuming that the underlying theory belongs to a class whose Lagrangian is defined solely by the Ricci scalar, one can conclude that even though GR is the simplest of such theories that admit a $\Lambda$CDM-like cosmological evolution as a solution, it is not a natural one. It is far more likely and natural for the $\Lambda$CDM-like evolution \eqref{eq:LCDM} to be a solution of gravity theories other than GR. 

\begin{figure}[H]
    \centering
    
    \begin{subfigure}[b]{0.45\linewidth}
    \includegraphics[width=\linewidth]{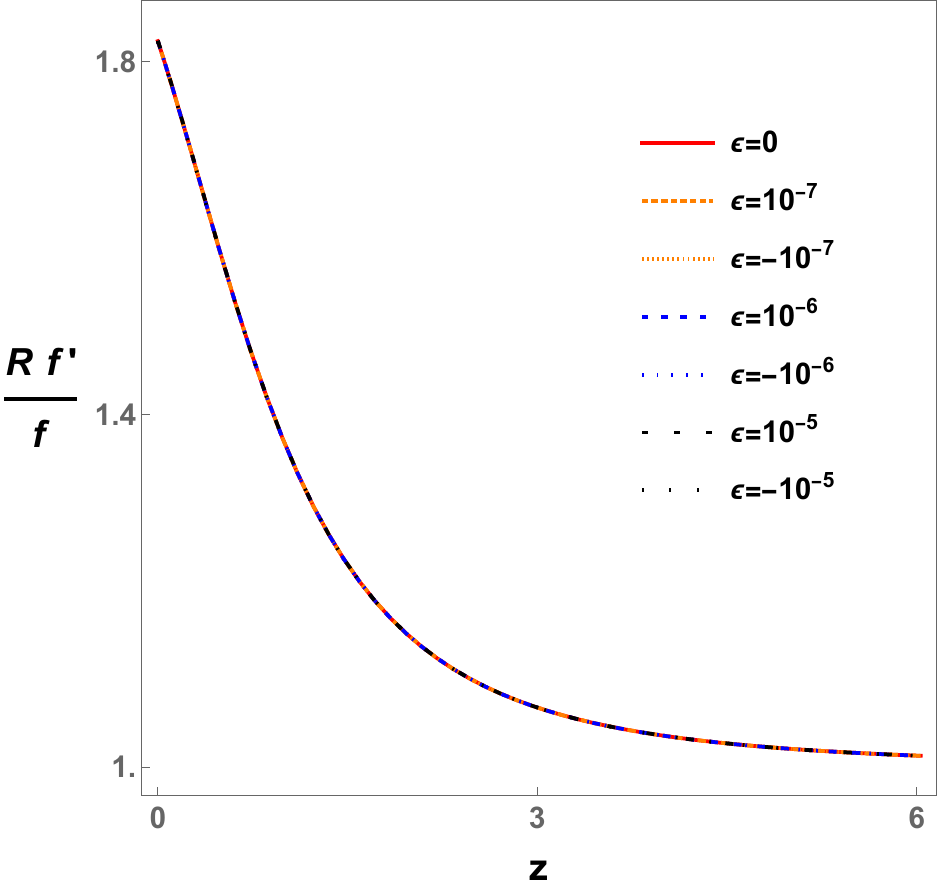}
     \caption{}
     \label{fig:r-z plot_1}
    \end{subfigure}
    \hspace{2mm}
    \vspace{0.5cm}
    \begin{subfigure}[b]{0.45\linewidth}
    \includegraphics[width=\linewidth]{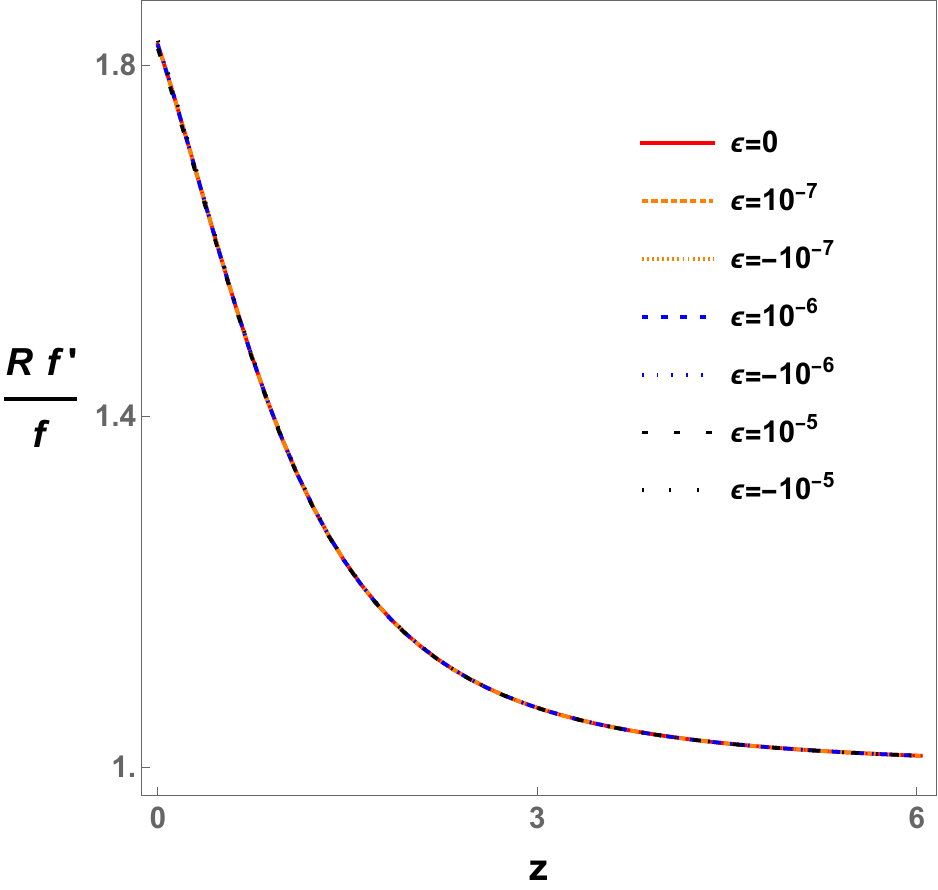}
     \caption{}
     \label{fig:r-z plot_2}
    \end{subfigure}
    
    \begin{subfigure}[b]{0.45\linewidth}
    \includegraphics[width=\linewidth]{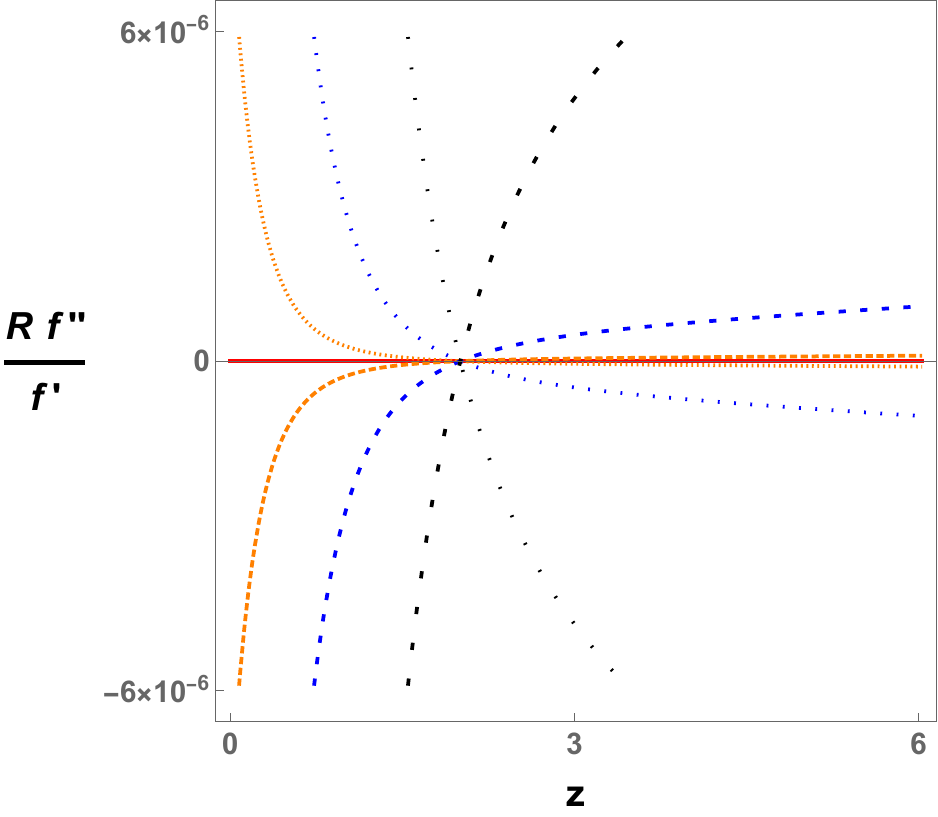}
     \caption{}
     \label{fig:m-z plot_1}
    \end{subfigure}
    \hspace{2mm}
    \vspace{0.5cm}
    \begin{subfigure}[b]{0.45\linewidth}
    \includegraphics[width=\linewidth]{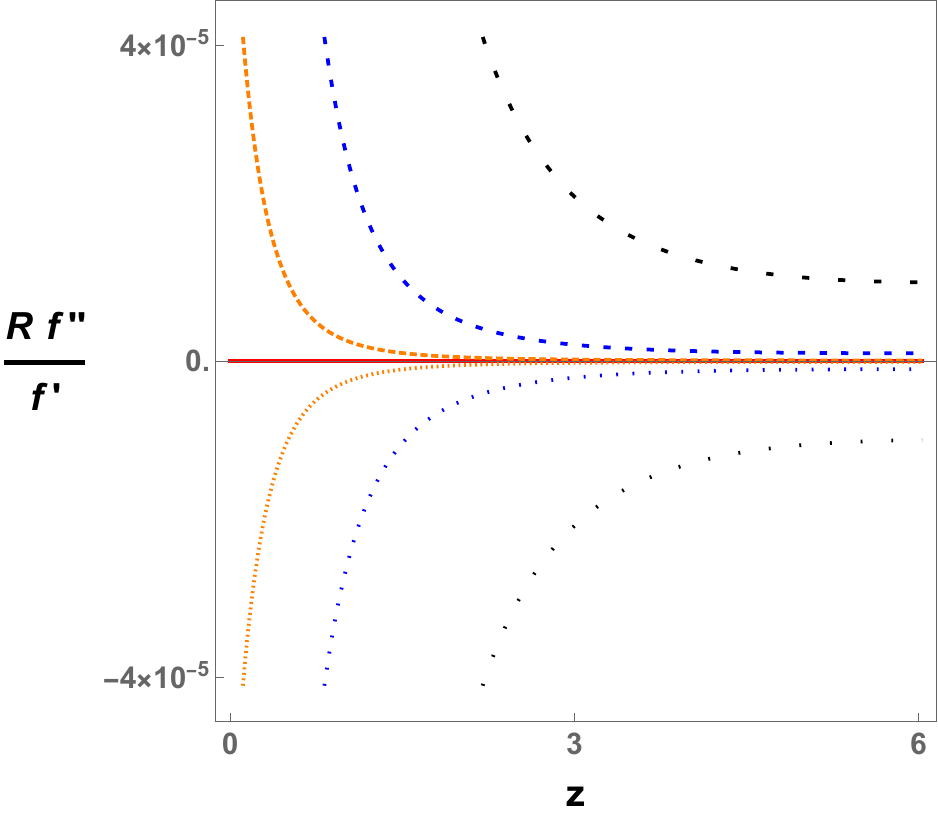}
     \caption{}
     \label{fig:m-z plot_2}
    \end{subfigure}
    
    \begin{subfigure}[b]{0.45\linewidth}
    \includegraphics[width=\linewidth]{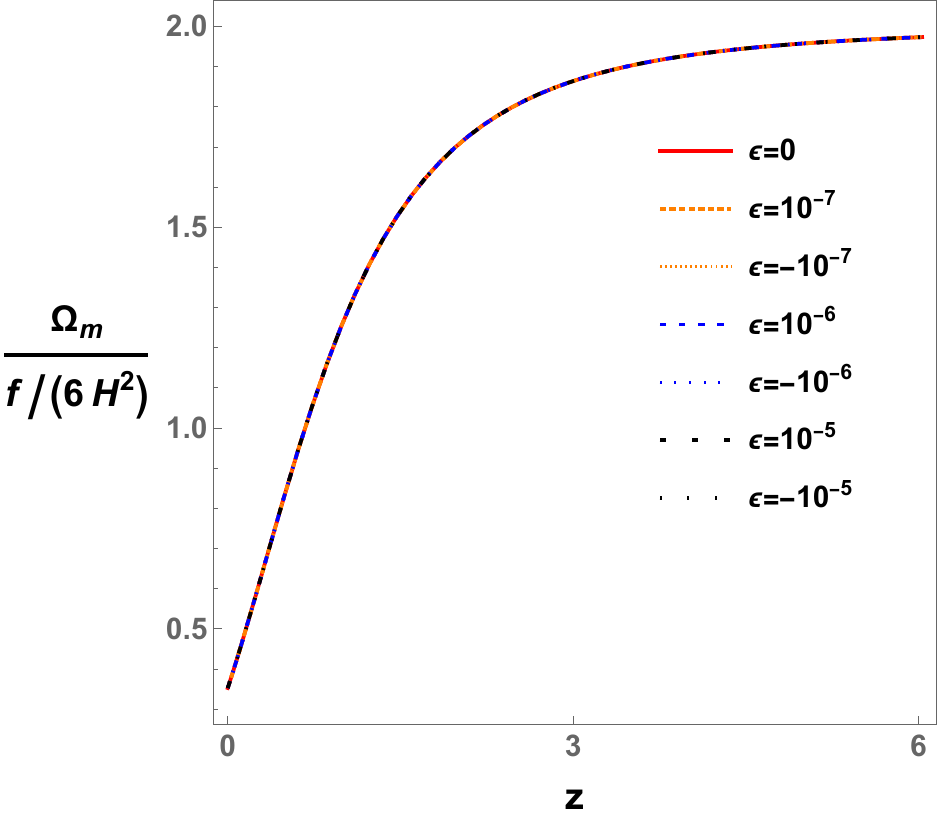}
     \caption{}
     \label{fig:Omega_m-z plot_1}
    \end{subfigure}
    \hspace{2mm}
    \vspace{0.5cm}
    \begin{subfigure}[b]{0.45\linewidth}
    \includegraphics[width=\linewidth]{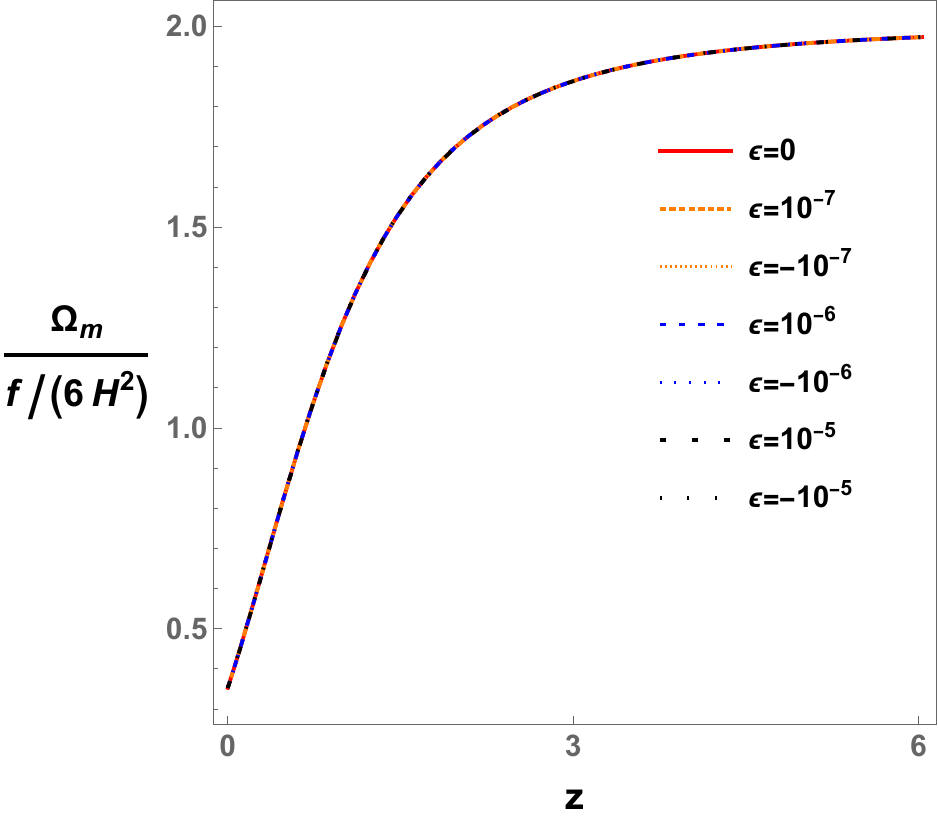}
     \caption{}
     \label{fig:Omega_m-z plot_2}
    \end{subfigure}
    
    \caption{The left and the right panels portray the evolution of the quantities $r\equiv\frac{Rf'}{f},\,m\equiv\frac{Rf''}{f'}$ and $\frac{\Omega_m}{f/(6H^2)}$ with respect to the redshift $z$ corresponding to the theory curves presented in Fig.\ref{fig:m-r plot_1} and Fig.\ref{fig:m-r plot_2} respectively. The red, dashed orange, dotted orange, dashed blue, dotted blue, dashed black, dotted black curves correspond to $\epsilon=0,10^{-7},-10^{-7},10^{-6},-10^{-6},10^{-5},-10^{-5}$ respectively (indicated in the figures of the top and the bottom panel). The plots are done within the redshift range from $z_{\rm in}=6.0316$ to $z=0$ (today).}
    \label{fig:r-m vs z forward}
\end{figure}

\begin{figure}[H]
    \centering
    
    \begin{subfigure}[b]{0.45\linewidth}
    \includegraphics[width=\linewidth]{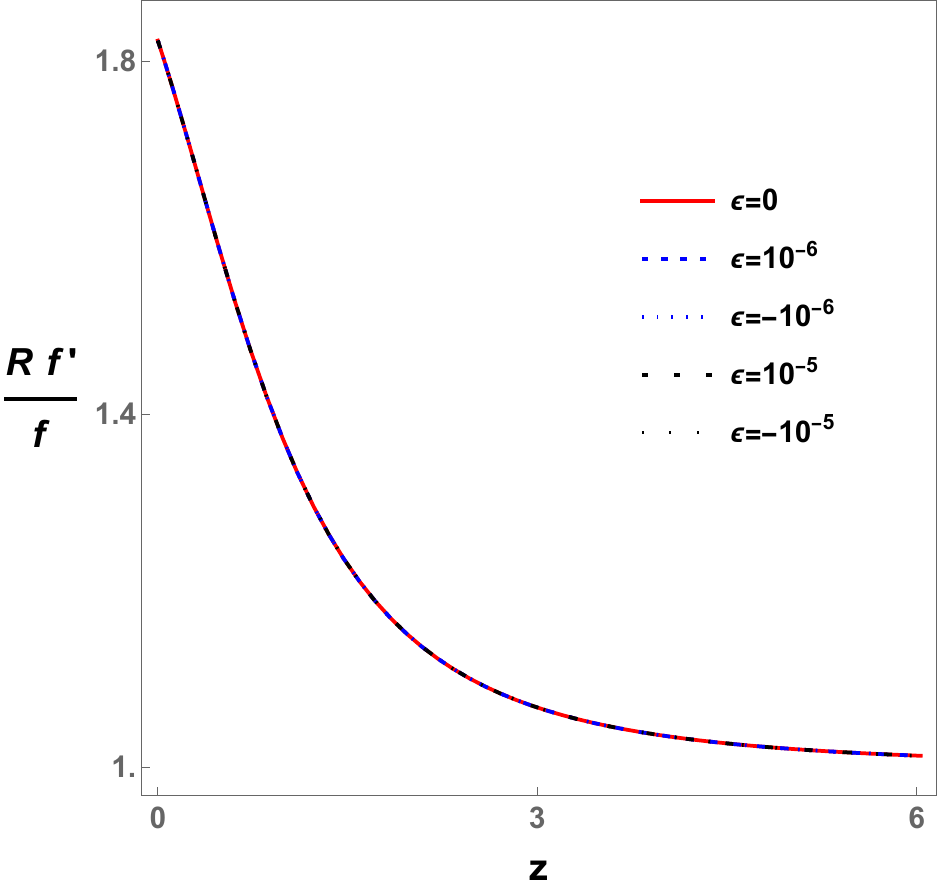}
     \caption{}
     \label{fig:r-z plot_3}
    \end{subfigure}
    \hspace{2mm}
    \vspace{0.5cm}
    \begin{subfigure}[b]{0.45\linewidth}
    \includegraphics[width=\linewidth]{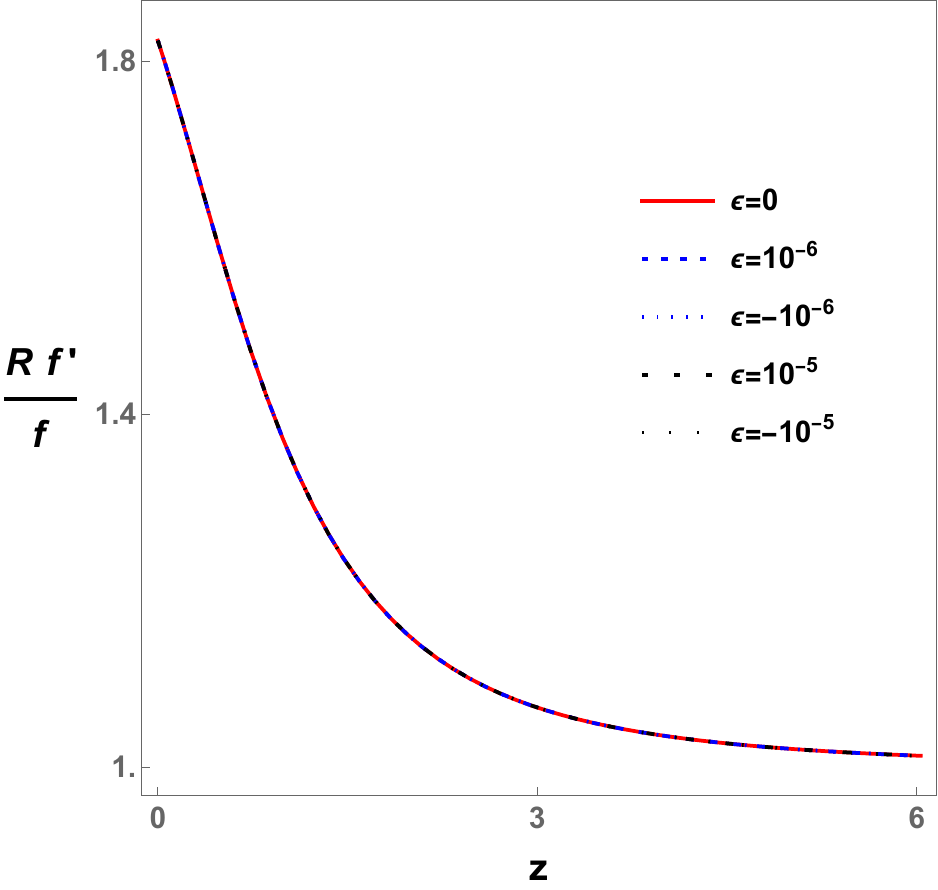}
     \caption{}
     \label{fig:r-z plot_4}
    \end{subfigure}
    
    \begin{subfigure}[b]{0.45\linewidth}
    \includegraphics[width=\linewidth]{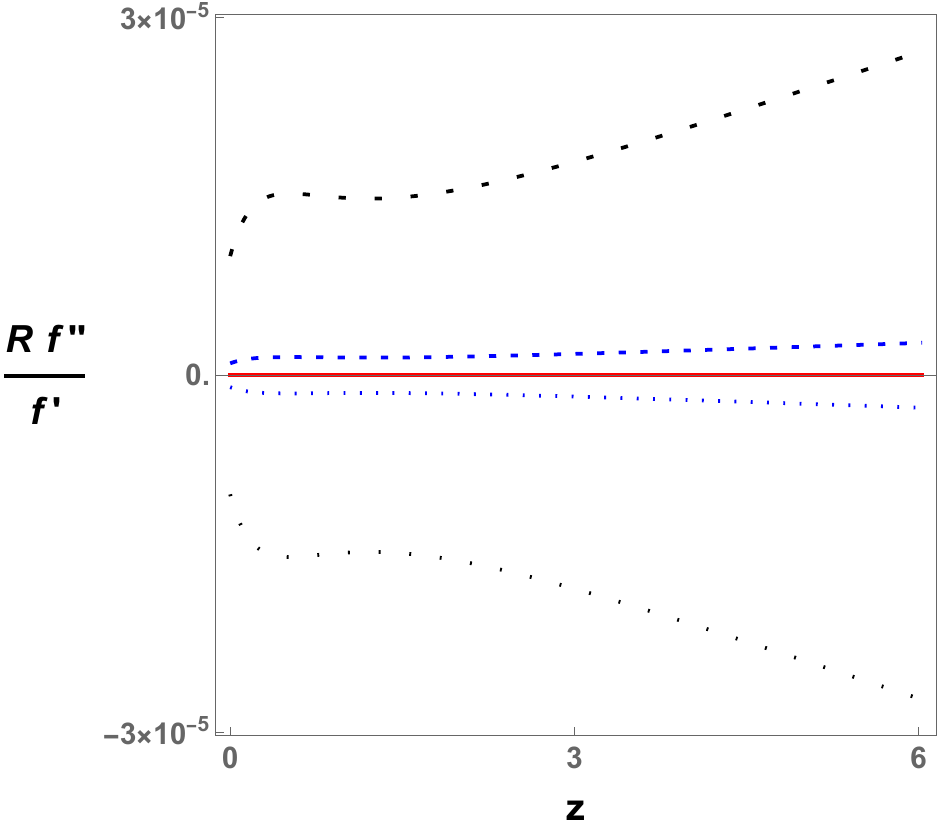}
     \caption{}
     \label{fig:m-z plot_3}
    \end{subfigure}
    \hspace{2mm}
    \vspace{0.5cm}
    \begin{subfigure}[b]{0.45\linewidth}
    \includegraphics[width=\linewidth]{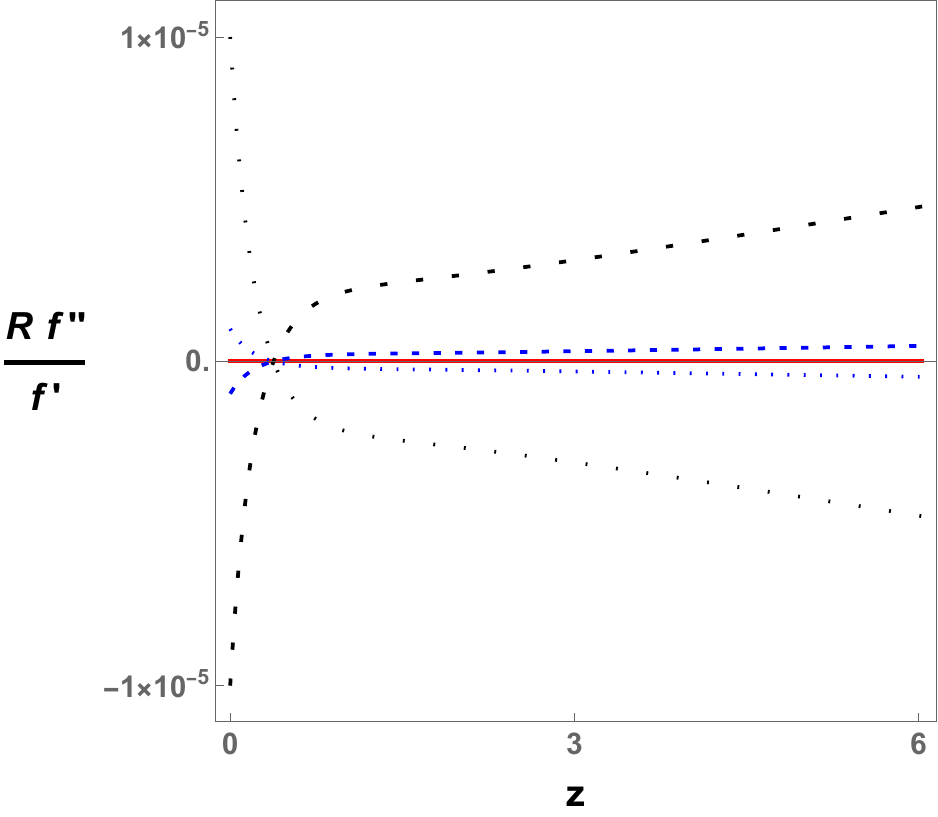}
     \caption{}
     \label{fig:m-z plot_4}
    \end{subfigure}
    
    \begin{subfigure}[b]{0.45\linewidth}
    \includegraphics[width=\linewidth]{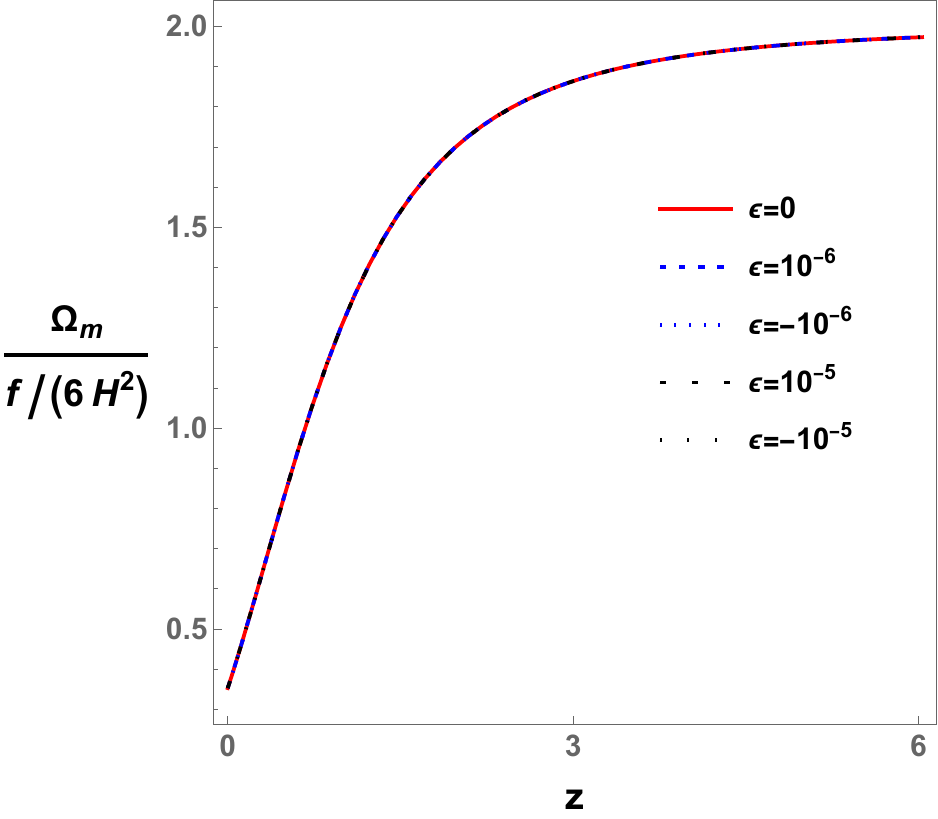}
     \caption{}
     \label{fig:Omega_m-z plot_3}
    \end{subfigure}
    \hspace{2mm}
    \vspace{0.5cm}
    \begin{subfigure}[b]{0.45\linewidth}
    \includegraphics[width=\linewidth]{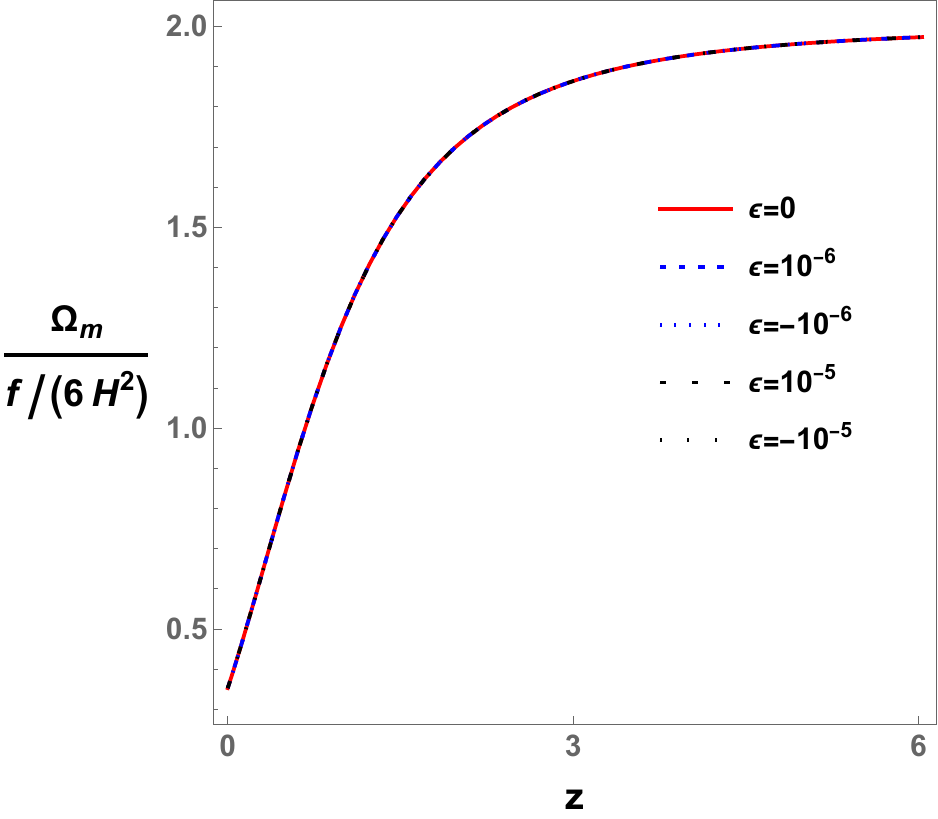}
     \caption{}
     \label{fig:Omega_m-z plot_4}
    \end{subfigure}
    
    \caption{The left and the right panels portray the evolution of the quantities $r\equiv\frac{Rf'}{f},\,m\equiv\frac{Rf''}{f'}$ and $\frac{\Omega_m}{f/(6H^2)}$ with respect to the redshift $z$ corresponding to the theory curves presented in Fig.\ref{fig:m-r plot_3} and Fig.\ref{fig:m-r plot_4} respectively. The red, dashed blue, dotted blue, dashed black, dotted black curves correspond to $\epsilon=0,10^{-6},-10^{-6},10^{-5},-10^{-5}$ respectively (indicated in the figures of the top and the bottom panel). The plots are done within the redshift range from $z_{\rm in}=6.0316$ to $z=0$ (today).}
    \label{fig:r-m vs z backward}
\end{figure}

For completeness, we also present in Figs.\ref{fig:r-m vs z forward} and \ref{fig:r-m vs z backward}, the redshift evolution of the quantities $r\equiv\frac{Rf'}{f},\,m\equiv\frac{Rf''}{f'}$ and the normalized density parameter $\frac{\Omega_m}{f/(6H^2)}$ corresponding to the theory curves presented in Figs.\ref{fig:LCDM_mimicking_m(r)s_forward} and \ref{fig:LCDM_mimicking_m(r)s_backward} respectively. Once the numerical solutions $r(z),\,m(z)$ is known, the quantity $\frac{\Omega_m}{f/(6H^2)}$ can be calculated from Eq.\eqref{eq:energy_eq_mr}

Keeping in mind that the arrows in the plots in Figs.\ref{fig:LCDM_mimicking_m(r)s_forward} and \ref{fig:LCDM_mimicking_m(r)s_backward} correspond towards decreasing values of redshift, one can verify that the plots in Figs.\ref{fig:r-m vs z forward} and \ref{fig:r-m vs z backward} presents a picture consistent with what is observed in Figs.\ref{fig:LCDM_mimicking_m(r)s_forward},\ref{fig:LCDM_mimicking_m(r)s_backward}. Whereas the evolution plots in Figs.\ref{fig:r-m vs z forward} and \ref{fig:r-m vs z backward} are a better portrayal of the time evolution of the two important quantities $r$ and $m$ as well as the quantity $\frac{\Omega_m}{f/(6H^2)}$ for an underlying $f(R)$, the solution curves in Figs.\ref{fig:LCDM_mimicking_m(r)s_forward} and \ref{fig:LCDM_mimicking_m(r)s_backward} gives a better intuitive visual of how the underlying $f(R)$ theory deviates given a certain initial deviation. 

Since $\Omega_m$ and $R=6H^2(1-q)$ are positive along the course of a $\Lambda$CDM-like cosmic evolution, the plots of the quantities $r=\frac{Rf'}{f}$ and $\frac{\Omega_m}{f/(6H^2)}$ in Figs.\ref{fig:r-m vs z forward} and \ref{fig:r-m vs z backward} imply that the quantities $f(R)$ and $f'(R)$ are always remain positive along the course of evolution, implying no appearance of a ghost instability. The negativity of the quantity $m=\frac{Rf''}{f'}$, whenever appears, therefore, indicates $f''(R)<0$, i.e. a tachyonic instability.

\subsection{Stability of the solution}

Let us consider a $\Lambda$CDM-like cosmological solution $h(N)$ satisfying Eq.\eqref{eq:master_eq_1_LCDM}, and a time-dependent homogeneous and isotropic perturbation $\delta h(N)$ on it. Using $j=1,\,s = -(2+3q)\,,l = 6q^2+14q+9$, the coefficients of the perturbation equation \eqref{eq:ptbn_f(R)} can be explicitly written as
\begin{subequations}\label{eq:a_orig}
\begin{align}
a_0(r,m,q) &= (q-1) \Bigl(r \bigl(m (q+1) (q (q (6 q+11)+8)-9) - 4 (q-2) (q-1)^2 (q+4)\bigr) \nonumber \\
            &\hspace{250pt} + 6 (q (2 q+5)-9) (q-1)^2\Bigr)\,, \label{eq:a0_orig} 
\\[10pt]
a_1(r,m,q) &= (q-1) \Bigl(m (q+1) (q (3 q+10)-12) r - 2 ((q-9) q+11) (q-1) r  + 6 (q-6) (q-1)^2\Bigr)\,, \label{eq:a1_orig} 
\\[10pt]
a_2(r,m,q) &= 2 (q-1) \Bigl(q r (m (q-2)+q-3) - 3 m r - 3 (q-2) q + 2 r - 3\Bigr)\,, \label{eq:a2_orig} 
\\[10pt]
a_3(r,m,q) &= -m \left(q^2-1\right) r\,. \label{eq:a3_orig}
\end{align}
\end{subequations}

For the General Relativistic $\Lambda$CDM solution with $f(R)=-2\Lambda+R$, the field equations are second order in metric, i.e., first order in Hubble parameter. Therefore, the first two coefficients in the perturbation equation \eqref{eq:ptbn_f(R)} should vanish when one sets the conditions $\{r,m\}\vert_{\rm GR} = \lbrace\frac{3q-3}{q-2},0\rbrace$ from Eq.\eqref{eq:GR}. As a consistency check, one can verify that this is indeed the case. The perturbation equation corresponding to the General Relativistic $\Lambda$CDM solution comes out to be
\begin{equation}\label{eq:ptbn_GR}
 \frac{d\delta h}{dN} + (2-q)\delta h = 0\,.  
\end{equation}
Since one always has $q<2$ during a $\Lambda$CDM-like cosmic evolution of the form \eqref{eq:LCDM}, Eq.\eqref{eq:ptbn_GR} implies that the General Relativistic $\Lambda$CDM solution is stable with respect to small homogeneous metric perturbations \emph{within the homogeneous isotropic solution space of GR}.

\begin{figure}[H]
    \centering
    \begin{subfigure}[b]{0.32\linewidth}
    \includegraphics[width=\linewidth]{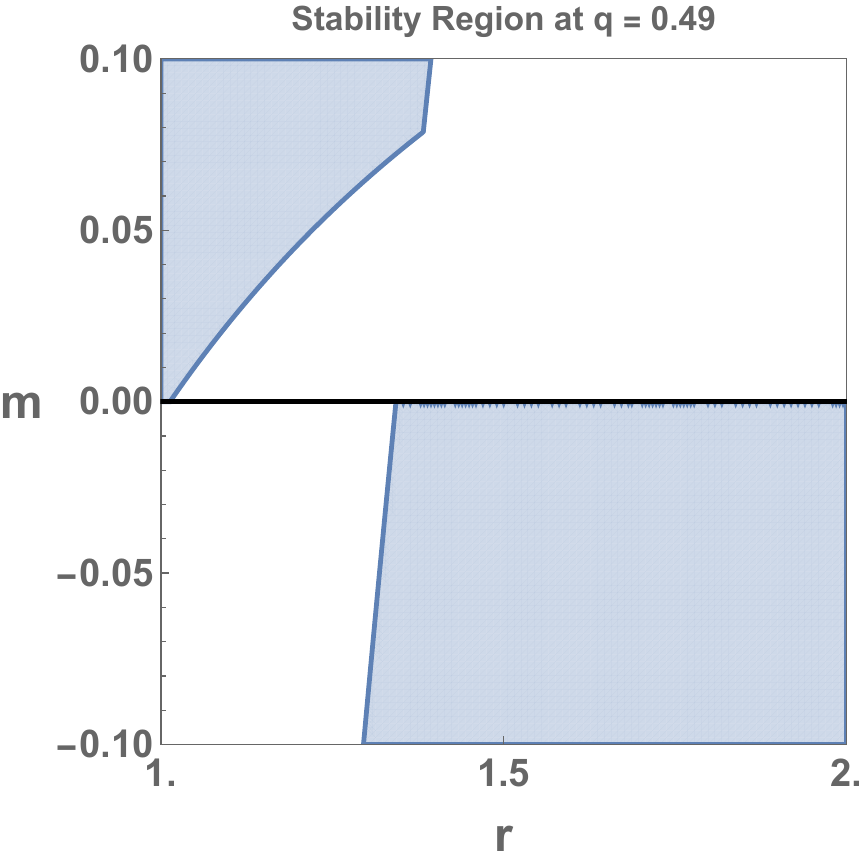}
     \caption{}
     \label{fig:stab_0.49}
    \end{subfigure}
    \hfill
    \begin{subfigure}[b]{0.32\linewidth}
    \includegraphics[width=\linewidth]{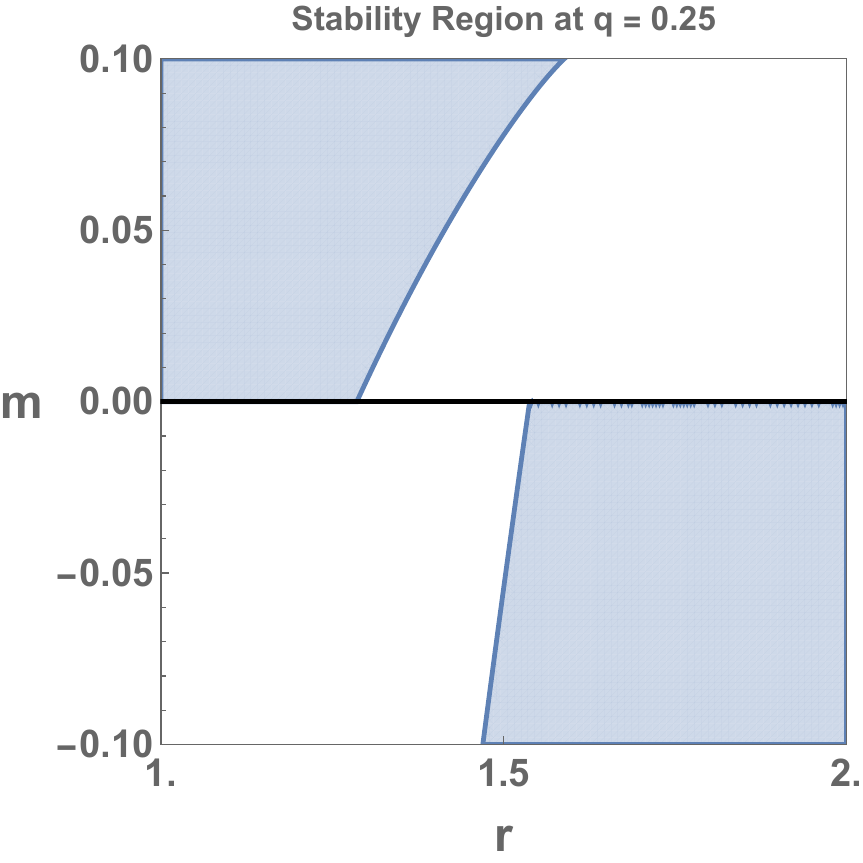}
     \caption{}
     \label{fig:stab_0.25}
    \end{subfigure}
    \hfill
    \vspace{0.5cm}
    \begin{subfigure}[b]{0.32\linewidth}
    \includegraphics[width=\linewidth]{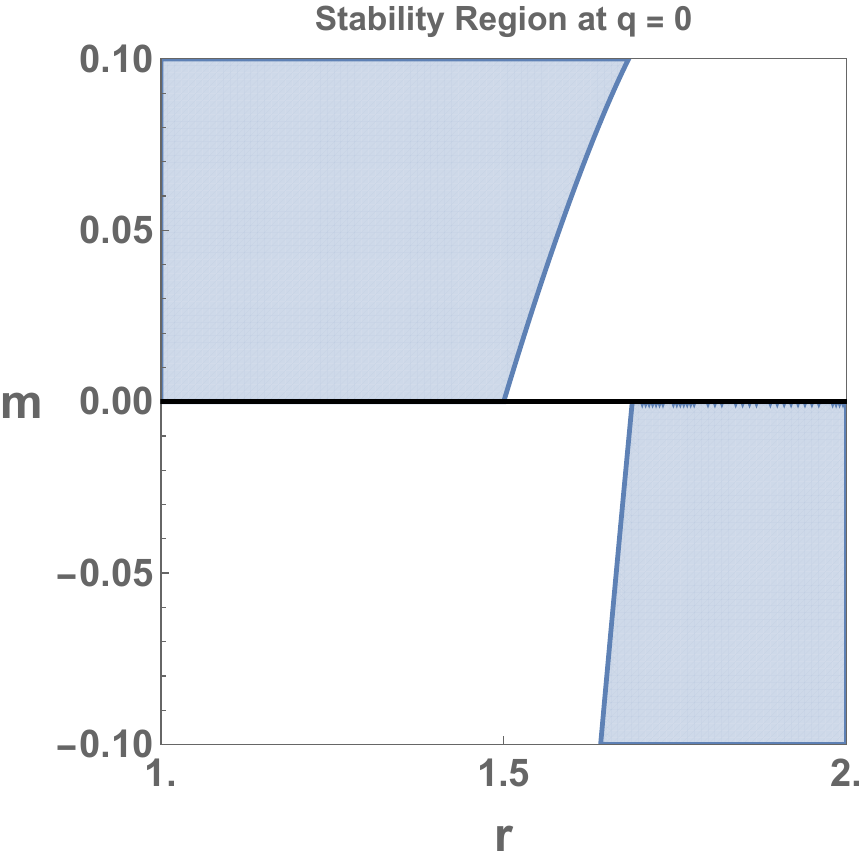}
     \caption{}
     \label{fig:stab_0}
    \end{subfigure}
    \begin{subfigure}[b]{0.32\linewidth}
    \includegraphics[width=\linewidth]{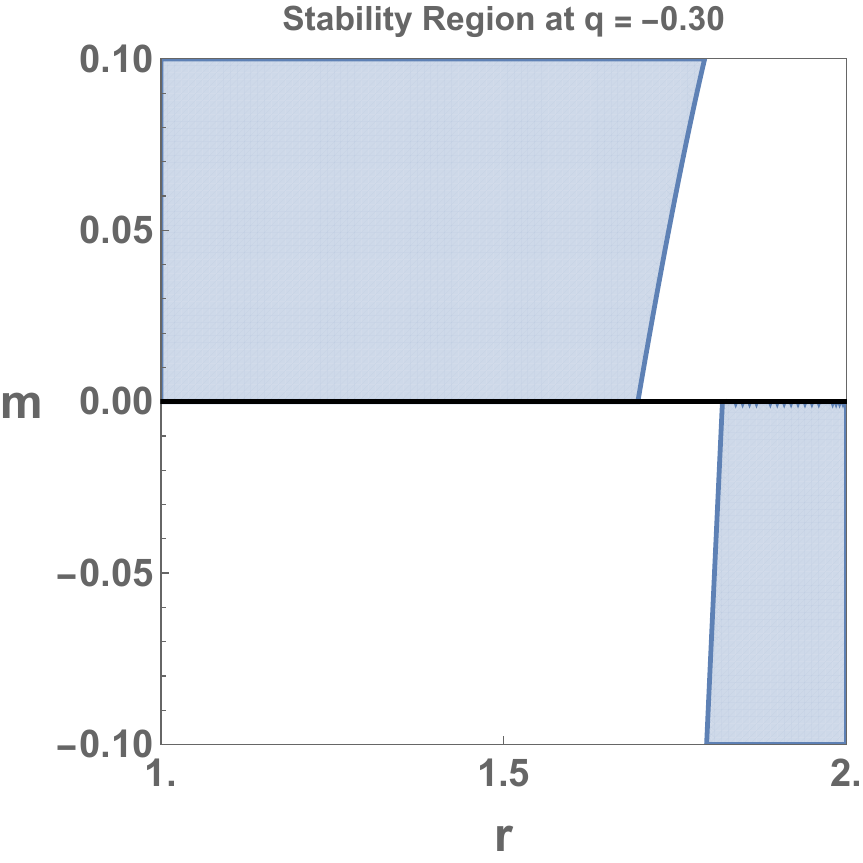}
     \caption{}
     \label{fig:stab_-0.3}
    \end{subfigure}
    \hfill
    \begin{subfigure}[b]{0.32\linewidth}
    \includegraphics[width=\linewidth]{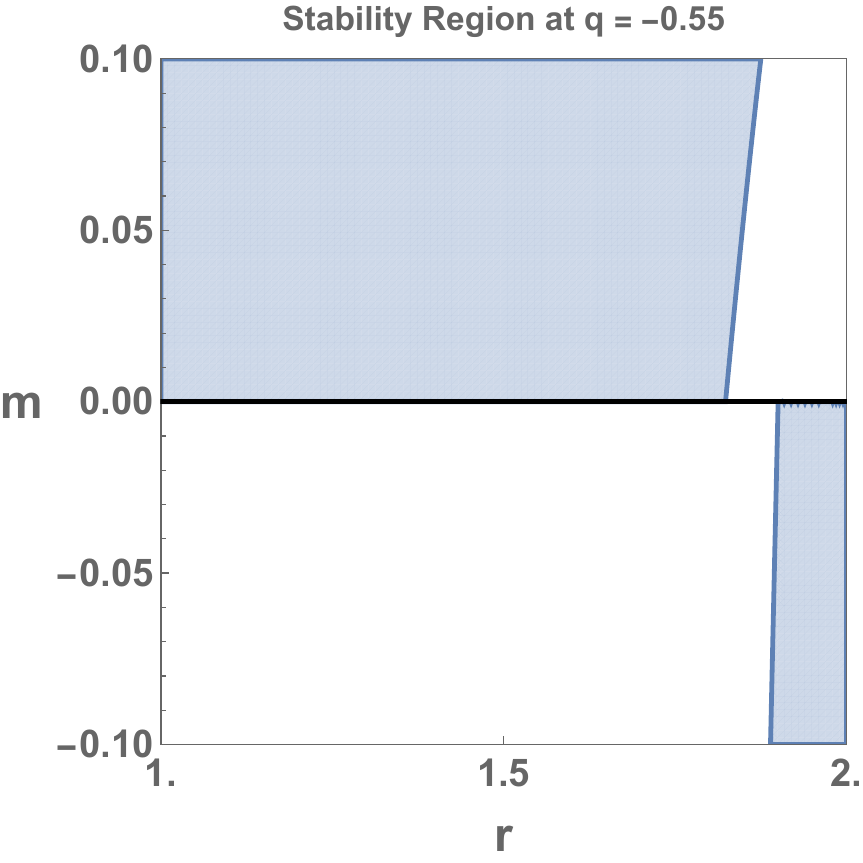}
     \caption{}
     \label{fig:stab_-0.55}
    \end{subfigure}
    \hfill
    \begin{subfigure}[b]{0.32\linewidth}
    \includegraphics[width=\linewidth]{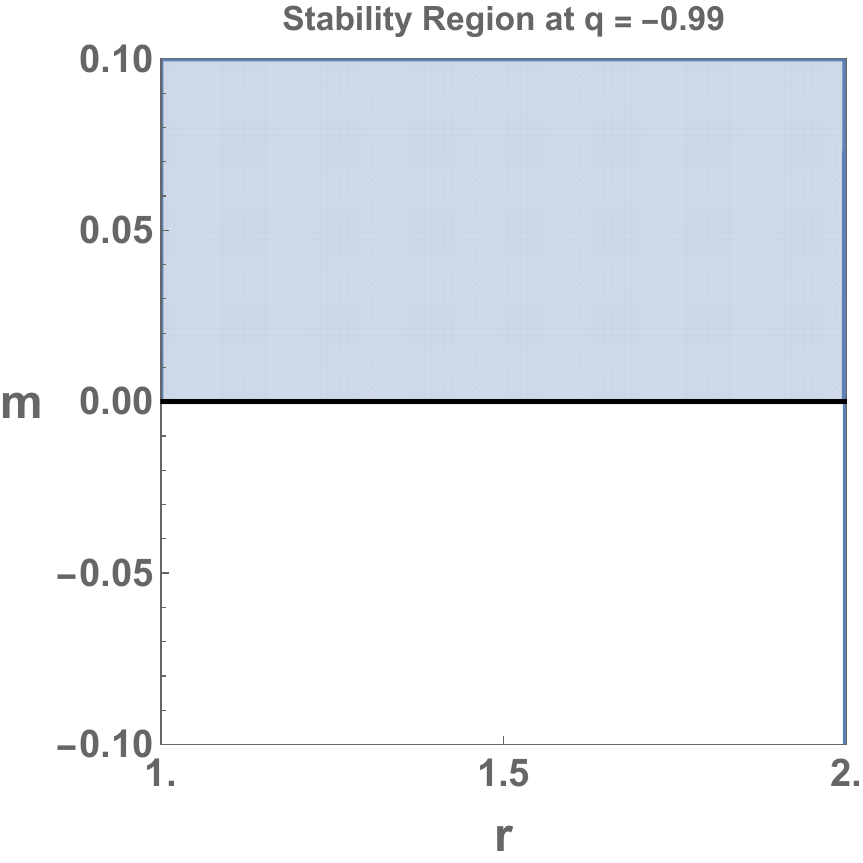}
     \caption{}
     \label{fig:stab_-0.99}
    \end{subfigure}
    \caption{2-dimensional screenshots of $r$-$m$-$q$ phase space at six different values of $q$, starting from $q=0.49$ to $q=-0.99$. The region of stability with respect to small homogeneous and isotropic perturbation (i.e. where the condition \eqref{coeff_conds} is satisfied) for a $\Lambda$CDM-like cosmological solution $j=1$ is shown by the shaded region, which changes with time.}
    \label{fig:stability_region}
\end{figure}

For the more generic case of $\Lambda$CDM-like cosmological solutions in $f(R)$, however, one needs to take into consideration the Routh-Hurwitz stability criterion \eqref{coeff_conds}. Since $q$ is a monotonic function of $N$, at each $q$-slice the stability condition \eqref{coeff_conds} singles out a 2-dimensional region in the $r$-$m$ plane. However, as we have mentioned before, the Routh-Hurwitz criteria, in our case, gives only an \emph{instantaneous} stability condition. Consequently, this 2-dimensional region in the $r-m$ plane changes with $q$. It is perhaps more illuminating to visualize the stability condition as how these 2-dimensional regions change over different $q$-slices. The 2-dimensional \emph{screenshots} for six different $q$-slices are shown in Fig.\ref{fig:stability_region}. One can interpret the figure as follows. Consider a particular phase trajectory in the $r$-$m$-$q$ phase space. At each cosmic \emph{moment}, which is given by a particular value of $q$, this particular phase trajectory passes through some point of the corresponding $r$-$m$ plane at that $q$. If the point is inside the shaded region, then nearby phase trajectories are being attracted towards it at that cosmic moment, and vice versa. For a solution to always possess this attractive feature, the union of all these points must always remain in the shaded region. 
Some interesting conclusions can already be drawn from the above figures. Firstly, the $\Lambda$CDM-mimicking $f(R)$ cosmologies that start in the $m<0$ region at high redshift already violate the stability condition\footnote{This might actually be an artefact of tachyonic instability $f''<0$, which is related to unbounded growth of perturbations.} However, even for the $\Lambda$CDM-mimicking $f(R)$ cosmologies that start in the $m>0$ region at high redshift, it is possible for the corresponding solution to be unstable with respect to small homogeneous and isotropic perturbations if it deviates from the General Relativistic $\Lambda$CDM solution (the $m=0$ line) very slowly. The latter behaviour is particularly true for those $\Lambda$CDM-mimicking $f(R)$ cosmologies which asymptote to GR at high redshift, e.g. the ones shown in Fig.\ref{fig:LCDM_mimicking_m(r)s} or Fig.\ref{fig:m-r plot_2}.

The above argument can be made slightly more sound analytically by particularly considering cosmologies in those $\Lambda$CDM-mimicking $f(R)$ theories that are close to GR in the limit $q\to\frac{1}{2}$. To consider such theories particularly, let us first express the coefficients \eqref{eq:a_orig} in terms of the quantity $\displaystyle{\bar{r}=r-\frac{3q-3}{q-2}>0}$ rather than $r$. Then, the perturbation equation \eqref{eq:ptbn_f(R)} becomes
\begin{equation}
 b_3(\bar{r},m,q)\frac{d^3 \delta h}{dN^3} + b_2(\bar{r},m,q)\frac{d^2 \delta h}{dN^2} + b_1(\bar{r},m,q)\frac{d\delta h}{dN} + b_0(\bar{r},m,q)\delta h = 0\,,  
\end{equation}
with
\begin{subequations}
\begin{align}
b_3(\bar{r},m,q) &= -m\,(q^2-1)\Bigl(\bar r + \tfrac{3}{q-2} + 3\Bigr)
\label{eq:b3} \\[1ex]
b_2(\bar{r},m,q) &= \frac{2\,(q-1)}{q-2}\Bigl[
      (q-2)\,\bar r\,\bigl(m\,(q-3)(q+1) + q^2 - 3q + 2\bigr) +\,3\,m\,(q-3)(q-1)(q+1)
      \Bigr]
\label{eq:b2} \\[1ex]
b_1(\bar{r},m,q) &= (q-1)\Bigl\{
      m\,(q+1)(3q^2+10q-12)\Bigl(\bar r + \tfrac{3}{q-2}+3\Bigr) -\,2\,(q^2-9q+11)(q-1)\Bigl(\bar r + \tfrac{3}{q-2}+3\Bigr)
      \nonumber\\
    &\quad\; +\,6\,(q-6)(q-1)^2
      \Bigr\}
\label{eq:b1} \\[1ex]
b_0(\bar{r},m,q) &= (q-1)\Bigl\{
      \left[m\,(q+1)(6q^3+11q^2+8q-9)
      - 4\,(q-2)(q-1)^2(q+4)\right]
      \Bigl(\bar r + \tfrac{3}{q-2}+3\Bigr)
      \nonumber\\
    &\quad\; +\,6\,(2q^2+5q-9)(q-1)^2
      \Bigr\}
\label{eq:b0}
\end{align}
\end{subequations}
The motivation behind this is that $\{\bar{r},m\}=\{0,0\}$ for GR. If a theory is close to GR in the limit $q\to\frac{1}{2}$, then both $\bar{r}$ and $m$ can be considered small ($\mathcal{O}(\epsilon)$) in this limit, and any product of them ($\mathcal{O}(\epsilon^2)$) can be neglected. Under this \emph{almost GR} assumption, the coefficients simplify to
\begin{subequations}
\begin{align}
b_3(\bar{r},m,q) &\simeq -\frac{3 m (q-1)^2 (q+1)}{q-2}
\label{eq:b3_final} \\[1ex]
b_2(\bar{r},m,q) &\simeq \frac{2 (q-1)^2 \left((q-2)^2 \bar{r}+3 m (q-3) (q+1)\right)}{q-2}
\label{eq:b2_final} \\[1ex]
b_1(\bar{r},m,q) &\simeq \frac{(q-1)^2 \left(3 (q+1) (m (q (3 q+10)-12)+2 (q-1))-2 (q-2) ((q-9) q+11) \bar{r}\right)}{q-2}
\label{eq:b1_final} \\[1ex]
b_0(\bar{r},m,q) &\simeq (q-1)^2 \left[-4 (q-2) (q-1) (q+4) \bar{r}+\frac{3 m (q+1) (q (q (6 q+11)+8)-9)}{q-2}-6 q^2+6\right]
\label{eq:b0_final}
\end{align}
\end{subequations}
One can then impose the limit $q\to\frac{1}{2}$ on the various ratios relevant to the Routh-Hurwitz stability criterion \eqref{coeff_conds}, and collect only the leading order terms (since $\mathcal{O}(\bar{r})=\mathcal{O}(m)=\mathcal{O}(\epsilon)$, $\frac{\mathcal{O}(\bar{r})}{\mathcal{O}(m)}=\mathcal{O}(1)$ is assumed)
\begin{equation}
\bigg\{ \frac{b_0}{b_3},\frac{b_1}{b_3},\frac{b_2}{b_3},\left(\frac{b_1}{b_3}\frac{b_2}{b_3}-\frac{b_0}{b_3}\right) \bigg\}\Bigg\vert_{q\to\frac{1}{2}} = \bigg\{ \frac{1}{2 m},\frac{1}{m},5-\frac{\bar{r}}{m},\frac{7}{2m}-\frac{\bar{r}}{m^2} \bigg\}\,.    
\end{equation}
For theories with $\bar{r},m>0$ in the limit $q\to\frac{1}{2}$, the criteria $\frac{b_0}{b_3}>0$ and $\frac{b_1}{b_3}>0$ can be met, but meeting the conditions $\frac{b_2}{b_3}>0$ and $\left(\frac{b_1}{b_3}\frac{b_2}{b_3}-\frac{b_0}{b_3}\right)>0$ requires additional conditions $\frac{m}{\bar{r}}>\frac{1}{5}$ and $\frac{m}{\bar{r}}>\frac{2}{7}$ respectively. Since at the vicinity of the point $(\bar{r},m)=(0,0)$, $\frac{m}{\bar{r}}\approx\frac{dm}{d\bar{r}}=\frac{dm}{dr}$, the lower limits on $\frac{m}{\bar{r}}$ translates into a lower bound on the slope of a theory curve near the point $(r,m)=(1,0)$. A $\Lambda$CDM-mimicking $f(R)$ solution curve that is asymptotic to the General Relativistic $\Lambda$CDM solution curve $m=0$ in the limit $q\to\frac{1}{2}$, definitely violates such a lower bound, thus giving rise to a cosmological solution that is unstable under small homogeneous and isotropic perturbations.

Physically, one can interpret the result in the following way. After setting $j=1,\,\Omega_k=0,\,w=0$ in Eq.\eqref{eq:master_eq_2}, the equation \emph{implicitly} specifies a family of $f(R)$ theories. Consider a particular theory belonging to this family that is asymptotic to GR ($f(R)=R$) at $R\to\infty$. The solution space of this theory can be imagined to be a phase space spanned by $\{h,h',h''\}$. If one wants to set initial conditions at a high redshift (say $z\approx6$) and obtain a cosmological solution $h(z)$ that is $\Lambda$CDM-like (\eqref{eq:LCDM}), one can get this for very fine-tuned initial conditions. Slight deviations in the initial conditions will result in an FLRW cosmology that can be very different from $\Lambda$CDM-like. 

\section{Phantom crossing $f(R)$ theories}\label{sec:phantom_crossing_f(R)}

Now that we have applied our framework to study $\Lambda$CDM-mimicking $f(R)$ models, let us now come to the somewhat more realistic cosmological solution specified by the cosmographic condition $j=1+3\epsilon\left(q-\frac{1}{2}\right)$. As we have shown in section \ref{subsec:phantom}, given the parameters satisfy the bound \eqref{eps_bound}, it is \emph{possible} for this toy model of cosmic evolution to admit a phantom crossing scenario, something that is indicated by DESI DR2 \cite{DESI:2025zgx}. 

At this point, let us clarify that what we are after is the family of $f(R)$ theories that give rise to a cosmological evolution of the form \eqref{eq:almost_LCDM}, and \emph{not} of the form $h^2(z)=\Omega_{m0}(1+z)^3+\Omega_{\rm{DE}0}$. The identification $\{\Omega_{m0},\Omega_{\rm{DE}0}\}=\{c_1,1-c_1\}$ is valid only for the $w$CDM model with $w_{\rm DE}=-1+\varepsilon$, which serves as the future asymptotic of all the models given by \eqref{eq:DDE_phantom}. Ironically, this latter model does not have any phantom crossing. The underlying reasoning is similar to that of the exact $\Lambda$CDM-like background $j=1$. In Eq.\eqref{eq:almost_LCDM}, $c_1$ is the coefficient of $(1+z)^3$, which can be identified with $\Omega_{m0}$ only when no part of the effective dark energy fluid is scaling as $\sim(1+z)^3$. As can be understood from Eq.\eqref{eq:rho_almost_LCDM}, this latter requirement is in general violated by a fluid with equation of state parameter \eqref{eq:DDE_phantom}, except for the very special $w$CDM case with $w_{\rm DE}=-1+\varepsilon$. If an $f(R)$ theory reproduces a cosmological evolution of the form \eqref{eq:almost_LCDM}, the energy density of the effective curvature fluid (\eqref{eq:rho_curv}), in general, scales partially as $\sim(1+z)^3$ and partially as $\sim(1+z)^{3\epsilon}$.

This particular case is a classic example of when the reconstruction method fails, whereas the theory space analysis still succeeds and provides a glimpse of the nature of the underlying theory. Before going to the theory space analysis, let us first show where exactly the reconstruction method fails. For $j=1+3\varepsilon\left(q-\frac{1}{2}\right)$, the second relation in \eqref{eq:CP_rel} gives the snap parameter
\begin{equation}
    s = -(2+3q) + \frac{3}{2}\varepsilon(3\varepsilon+q)(1-2q)
\end{equation}
Substituting these expressions for $j=j(q),\,s=s(q)$, as well as $w=0,\,\Omega_k=0$ into Eq.\eqref{eq:master_eq_1} gives 
\begin{subequations}\label{eq:master_eq_1_phantom}
\begin{align}\label{}
& -(q+1)^2\left[1 - 3\varepsilon\left(\frac{q-1/2}{q+1}\right)\right]^2 R^3 f^{(3)}(R) \nonumber\\
&\qquad\qquad  + (q+1)(q+2)(q-1)\left[1 - \frac{3}{2}\varepsilon\left(\frac{q-1/2}{q+1}\right)\left(\frac{q-2+6\varepsilon}{q+2}\right)\right] R^2 f''(R)  \nonumber\\
&\qquad\qquad\qquad\qquad\qquad\qquad  - (q-2)(q-1)^2 R f'(R) + 3(q-1)^3 f(R)=0\,.  
\end{align}    
\end{subequations}
One can check that Eq.\eqref{eq:master_eq_1_phantom} correctly reduces to Eq.\eqref{eq:master_eq_1_LCDM} for $\varepsilon\to0$, ensuring a consistency. The next step is to express $q$ as a function of $R$. For the cosmic evolution \eqref{eq:almost_LCDM}, one can calculate that
\begin{equation}\label{eq:dec_almost_LCDM}
    q(z) = -1 + (1+z)\frac{h'(z)}{h(z)} = \frac{1}{2} - \frac{3}{2}\left(\frac{1-c_1}{h^2}\right)(1-3\varepsilon)(1+z)^{3\varepsilon}\,.
\end{equation}
The next few steps in obtaining a reconstruction equation similar to \eqref{eq:recon_LCDM_1} were to express $h=h(q)$, substitute that into $R=6H_0^2h^2(1-q)$ to obtain $R=R(q)$, and invert this to obtain $q=q(R)$. However, for $\epsilon\neq0$, the relation \eqref{eq:dec_almost_LCDM} cannot be inverted to obtain $h=h(q)$, because of the explicit appearance of $z$ in \eqref{eq:dec_almost_LCDM}. This is exactly where the reconstruction method fails to reconstruct the underlying $f(R)$, even numerically, for the cosmic evolution corresponding to $j=1+3\varepsilon\left(q-1/2\right)$. 

The failure of the reconstruction method due to some kind of an invertibility issue is actually quite common in the reconstruction methods, and we run into a similar situation when we even slightly deviate from an exact $\Lambda$CDM-like evolution.
The above relation cannot be inverted to obtain $h=h(q)$. This is one of the big motivations behind introducing the theory space analysis, which, as we will show in the next subsection, allows us to get a glimpse at the nature of the underlying theory.

\subsection{Theory space analysis}

For $f(R)$ theories reproducing the cosmological solution \eqref{eq:almost_LCDM}, the autonomous system \eqref{eq:autonomous} can be reduced by substituting $\Omega_k=0,\,w=0,\,j=1+3\varepsilon(q-1/2),\,s=-(2+3q)+\frac{3}{2}\varepsilon(3\varepsilon+q)(1-2q)$:
\begin{subequations}\label{eq:autonomous_almost_LCDM}
    \begin{eqnarray}
       \frac{dr}{dN} &=& -r(m-r+1)\left(\frac{1+q}{1-q}\right)\left[1-\frac{3}{2}\varepsilon\left(\frac{q-1/2}{q+1}\right)\right]\,,
       \\
       \frac{dm}{dN} &=& \frac{r \left(m^2 (q+1)^2+m (q+1) \left(-q^2-2 q+1\right)-(2-q) (1-q)^2\right)+3 (1-q)^3}{(1+q)(1-q)r\left[1-\frac{3}{2}\varepsilon\left(\frac{q-1/2}{q+1}\right)\right]}\nonumber\\
       && + \frac{\frac{9}{4}\varepsilon ^2 m (1-2 q) r (m (1-2 q)+1) - \frac{3}{2}\varepsilon m (1-2 q) r \left(-(2 m+1) q+(4-2 m)+q^2\right)}{(1+q)(1-q)r\left[1-\frac{3}{2}\varepsilon\left(\frac{q-1/2}{q+1}\right)\right]}
       \\
       \frac{dq}{dN} &=& (2q-1)\left(q+1-\frac{3}{2}\varepsilon\right)\,.
    \end{eqnarray}
\end{subequations}
One can see from the above that $q=1/2$ acts as a repelling plane and $q=-1+\frac{3}{2}\varepsilon$ acts as an attracting plane, and these two planes cannot be crossed by any trajectory. The entire relevant cosmological dynamics occurs between these two planes, with $q$ monotonically decreasing from the value $1/2$ to the value $-1+\frac{3}{2}\varepsilon$ along the course of evolution. This monotonically decreasing nature of $q$ allows one to treat $-q$ as a proxy time variable and define the non-autonomous system
\begin{subequations}\label{eq:nonautonomous_almost_LCDM}
    \begin{align}
        \frac{dr}{d(-q)} = - \frac{dr/dN}{dq/dN} =&\,\,\, r(m-r+1)\left(\frac{1+q}{1-q}\right)\frac{\left[1-\frac{3}{2}\varepsilon\left(\frac{q-1/2}{q+1}\right)\right]}{(2q-1)\left(q+1-\frac{3}{2}\varepsilon\right)}\,,
        \\
        \frac{dm}{d(-q)} = - \frac{dm/dN}{dq/dN} =& -\frac{r \left(m^2 (q+1)^2+m (q+1) \left(-q^2-2 q+1\right)-(2-q) (1-q)^2\right)+3 (1-q)^3}{(1+q)(1-q)r\left[1-\frac{3}{2}\varepsilon\left(\frac{q-1/2}{q+1}\right)\right](2q-1)\left(q+1-\frac{3}{2}\varepsilon\right)}\nonumber\\
       & - \frac{\frac{9}{4}\varepsilon ^2 m (1-2 q) r (m (1-2 q)+1) - \frac{3}{2}\varepsilon m (1-2 q) r \left(-(2 m+1) q+(4-2 m)+q^2\right)}{(1+q)(1-q)r\left[1-\frac{3}{2}\varepsilon\left(\frac{q-1/2}{q+1}\right)\right](2q-1)\left(q+1-\frac{3}{2}\varepsilon\right)}
    \end{align}
\end{subequations}
The domain of validity of $q$ in which the system \eqref{eq:nonautonomous_almost_LCDM} is regular is given by
\begin{equation}
    \mathcal{D} = \{q\in\mathbb{R}:q\neq1 \land q\neq\frac{1}{2} \land q\neq-1 \land q\neq-1+\frac{3}{2}\varepsilon \land r(q)\neq0\}\,.
\end{equation}
The system \eqref{eq:nonautonomous_almost_LCDM} can be solved to find meaningful results within a range of $q$ where the above validity condition is always satisfied. As we will see later on, we will take a small value of $\varepsilon$ ($\sim\mathcal{O}(10^{-2})$), and solve the system within a range $-0.55\leq q\leq0.49$. As we will see, for the chosen value of $\varepsilon$, and within this range of $q$, $r(q)>1$ always. Therefore, we can safely say that the system \eqref{eq:nonautonomous_almost_LCDM} remains regular in the domain of $q$ that we consider, and the subsequent results remain meaningful.

In the context of \eqref{eq:almost_LCDM}, the deviation parameter $\varepsilon$ must be small ($0<\varepsilon<1$) for a standard General Relativistic matter-dominated limit to exist at the high-redshift limit. When $\varepsilon$ is small, the cosmology of the form \eqref{eq:almost_LCDM} can be termed as an \emph{almost} $\Lambda$CDM-like cosmic evolution. It has been recently shown that such almost $\Lambda$CDM-like cosmic evolutions can actually be significantly different from the $\Lambda$CDM model physically, in the sense that they give rise to a dark energy equation of state that deviates from $-1$ quite significantly \cite{Chakraborty:2025rvc}. Along the same line, for small $\varepsilon$, the cosmic evolution of the form \eqref{eq:almost_LCDM} is an almost $\Lambda$CDM-like cosmic solution that is designed to asymptote to the exact $\Lambda$CDM-like evolution $j=1$ at high redshift and allow for a phantom crossing at a lower redshift.

\begin{figure}[H]
    \centering
    
    \begin{subfigure}[b]{0.45\linewidth}
    \includegraphics[width=\linewidth]{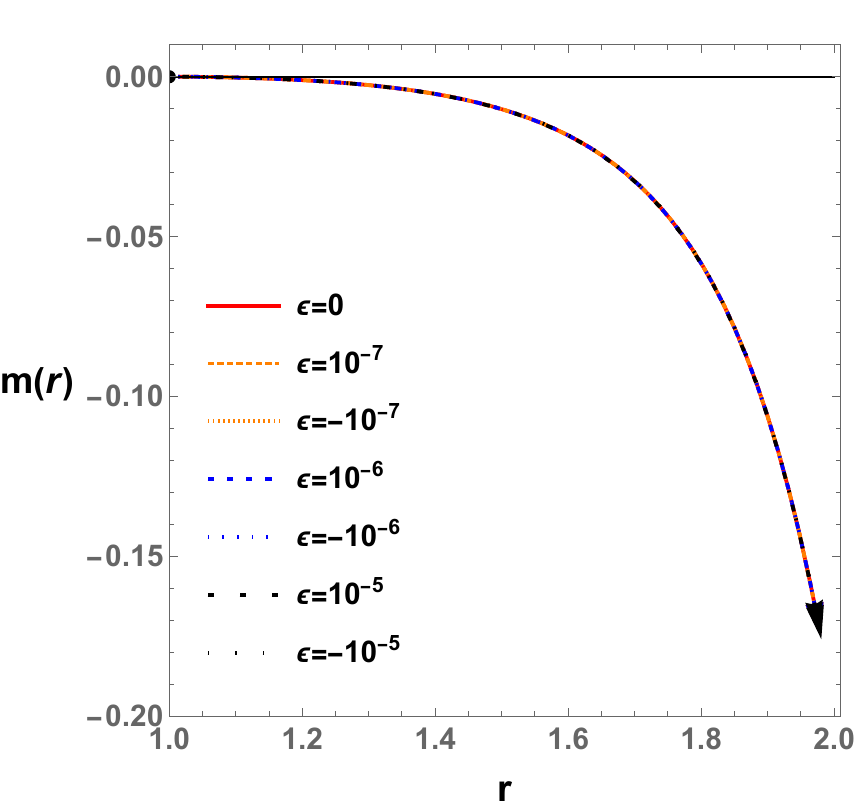}
     \caption{}
     \label{fig:m-r plot_1_almost}
    \end{subfigure}
    \hfill
    \begin{subfigure}[b]{0.45\linewidth}
    \includegraphics[width=\linewidth]{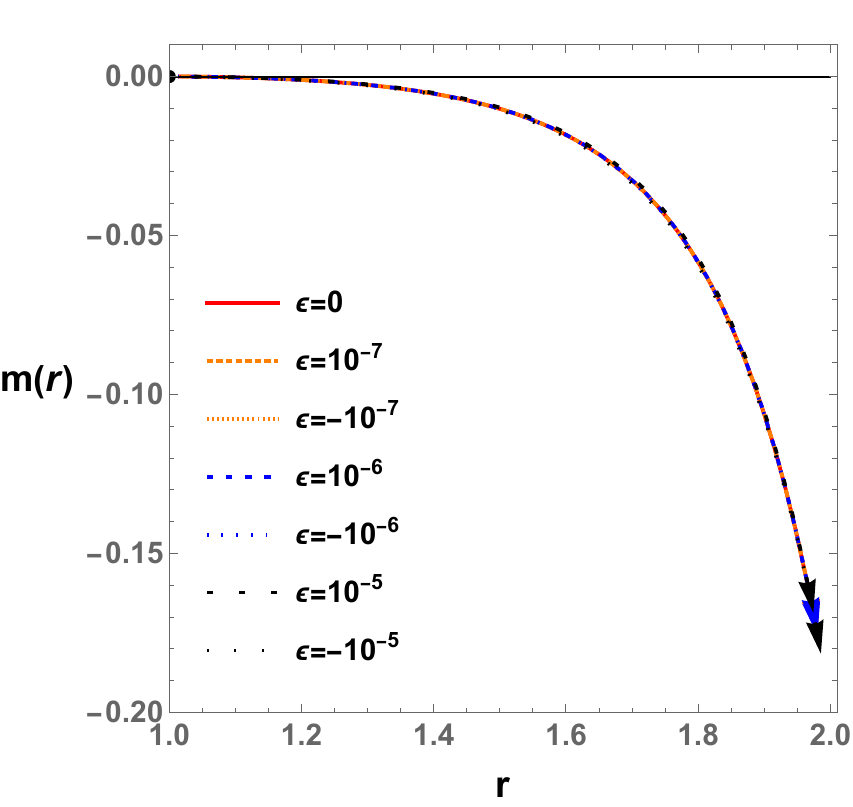}
     \caption{}
     \label{fig:m-r plot_2_almost}
    \end{subfigure}
    
    \begin{subfigure}[b]{0.45\linewidth}
    \includegraphics[width=\linewidth]{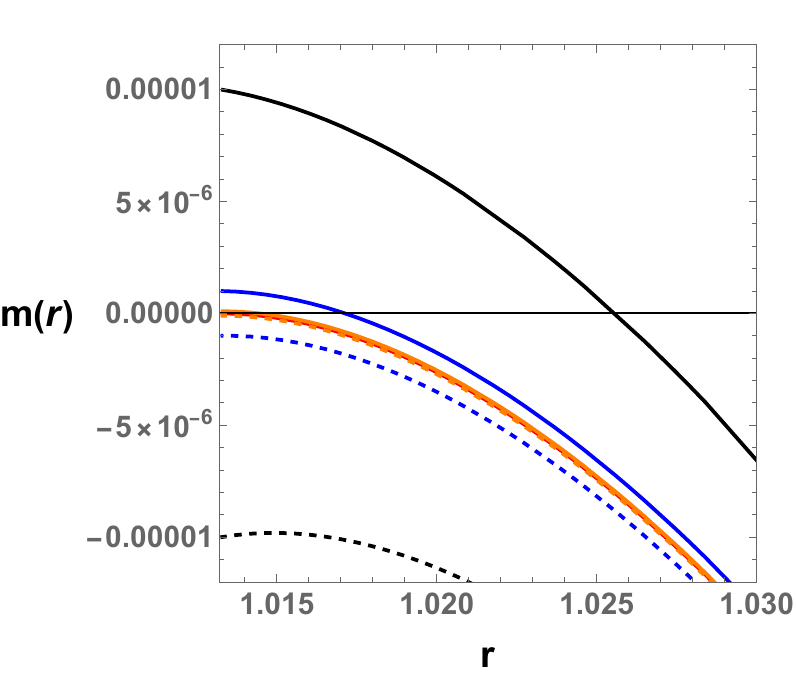}
     \caption{}
     \label{fig:m-r plot_1_almost_zoomed}
    \end{subfigure}
    \hfill 
    \begin{subfigure}[b]{0.45\linewidth}
    \includegraphics[width=\linewidth]{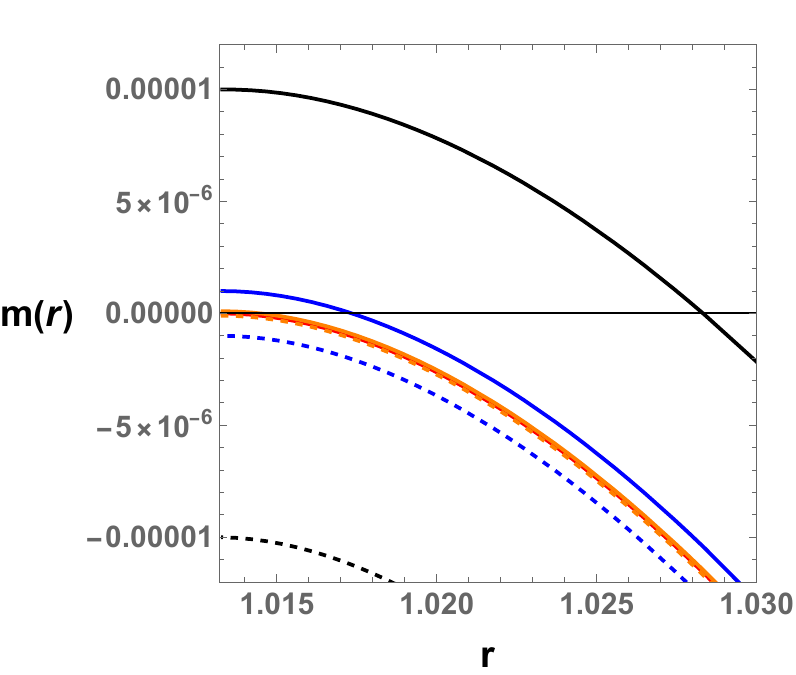}
     \caption{}
     \label{fig:m-r plot_2_almost_zoomed}
    \end{subfigure}
    
    \begin{subfigure}[b]{0.45\linewidth}
    \includegraphics[width=\linewidth]{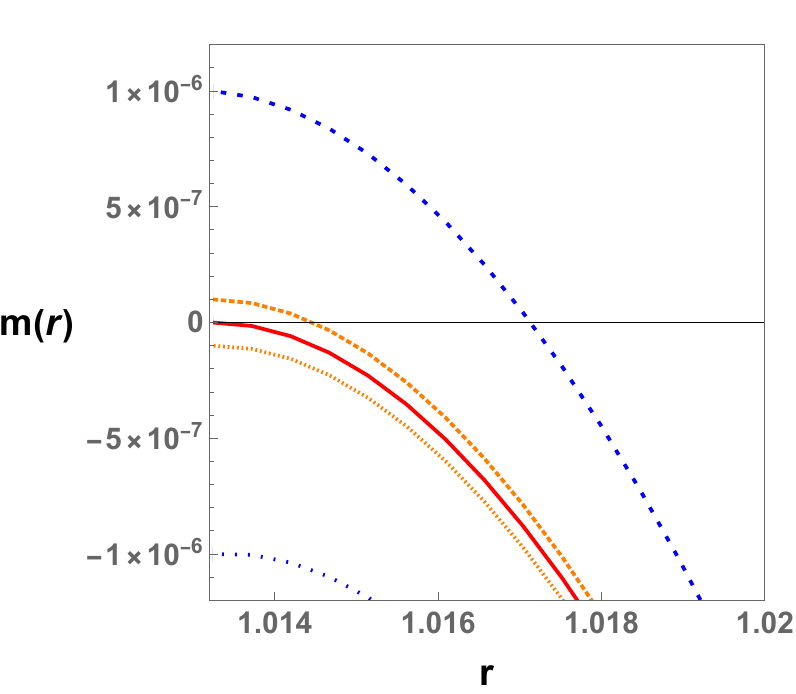}
     \caption{}
     \label{fig:m-r plot_1_almost_zoomed_zoomed}
    \end{subfigure}
    \hfill
    \begin{subfigure}[b]{0.45\linewidth}
    \includegraphics[width=\linewidth]{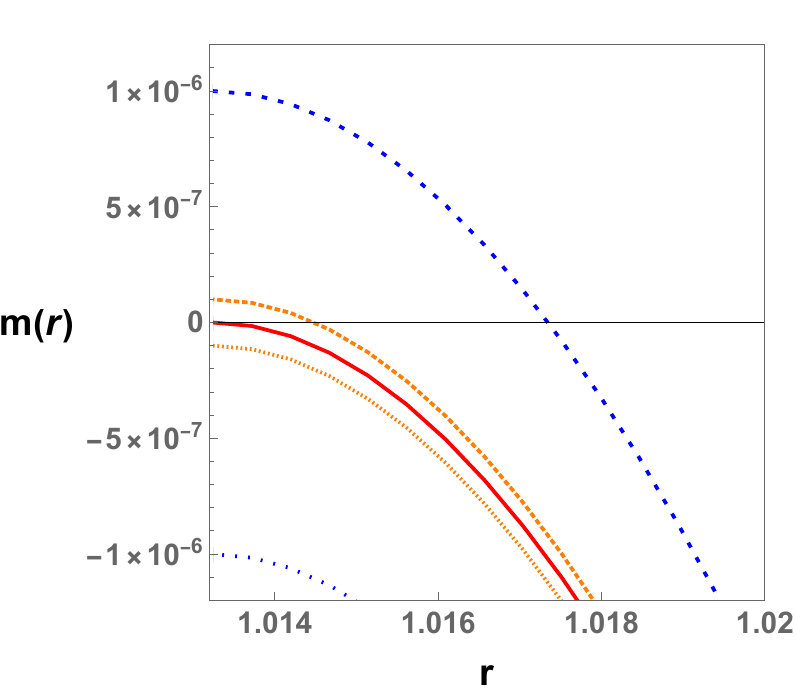}
     \caption{}
     \label{fig:m-r plot_2_almost_zoomed_zoomed}
    \end{subfigure}
    
    \caption{Parametric curves $\{r(q), m(q)\}$ show $f(R)$ cosmological dynamics matching $j = 1 + 3\varepsilon(q - 1/2)$ ($\varepsilon = 10^{-2}$). Panels \eqref{fig:m-r plot_1_almost} and \eqref{fig:m-r plot_2_almost} present solutions of system \eqref{eq:nonautonomous_almost_LCDM} with initial conditions $\big\{ r(0.49), m(0.49) \big\} = \big\{ \frac{3q-3}{q-2}\vert_{q=0.49} \pm \epsilon, \epsilon \big\}$. The red, dashed orange, dotted orange, dashed blue, dotted blue, dashed black, dotted black curves correspond to $\epsilon=0,10^{-7},-10^{-7},10^{-6},-10^{-6},10^{-5},-10^{-5}$ respectively (indicated in the figures of the top panel). Panels \ref{fig:m-r plot_1_almost_zoomed} and \ref{fig:m-r plot_2_almost_zoomed} (respectively \ref{fig:m-r plot_2_almost_zoomed} and \ref{fig:m-r plot_2_almost_zoomed_zoomed}) provide successive zooms near $\{r,m\}=\{1,0\}$ of panel \ref{fig:m-r plot_1_almost} (respectively \ref{fig:m-r plot_2_almost}). All plots span $q=0.49$ to $q_0 \approx -0.55$. Initial conditions match Fig.~\ref{fig:LCDM_mimicking_m(r)s_forward} to compare exact and near-$\Lambda$CDM $f(R)$ cosmologies, which coincide as $q \to 1/2$.}
    \label{fig:almost_LCDM_mimicking_m(r)s_forward}
\end{figure}

It is interesting to compare how the underlying gravity theories driving an almost $\Lambda$CDM-like evolution $j=1+3\varepsilon(q-1/2)$ evolutions compare with the underlying gravity theories driving an exact $\Lambda$CDM-like evolution $j=1$. Since the merge with each other in the limit $q\to1/2$, it would be best to solve the non-autonomous system \eqref{eq:nonautonomous_almost_LCDM} by setting the same initial conditions at $q=0.49$ as what was used to solve the system \eqref{eq:nonautonomous_LCDM}. The results are shown in Fig.\ref{fig:almost_LCDM_mimicking_m(r)s_forward} for $\varepsilon=10^{-2}$.
Comparing Fig.\ref{fig:LCDM_mimicking_m(r)s_forward} and Fig.\ref{fig:almost_LCDM_mimicking_m(r)s_forward} allows us to unlock a distinctive feature between the $f(R)$ theories that reproduce an exact $\Lambda$CDM-like {\it vs} those that reproduce an almost $\Lambda$CDM-like cosmological solutions. Recall that the region $m\equiv\frac{Rf''}{f'}<0$ is characterized by some kind of a theoretical instability (either the ghost instability $f'<0$ or the tachyonic instability $f''<0$). For the theories that reproduce an exact $\Lambda$CDM-like evolution $j=1$, the underlying theories may or may not become plagued by a theoretical instability along the course of evolution, depending on the initial conditions. On the contrary, for the theories that reproduce an almost $\Lambda$CDM-like evolution $j=1+3\varepsilon(q-1/2)$, the underlying theories certainly become plagued by instability, as \emph{all} the solution curves in Fig.\ref{fig:almost_LCDM_mimicking_m(r)s_forward} end up in the region $m<0$ at $q=-0.55$ (today).

\begin{figure}[H]
    \centering
    \begin{subfigure}[b]{0.49\linewidth}
    \includegraphics[width=\linewidth]{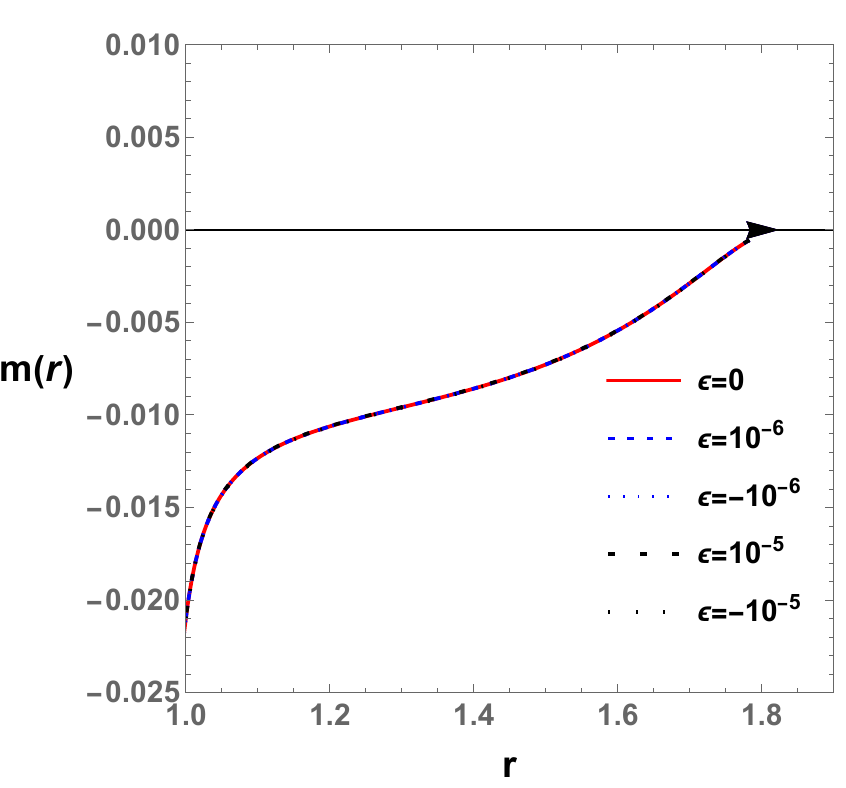}
     \caption{}
     \label{fig:m-r plot_3_almost}
    \end{subfigure}
    \hspace{1.6mm}
    \begin{subfigure}[b]{0.49\linewidth}
    \includegraphics[width=\linewidth]{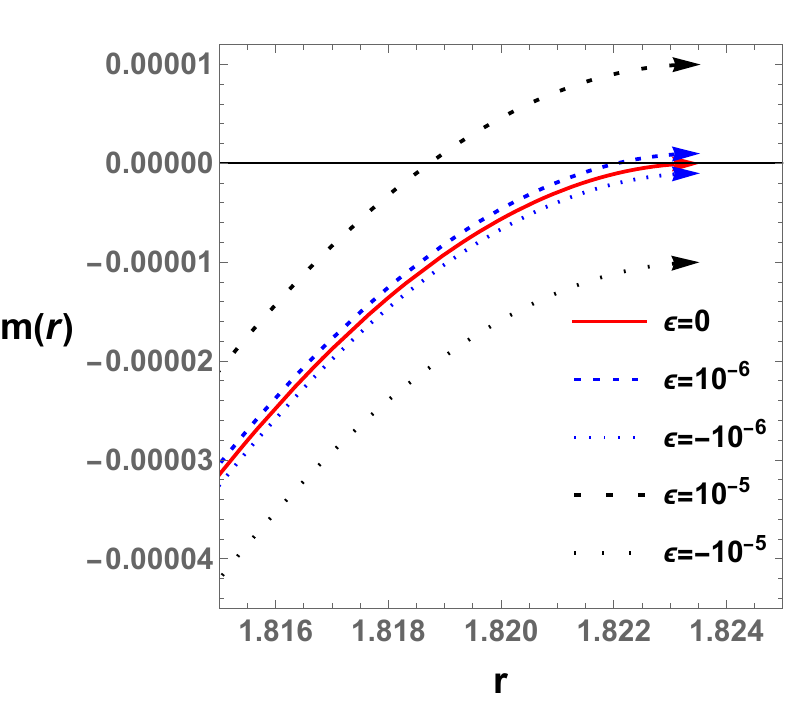}
     \caption{}
     \label{fig:m-r plot_3_almost_zoomed}
    \end{subfigure}
    \begin{subfigure}[b]{0.49\linewidth}
    \includegraphics[width=\linewidth]{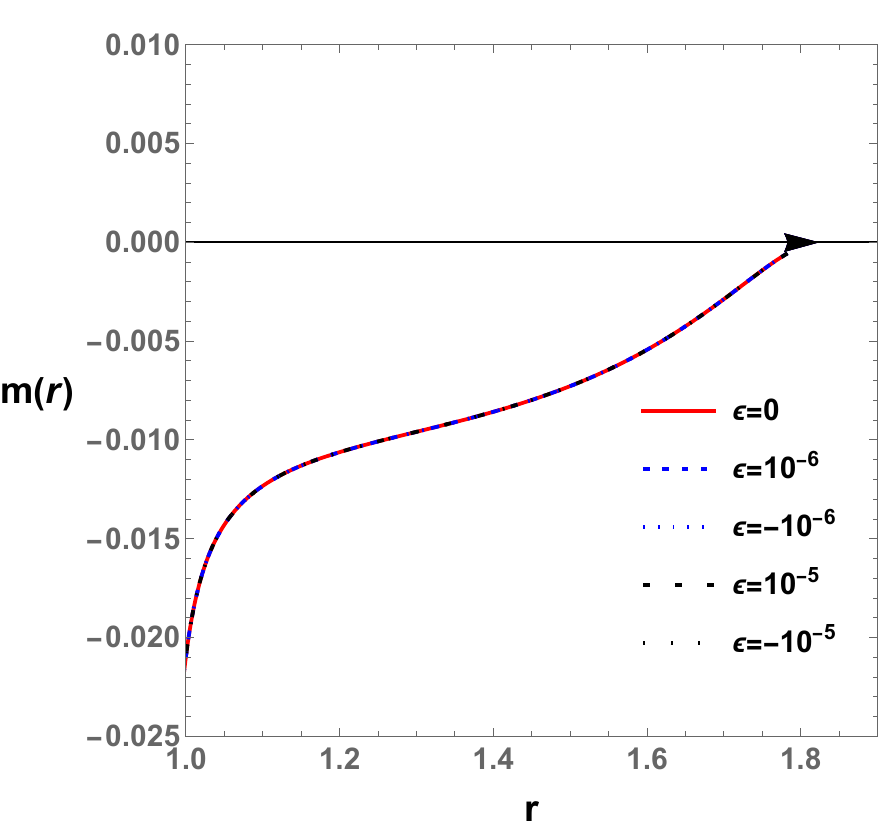}
     \caption{}
     \label{fig:m-r plot_4_almost}
    \end{subfigure}
    \hspace{1.6mm}
    \begin{subfigure}[b]{0.49\linewidth}
    \includegraphics[width=\linewidth]{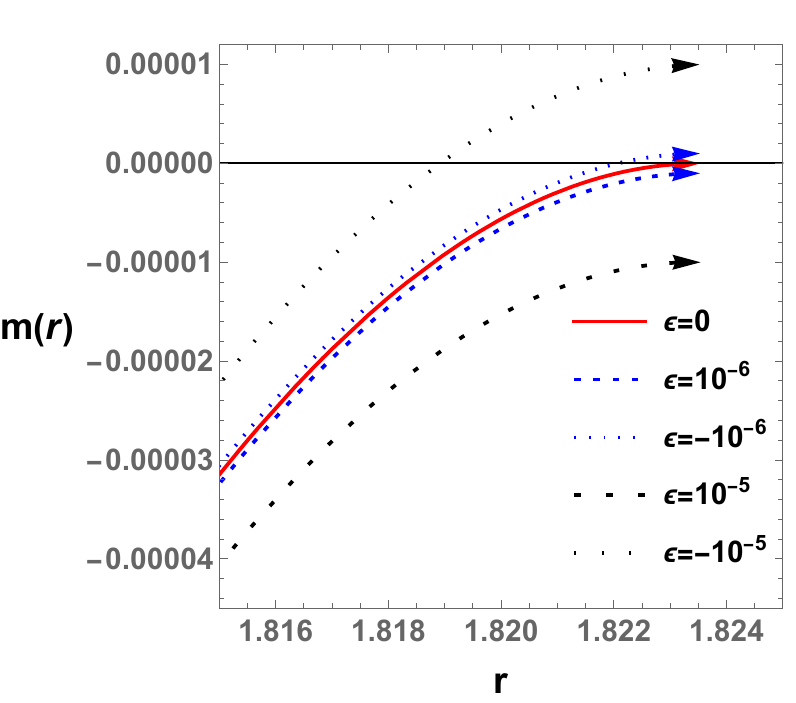}
     \caption{}
     \label{fig:m-r plot_4_almost_zoomed}
    \end{subfigure}
    \caption{The curves in the panel \eqref{fig:m-r plot_3_almost} corresponds to solutions of the nonautonomous system \eqref{eq:nonautonomous_almost_LCDM} with the initial condition $\lbrace r(-0.55),m(-0.55) \rbrace = \lbrace \frac{3q-3}{q-2}\vert_{q=-0.55} + \epsilon,\epsilon\rbrace$, whereas that in the panel \eqref{fig:m-r plot_4_almost} corresponds to the initial conditions $\lbrace r(-0.55),m(-0.55) \rbrace = \lbrace \frac{3q-3}{q-2}\vert_{q=-0.55} + \epsilon,-\epsilon\rbrace$. The red, dashed blue, dotted blue, dashed black, dotted black curves correspond to $\epsilon=0,10^{-6},-10^{-6},10^{-5},-10^{-5}$ respectively. Panels \ref{fig:m-r plot_3_almost_zoomed} and \ref{fig:m-r plot_4_almost_zoomed} are zooms of the panels \ref{fig:m-r plot_3_almost} and \ref{fig:m-r plot_2_almost_zoomed} respectively near the tip of the solution curves. All the parametric plots are made within the range $q=0.49$ to $q=q_0\approx-0.55$ (today). The initial conditions chosen for these plots are exactly the same as those in Fig.\ref{fig:LCDM_mimicking_m(r)s_backward}, and same colour coding is maintained, to facilitate easy comparison.}
    \label{fig:almost_LCDM_mimicking_m(r)s_backward}
\end{figure}

For completeness, one can again ask, what if the underlying theory is very close to GR at the current epoch? To investigate that picture, one must set \emph{almost GR} initial conditions at the present epoch $q=-0.55$. Since, for a small value of $\varepsilon$ like $\varepsilon=10^{-2}$, the cosmic evolution is almost $\Lambda$CDM-like, let us take exactly the same initial conditions as were taken to produce the plots of Fig.\ref{fig:LCDM_mimicking_m(r)s_backward}, and use them to solve the system \eqref{eq:nonautonomous_almost_LCDM}. The results are shown in Fig.\ref{fig:almost_LCDM_mimicking_m(r)s_backward}.
Comparing Fig.\ref{fig:LCDM_mimicking_m(r)s_backward} and Fig.\ref{fig:almost_LCDM_mimicking_m(r)s_backward}, one can notice that, whereas it is possible for some $\Lambda$CDM-mimicking $f(R)$ cosmological solutions to always above the $m=0$ line, for $f(R)$ cosmologies reproducing an \emph{almost} $\Lambda$CDM-like solution $j=1+3\varepsilon(q-1/2)$ \emph{always} ends up below the $m=0$ line in the past. 

\begin{figure}[H]
    \centering

    \begin{subfigure}[b]{0.45\linewidth}
    \includegraphics[width=\linewidth]{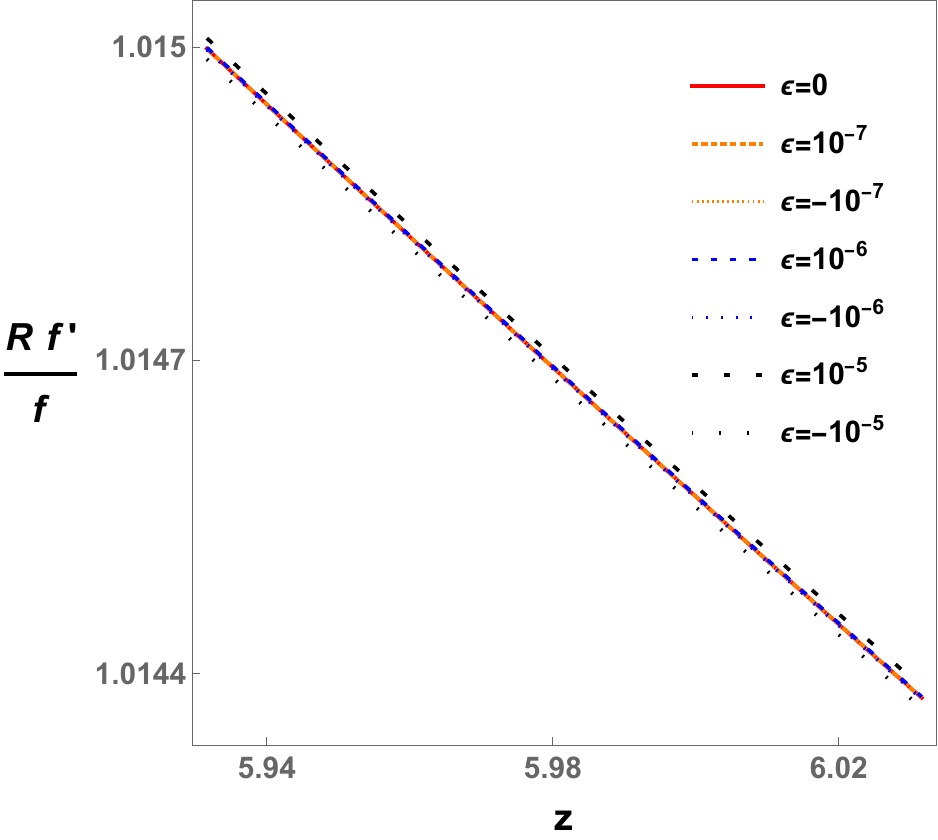}
     \caption{}
     \label{fig:r-z plot_1 (almost)}
    \end{subfigure}
    \hspace{2mm}
    \vspace{0.5cm}
    \begin{subfigure}[b]{0.45\linewidth}
    \includegraphics[width=\linewidth]{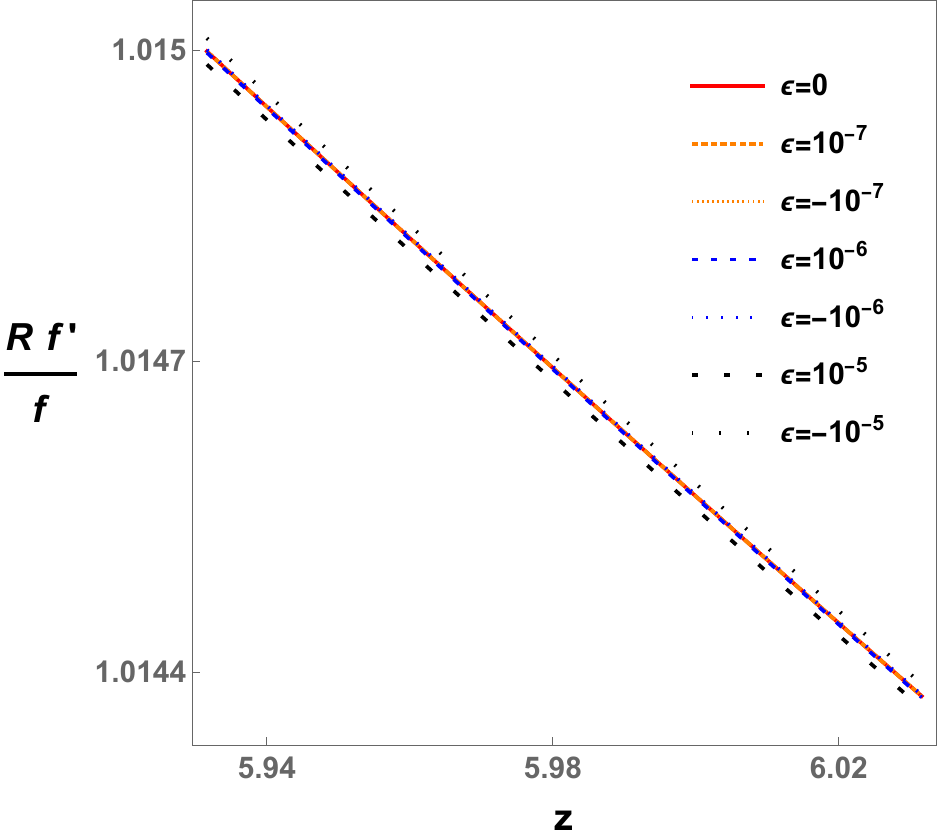}
     \caption{}
     \label{fig:r-z plot_2 (almost)}
    \end{subfigure}
    
    \begin{subfigure}[b]{0.45\linewidth}
    \includegraphics[width=\linewidth]{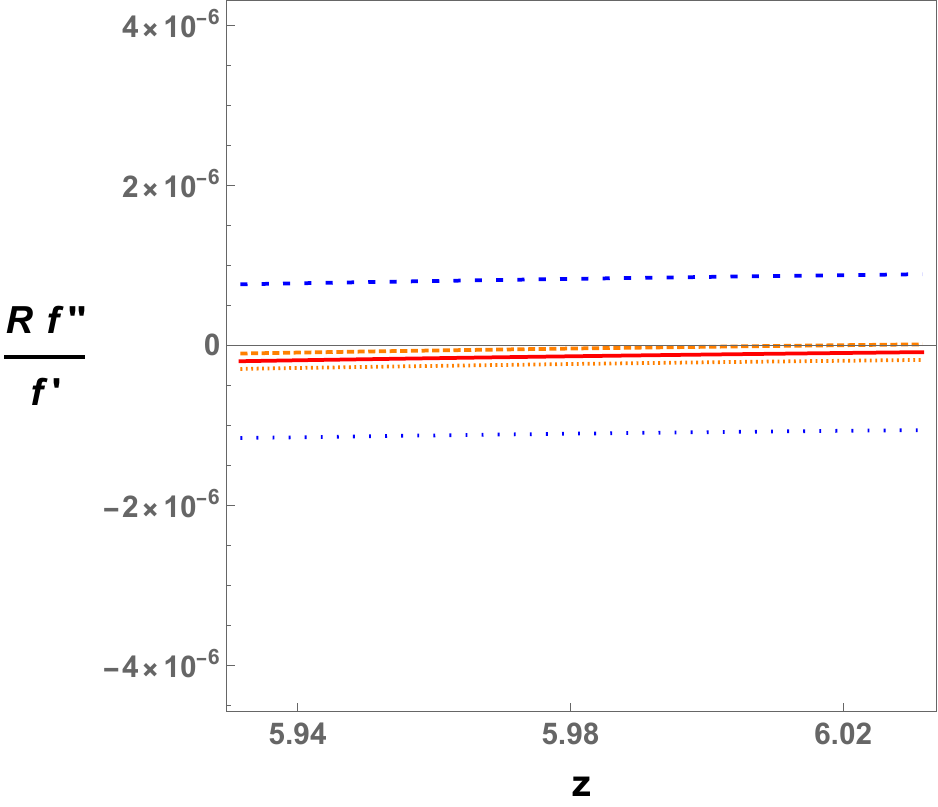}
     \caption{}
     \label{fig:m-z plot_1 (almost)}
    \end{subfigure}
    \hspace{2mm}
    \vspace{0.5cm}
    \begin{subfigure}[b]{0.45\linewidth}
    \includegraphics[width=\linewidth]{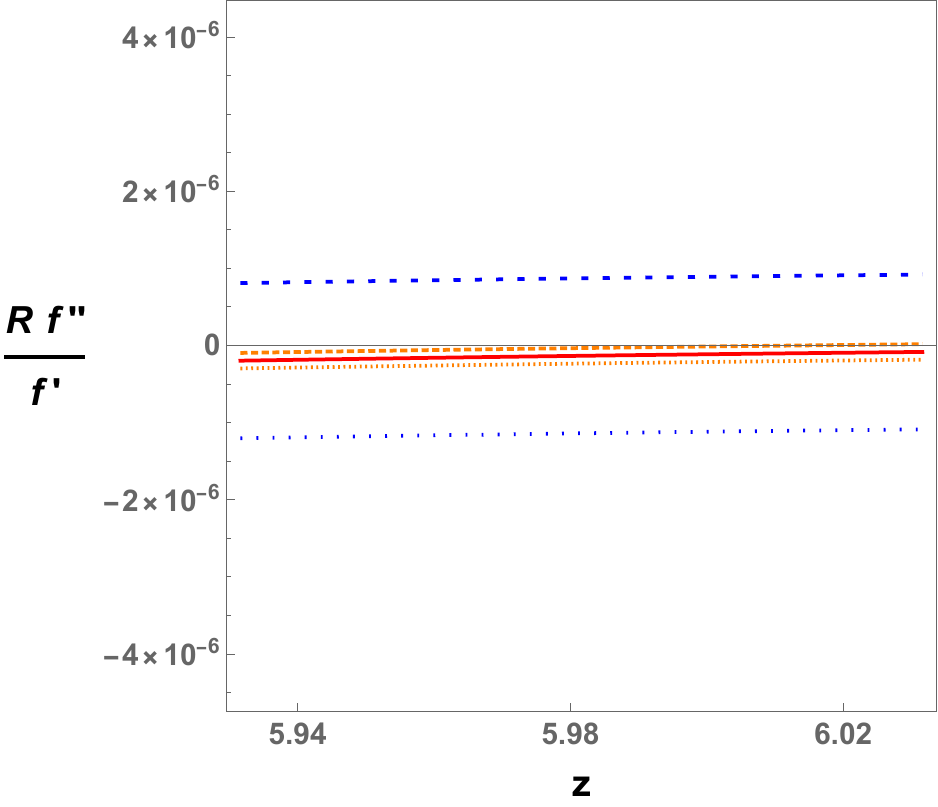}
     \caption{}
     \label{fig:m-z plot_2 (almost)}
    \end{subfigure}
    
    \begin{subfigure}[b]{0.45\linewidth}
    \includegraphics[width=\linewidth]{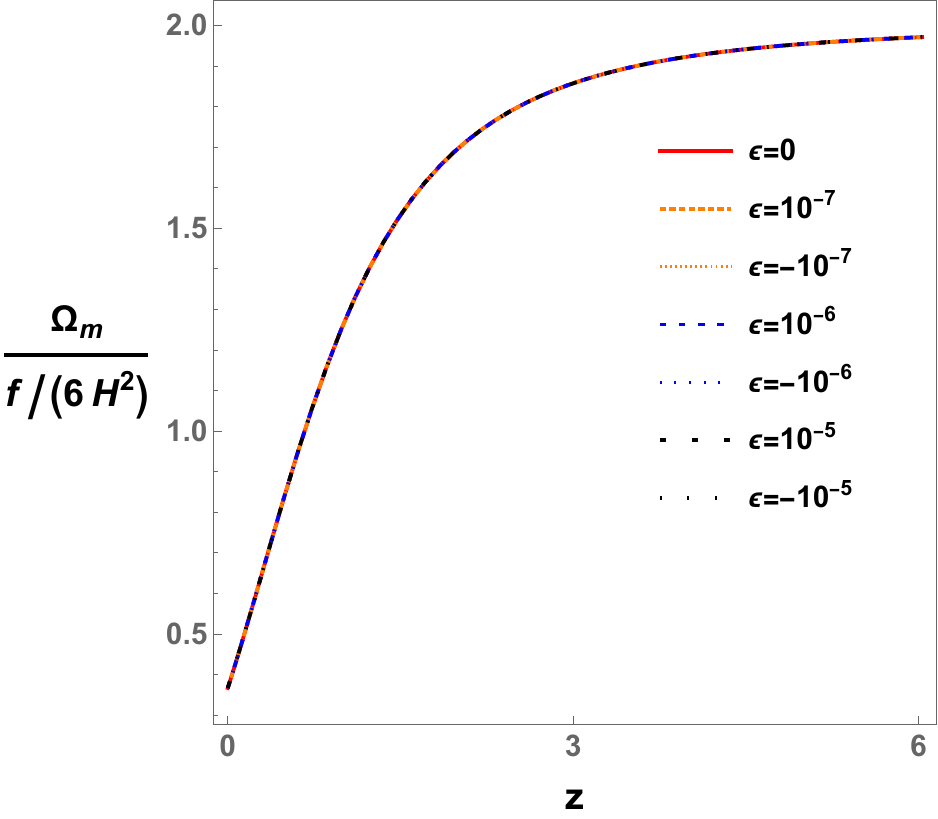}
     \caption{}
     \label{fig:Omega_m-z plot_1 (almost)}  
    \end{subfigure}
    \hspace{2mm}
    \vspace{0.5cm}
    \begin{subfigure}[b]{0.45\linewidth}
    \includegraphics[width=\linewidth]{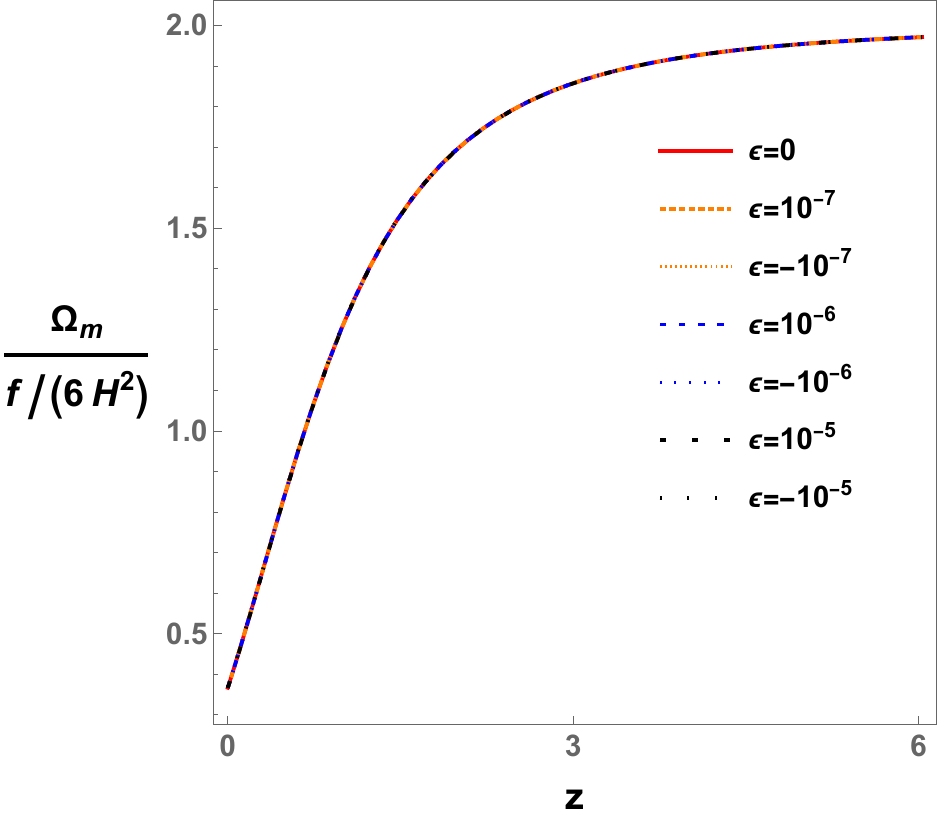}
     \caption{}
     \label{fig:Omega_m-z plot_2 (almost)}  
    \end{subfigure}
    
    \caption{The left and the right panels portray the evolution of the quantities $r\equiv\frac{Rf'}{f},\,m\equiv\frac{Rf''}{f'}$ and $\frac{\Omega_m}{f/(6H^2)}$ with respect to the redshift $z$ corresponding to the solution curves presented in Fig.\ref{fig:m-r plot_1_almost} and Fig.\ref{fig:m-r plot_2_almost} respectively. The red, dashed orange, dotted orange, dashed blue, dotted blue, dashed black, dotted black curves correspond to $\epsilon=0,10^{-7},-10^{-7},10^{-6},-10^{-6},10^{-5},-10^{-5}$ respectively (indicated in the figures of the top and bottom panels). The behaviours of $r(z)$ and $m(z)$ are so close to each other that, to show them distinctively, we plot them in the redshift range from $z_{\rm in}=6.0316$ to $z=z_{\rm in}-1$. The plot of $\frac{\Omega_m}{f/(6H^2)}$ is done for the full range $z_{\rm in}$ to $z=0$.}
    \label{fig:r-m vs z forward (almost)}
\end{figure}

\begin{figure}[H]
    \centering
    
    \begin{subfigure}[b]{0.45\linewidth}
    \includegraphics[width=\linewidth]{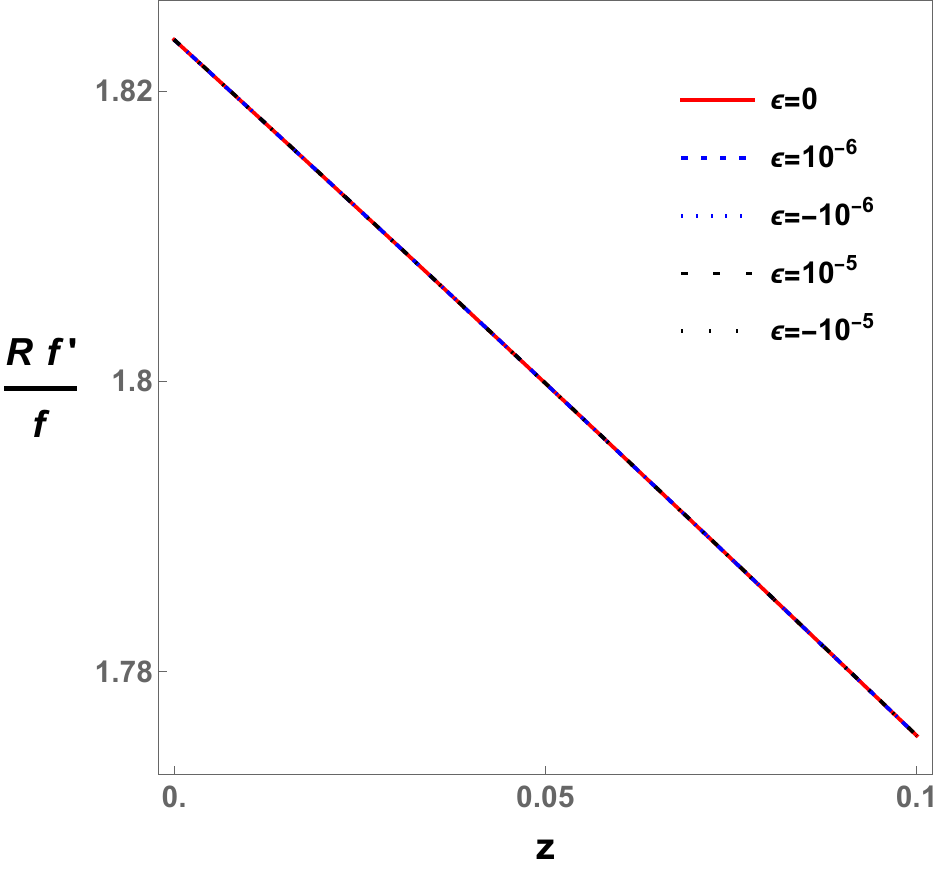}
     \caption{}
     \label{fig:r-z plot_3 (almost)}
    \end{subfigure}
    \hspace{2mm}
    \vspace{0.5cm}
    \begin{subfigure}[b]{0.45\linewidth}
    \includegraphics[width=\linewidth]{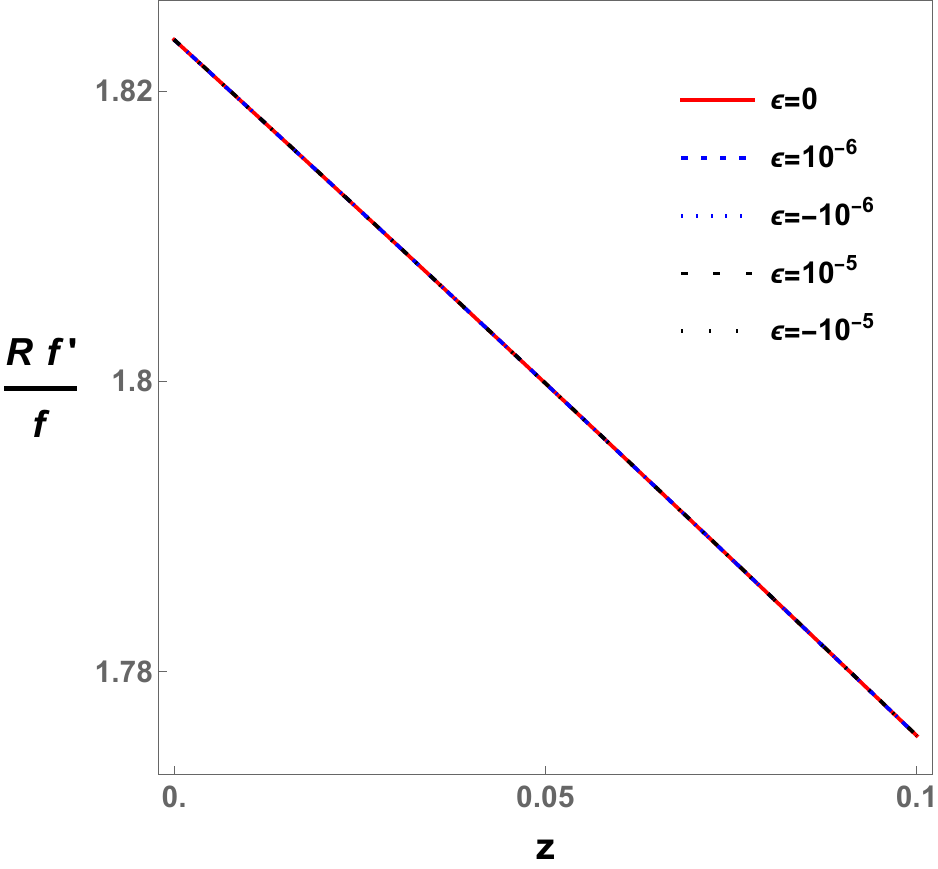}
     \caption{}
     \label{fig:r-z plot_4 (almost)}
    \end{subfigure}
    
    \begin{subfigure}[b]{0.45\linewidth}
    \includegraphics[width=\linewidth]{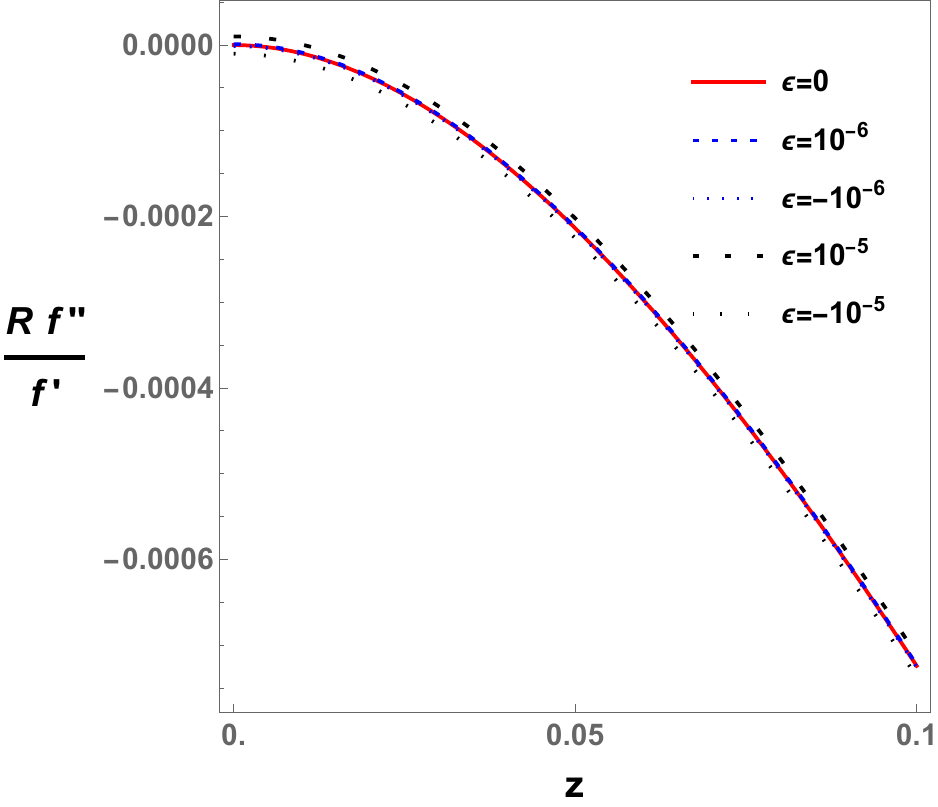}
     \caption{}
     \label{fig:m-z plot_3 (almost)}
    \end{subfigure}
    \hspace{2mm}
    \vspace{0.5cm}
    \begin{subfigure}[b]{0.45\linewidth}
    \includegraphics[width=\linewidth]{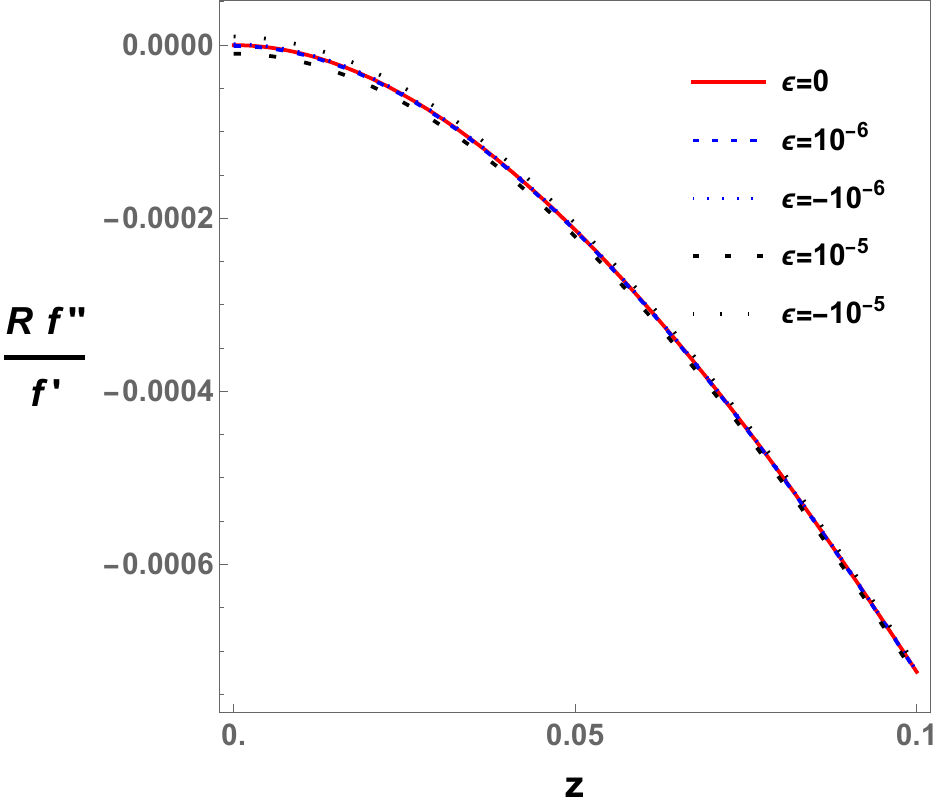}
     \caption{}
     \label{fig:m-z plot_4 (almost)}
    \end{subfigure}
    
    \begin{subfigure}[b]{0.45\linewidth}
    \includegraphics[width=\linewidth]{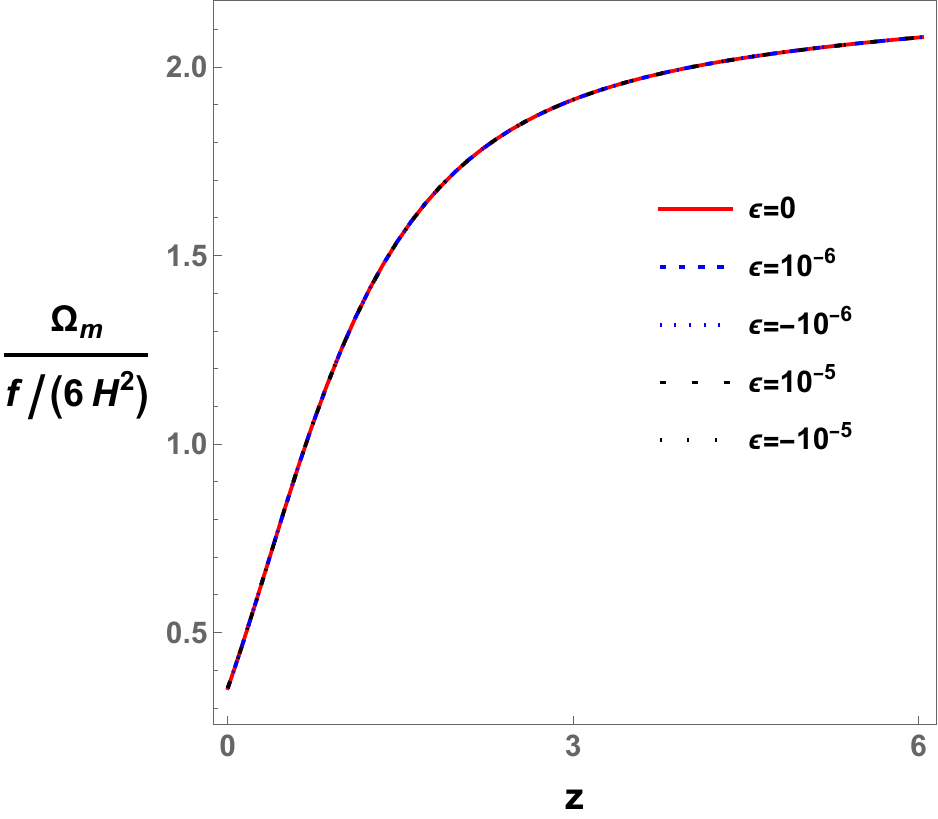}
     \caption{}
     \label{fig:Omega_m-z plot_3 (almost)}  
    \end{subfigure}
    \hspace{2mm}
    \vspace{0.5cm}
    \begin{subfigure}[b]{0.45\linewidth}
    \includegraphics[width=\linewidth]{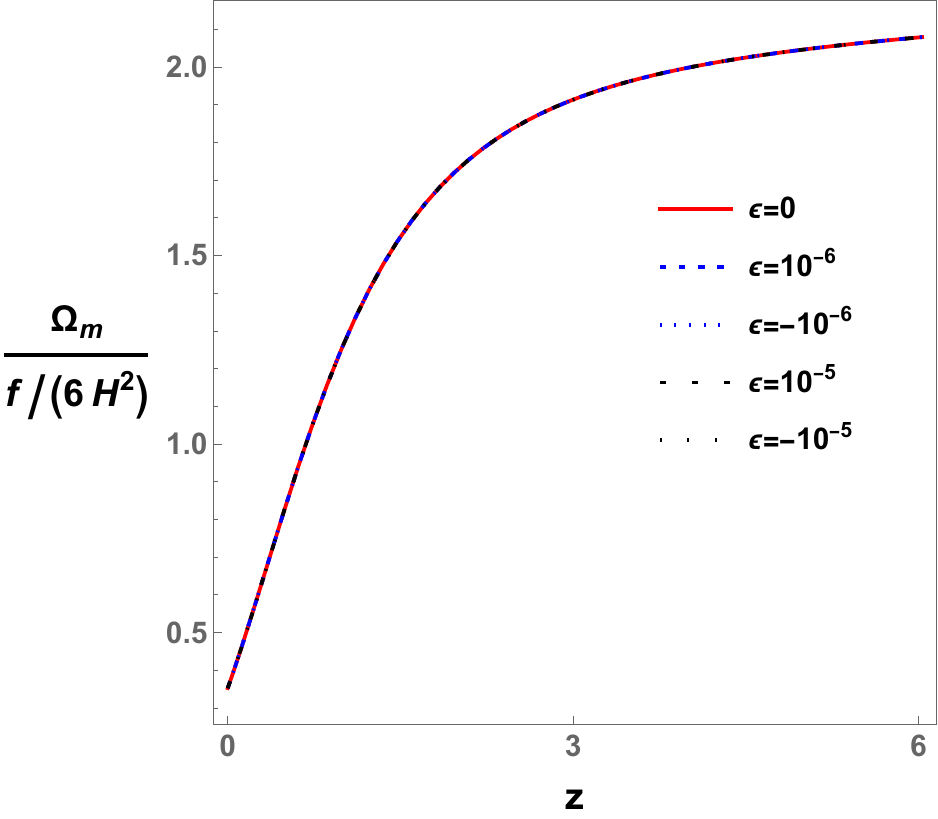}
     \caption{}
     \label{fig:Omega_m-z plot_4 (almost)}  
    \end{subfigure}
    
    \caption{The left and the right panels portray the evolution of the quantities $r\equiv\frac{Rf'}{f},\,m\equiv\frac{Rf''}{f'}$ and $\frac{\Omega_m}{f/(6H^2)}$ with respect to the redshift $z$ corresponding to the solution curves presented in Fig.\ref{fig:m-r plot_3_almost} and Fig.\ref{fig:m-r plot_4_almost} respectively. The red, dashed blue, dotted blue, dashed black, dotted black curves correspond to $\epsilon=0,10^{-7},-10^{-7},10^{-6},-10^{-6},10^{-5},-10^{-5}$ respectively. The behaviours of $r(z)$ and $m(z)$ are so close to each other that, to show them distinctively, we plot them in the redshift range from $z=0.1$ to $z=0$ (today).  The plot of $\frac{\Omega_m}{f/(6H^2)}$ is done for the full range $z_{\rm in}$ to $z=0$.}
    \label{fig:r-m vs z backward (almost)}
\end{figure}

To finish our comparison with the exact $\Lambda$CDM-like scenario, we also present in Figs.\ref{fig:r-m vs z forward (almost)} and \ref{fig:r-m vs z backward (almost)}, the redshift evolution of the quantities $r\equiv\frac{Rf'}{f},\,m\equiv\frac{Rf''}{f'}$ and $\frac{\Omega_m}{f/(6H^2)}$ corresponding to the solution curves presented in Figs.\ref{fig:almost_LCDM_mimicking_m(r)s_forward} and \ref{fig:almost_LCDM_mimicking_m(r)s_backward} respectively. Once the numerical solutions $r(z),\,m(z)$ is known, the quantity $\frac{\Omega_m}{f/(6H^2)}$ can be calculated from Eq.\eqref{eq:energy_eq_mr}

As in the case of the $f(R)$ models mimicking a $\Lambda$CDM-like cosmic evolution, one can notice that, for $f(R)$ models mimicking an almost $\Lambda$CDM-like phantom crossing evolution as well, the positivity of the quantities $r=\frac{Rf'}{f}$ and $\frac{\Omega_m}{f/(6H^2)}$ in Figs.\ref{fig:r-m vs z forward (almost)} and \ref{fig:r-m vs z backward (almost)} imply that the $f(R)$ and $f'(R)$ are always remain positive along the course of evolution, so that no ghost instability appears. The negativity of the quantity $m=\frac{Rf''}{f'}$ indicates $f''(R)<0$, i.e. an inevitable appearance of a tachyonic instability.

\subsection{Comparison of $\Lambda$CDM-mimicking and almost $\Lambda$CDM-mimicking phantom crossing $f(R)$ cosmologies}

Our study reveals an interesting distinction between the underlying $f(R)$ theories mimicking an \emph{exact} $\Lambda$CDM-like cosmology $j=1$ and those mimicking an \emph{almost} $\Lambda$CDM-like phantom crossing cosmology $j=1+3\varepsilon(q-1/2)$ with small positive $\varepsilon$. If one looks at the solution curves of Figs.\ref{fig:LCDM_mimicking_m(r)s_forward} or \ref{fig:LCDM_mimicking_m(r)s_backward}, one finds that initial conditions (whether set at a high redshift or at the present epoch) of different orders of magnitude produce distinct solution curves, with the solutions diverging away from each other along the direction the system is solved (towards the future or the past). The distinct evolution in the $m$-$r$ plot for the case of $\Lambda$CDM-mimicking $f(R)$ cosmologies actually stems from the distinct evolution of the quantity $m(z)$ even though $r(z)$ evolves almost parallelly for all the initial conditions (see Figs.\ref{fig:r-m vs z forward} and \ref{fig:r-m vs z backward}). Mathematically, one can say that the solution curves on the $r$-$m$ plane are not stable; slight variation in the initial conditions in $r,\,m$ results in very different solution curves. Physically, this means that if we consider two possible $\Lambda$CDM-mimicking $f(R)$ cosmological solutions, whose underlying driving $f(R)$'s are close to each other in the past (or present), the theories actually become quite different from each other at present (or in the past).

On the contrary, if one looks at the solution curves of Figs.\ref{fig:almost_LCDM_mimicking_m(r)s_forward} or \ref{fig:almost_LCDM_mimicking_m(r)s_backward}, one finds that initial conditions of different orders of magnitude produce solution curves that evolve almost parallelly along the direction of solving the system (either past to present or present to past). This gives rise to the highly overlapped and almost indistinguishable solution curves in Figs.\ref{fig:m-r plot_1_almost},\ref{fig:m-r plot_2_almost},\ref{fig:m-r plot_3_almost},\ref{fig:m-r plot_4_almost}. This is consistent with the almost parallel evolution of both $r(z)$ and $m(z)$ for all the initial conditions (see Figs.\ref{fig:r-m vs z forward (almost)} and \ref{fig:r-m vs z backward (almost)}). Mathematically, one can say that the solution curves on the $r$-$m$ plane are stable; slight variation in the initial conditions in $r,\,m$ does not result in drastically different solution curves in the $r$-$m$ plane.  Physically, this means that if we consider two possible almost $\Lambda$CDM-mimicking phantom crossing $f(R)$ cosmological solutions, whose underlying driving $f(R)$'s are close to each other in the past (or present), the theories remain close to each other all the way until the present (or past).

As a disclaimer, let us mention that to produce the plots in Figs.\ref{fig:almost_LCDM_mimicking_m(r)s_forward} and \ref{fig:almost_LCDM_mimicking_m(r)s_backward}, one needs to choose a value of the deviation parameter $\varepsilon$, and we have chosen $\varepsilon=10^{-2}$. At the limit $\varepsilon\to0$, the cosmic evolution is exactly $\Lambda$CDM-like, and one should get back the same behaviour as in Figs.\ref{fig:LCDM_mimicking_m(r)s_forward} and \ref{fig:LCDM_mimicking_m(r)s_backward}. The deviation parameter $\varepsilon$, which characterizes the deviation from an \emph{exact} $\Lambda$CDM-like evolution, serves mathematically as a bifurcation parameter in the autonomous dynamical system \eqref{eq:autonomous_almost_LCDM}. The precise value (or at least the order of magnitude) of $\varepsilon$, at which the qualitative feature of the 3-dimensional phase portrait in $r$-$m$-$q$ change, can possibly be obtained by a more thorough mathematical treatment of the role of the parameter $\varepsilon$ in the autonomous system \eqref{eq:autonomous_almost_LCDM}. We do not perform such a detailed mathematical analysis here.

\subsection{Stability of the solution}

Lastly, let us come to the analysis of the stability of the cosmological solution $j=1+3\varepsilon(q-1/2)$ with respect to small homogeneous and isotropic perturbations within the solution space of the underlying $f(R)$. Consider a phantom crossing cosmological solution $h(N)$ satisfying Eq.\eqref{eq:master_eq_1_phantom}, and a time-dependent homogeneous and isotropic perturbation $\delta h(N)$ on it. Using $j=1+3\varepsilon(q-1/2),\,s=-(2+3q)+\frac{3}{2}\varepsilon(3\varepsilon+q)(1-2q)$, one can calculate the cosmographic lerk parameter as
\begin{equation}
    l = 6q^2 + 14q + 9 + \varepsilon(12q-6) + \frac{9}{4}\varepsilon^2 (2q-1)(8q+1) + \frac{27}{2}\varepsilon^3 (2q-1)\,.
\end{equation}
Substituting all these expressions into the perturbation equation \eqref{eq:ptbn_f(R)}, its coefficients come out with a rather cumbersome form
\begin{subequations}\label{eq:ai}
\begin{align}
a_0(r,m,q) &= (q-1)\Bigl[
  18\varepsilon^2(2q-1)\bigl(
    r\bigl(m(q(q(2q(6q+5)-17)-14)+85) - 4q^3 + 22q - 18\bigr)
    + 12(q+2)(q-1)^2
  \bigr) \nonumber\\
  &\quad -12\varepsilon(2q-1)\bigl(
    r\bigl(m(q(4(3q+7)q^2 + q + 54) + 69)
    - 2(q-1)(2q^3 - q - 11)\bigr)
    + 6(q(2q+3)+5)(q-1)^2
  \bigr) \nonumber\\
  &\quad -405\varepsilon^4 m r (1-2q)^2 
         -27\varepsilon^3 m r (2q-1)\bigl(2q(q(6q - 11) - 24) + 43\bigr) \nonumber\\
  &\quad +8(q+1)\bigl(
    r\bigl(m(q+1)(q(q(6q+11)+8) - 9)
    - 4(q-2)(q-1)^2(q+4)\bigr)
    + 6(q(2q+5)-9)(q-1)^2
  \bigr)
  \Bigr], \\[6pt]
a_1(r,m,q) &= (q-1)\Bigl[
  -162\varepsilon^4 m r (1-2q)^2 
   +27\varepsilon^3 m r (4(q - 4)q + 7) \nonumber\\
  &\quad +18\varepsilon^2(2q-1)\bigl(
    m(2q+1)(q(3q - 2) + 22)r
    - 4(q - 2)(q - 1)r
    + 12(q - 1)^2
  \bigr) \nonumber\\
  &\quad -12\varepsilon(2q-1)\bigl(
    m(q+1)(q(6q+11)+27)r
    - 2((q - 5)q + 3)(q - 1)r
    + 6(q - 2)(q - 1)^2
  \bigr) \nonumber\\
  &\quad +8(q+1)\bigl(
    m(q+1)(q(3q+10) - 12)r
    - 2((q - 9)q + 11)(q - 1)r
    + 6(q - 6)(q - 1)^2
  \bigr)
  \Bigr], \\[6pt]
a_2(r,m,q) &= 4(q-1)\bigl(-3\varepsilon + (6\varepsilon -2)q -2\bigr)\Bigl[
  3\varepsilon m (2q^2 + q - 1) r \nonumber\\
  &\quad -2r\bigl(m(q - 3)(q + 1) + q^2 - 3q + 2\bigr)
  + 9\varepsilon^2 m (1 - 2q) r
  + 6(q - 1)^2
  \Bigr], \\[6pt]
a_3(r,m,q) &= -2m(q - 1)r\bigl(3\varepsilon + (2 - 6\varepsilon)q + 2\bigr)^2
\end{align}
\end{subequations}
Nonetheless, one can carefully check that the perturbation equation for $j=1+3\varepsilon(q-1/2)$ cosmological solution correctly reduces to the perturbation equation for $j=1$ cosmological solution when one substitutes $\varepsilon=0$, ensuring consistency.

\begin{figure}[H] 
    \centering
    \begin{subfigure}[b]{0.32\linewidth}
    \includegraphics[width=\linewidth]{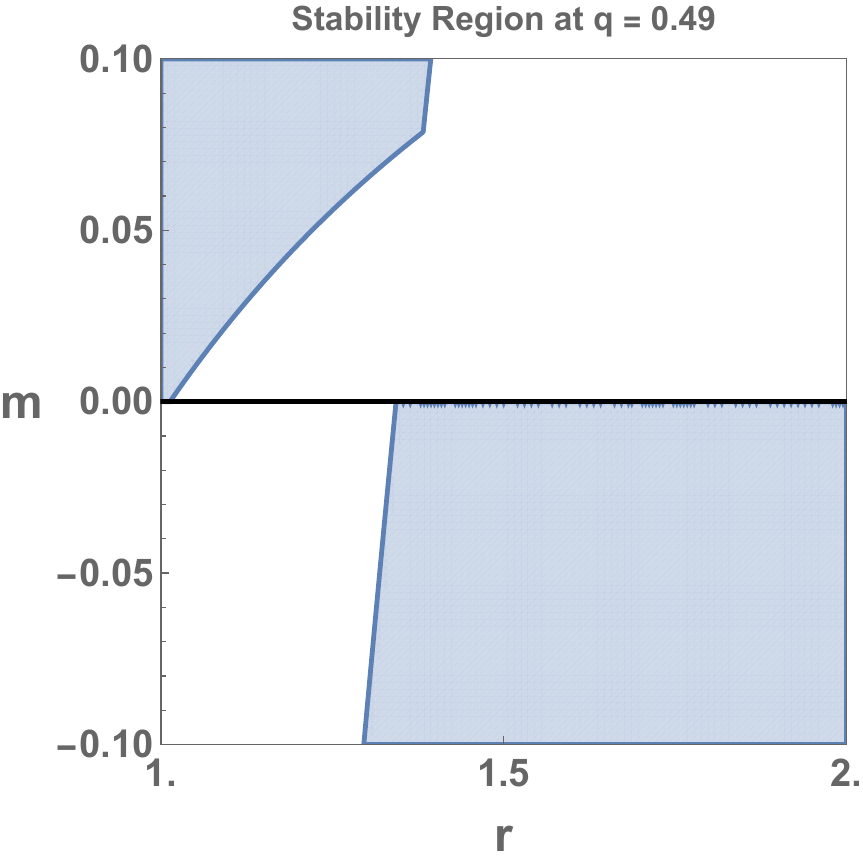}
     \caption{}
     \label{fig:stab_0.49 (almost)}
    \end{subfigure}
    \hfill
    \begin{subfigure}[b]{0.32\linewidth}
    \includegraphics[width=\linewidth]{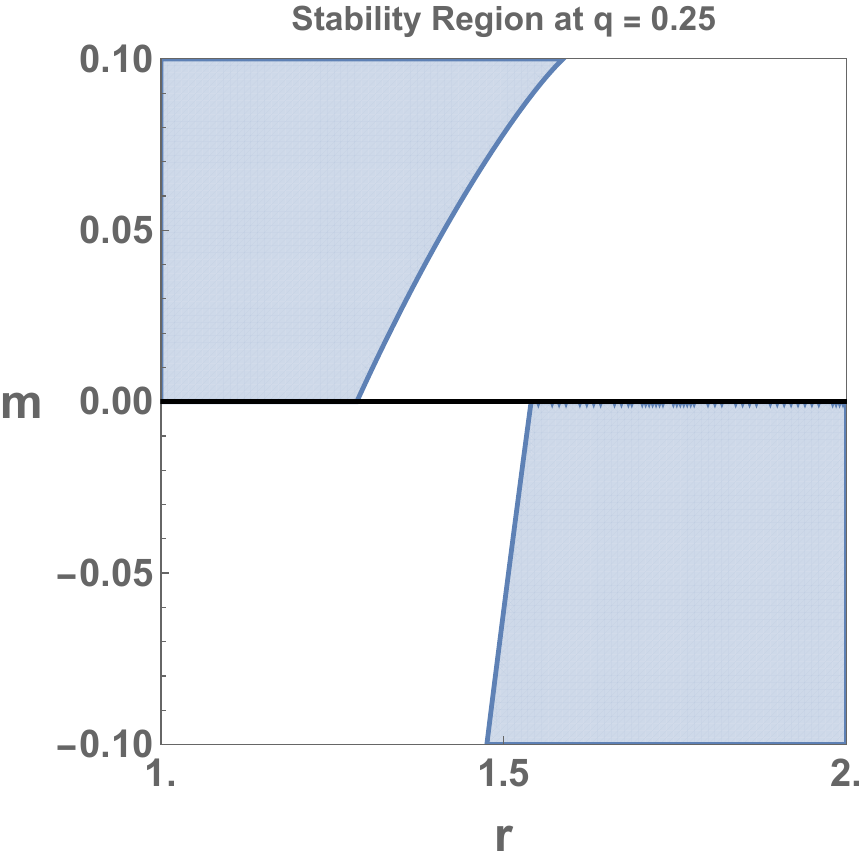}
     \caption{}
     \label{fig:stab_0.25 (almost)}
    \end{subfigure}
    \hfill
    \begin{subfigure}[b]{0.32\linewidth}
    \includegraphics[width=\linewidth]{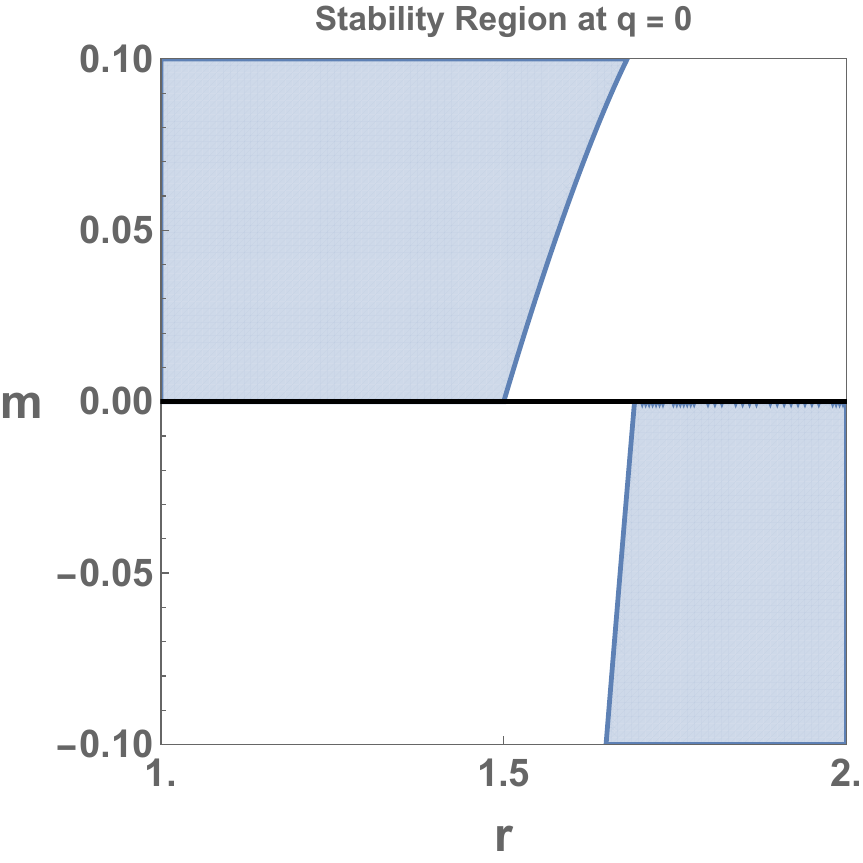}
     \caption{}
     \label{fig:stab_0 (almost)}
    \end{subfigure}
    \begin{subfigure}[b]{0.32\linewidth}
    \includegraphics[width=\linewidth]{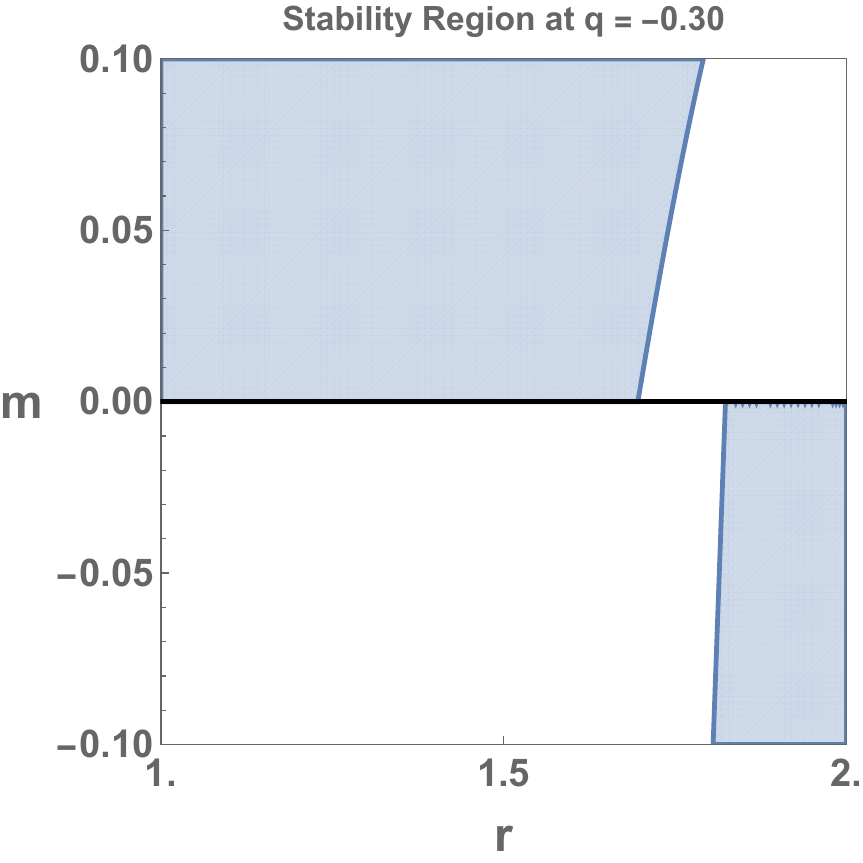}
     \caption{}
     \label{fig:stab_-0.3 (almost)}
    \end{subfigure}
    \hfill
    \begin{subfigure}[b]{0.32\linewidth}
    \includegraphics[width=\linewidth]{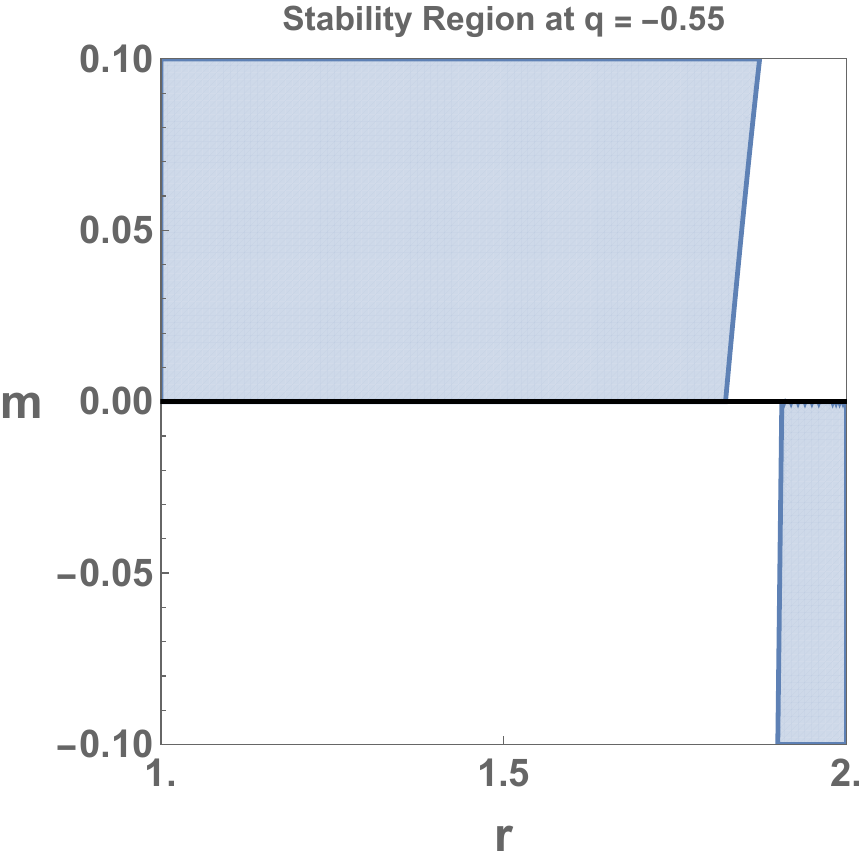}
     \caption{}
     \label{fig:stab_-0.55 (almost)}
    \end{subfigure}
    \hfill
    \begin{subfigure}[b]{0.32\linewidth}
    \includegraphics[width=\linewidth]{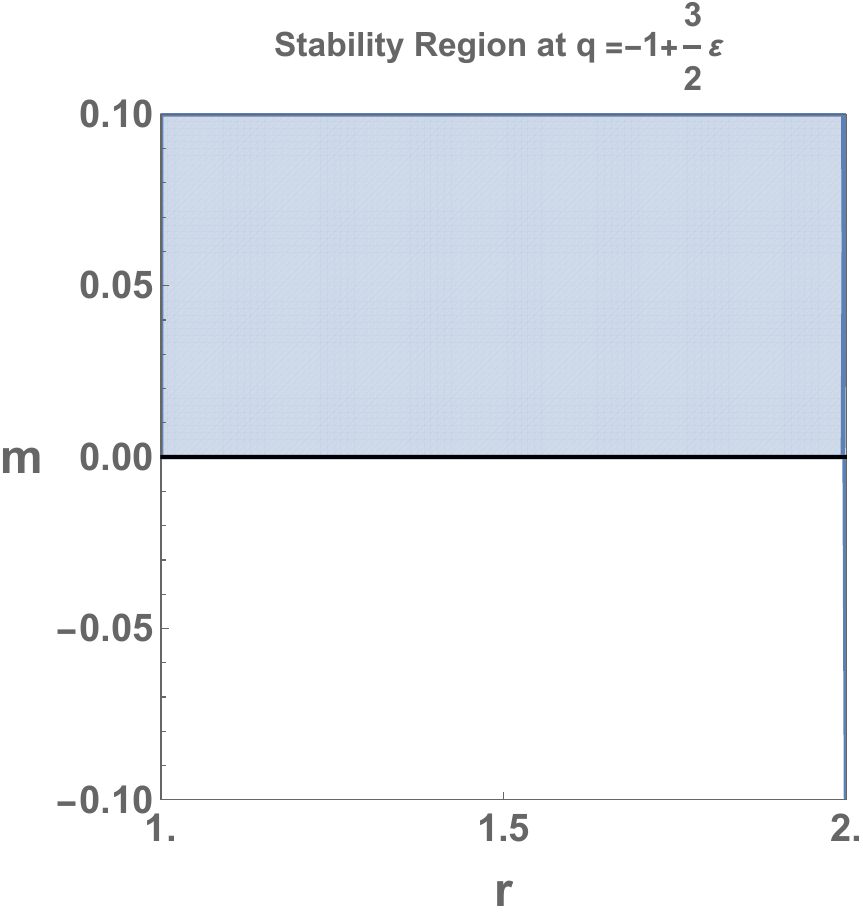}
     \caption{}
     \label{fig:stab_-0.985 (almost)}
    \end{subfigure}
    \caption{2-dimensional screenshots of $r$-$m$-$q$ phase space at six different values of $q$, starting from $q=0.49$ to $q=-1+\frac{3}{2}\varepsilon$ for $\varepsilon=10^{-2}$. The region of stability with respect to small homogeneous and isotropic perturbation (i.e. where the condition \eqref{coeff_conds} is satisfied) for an almost $\Lambda$CDM-like phantom crossing cosmological solution $j=1+3\varepsilon(q-1/2)$ is shown by the shaded region, which changes with time.}
    \label{fig:stability_region (almost)}
\end{figure}

We remind the reader again about the fact that the application of the Routh-Hurwitz stability criteria, in our approach, provides only an \emph{instantaneous} stability criteria. At each $q$-slice, the Routh-Hurwitz criteria singles out a region in the $r-m$ plane, which changes over different $q$-slices. The region in the $r-m$ plane at various $q$-slices, for a small value of $\varepsilon$ like $\varepsilon=10^{-2}$, comes out to be almost the same as that of the exact $\Lambda$CDM-like cosmology. Nonetheless, we present the 2-dimensional screenshots for six different $q$-slices in Fig.\ref{fig:stability_region (almost)}. 
However, the very different behaviour in the theory space $r$-$m$ of the almost $\Lambda$CDM-like phantom crossing $f(R)$ cosmological solutions, as compared to the exact $\Lambda$CDM-mimicking $f(R)$ cosmological solutions, implies that they respond differently to small homogeneous and isotropic perturbations. Firstly, for almost $\Lambda$CDM-like phantom crossing $f(R)$ cosmological solutions in the theory space that starts below the $m=0$ line is already unstable with respect to small homogeneous and isotropic perturbations at $q\gtrsim0.5$. However, as seen from Fig.\ref{fig:almost_LCDM_mimicking_m(r)s_forward}, even if a solution starts above the $m=0$ line, it ends up below the $m=0$ line quickly. From Fig.\ref{fig:stability_region (almost)}, one can see that the region of the phase space $r$-$m$-$q$ below the $m=0$ line, which is characterized by stability with respect to small homogeneous and isotropic perturbations, shrinks with decreasing $q$. Qualitatively, one can conclude that it is highly likely for such a solution to run into a region of the phase space characterized by instability with respect to small homogeneous and isotropic perturbations.

Physically, one can interpret the result in the following way. After setting $j=1+3\varepsilon(q-1/2),\,\Omega_k=0,\,w=0$ in Eq.\eqref{eq:master_eq_2}, the equation \emph{implicitly} specifies a family of $f(R)$ theories. Pick any particular theory at random from this family. The solution space of this theory can be imagined to be a phase space spanned by $\{h,h',h''\}$. If one wants to set initial conditions at a high redshift (say, $z\approx6$) and obtain a cosmological solution $h(z)$ that is almost $\Lambda$CDM-like (\eqref{eq:almost_LCDM}), it is most probable that one can achieve this for very fine-tuned initial conditions. Slight deviations in the initial conditions will result in an FLRW cosmology that can be very different.

\section{Summary, Discussion and Future Outlook}\label{sec:concl}

The goal of this paper is to present a systematic approach for studying aspects of a given General Relativistic cosmological solution, as specified by a cosmographic condition $j=j(q)$, within the framework of $f(R)$ gravity plus nonrelativistic fluid. Note that, when one does not specify the underlying $f(R)$ theory, one can always, in principle, reconstruct the $f(R)$ \cite{Nojiri:2009kx,Nojiri:2010oco,Nojiri:2010wj}, so that the existence of the solution within the $f(R)$ framework is almost guaranteed. However, the more sensible question to ask is whether the reconstructed $f(R)$ is free from theoretical pathologies like the ghost instability ($f'<0$) or the tachyonic instability ($f''<0$), and whether the given solution is robust against small deviation in the initial conditions in the solution space of the underlying $f(R)$. The difficulty here is that, more often than not, the reconstruction method fails in practice. To alleviate this problem, we have developed the \emph{theory space} approach. Portrayal of the $f(R)$ cosmological solutions as solution curves in the 2-dimensional space $\{r,m\}=\lbrace\frac{R f'}{f},\frac{R f''}{f'}\rbrace$ enables us to answer the above questions even when the reconstruction method fails. Thus, the theory space approach has a much broader scope of applicability. One disadvantage of the theory space analysis, however, is that one cannot numerically calculate $\Omega_m(z),\,w_{\rm DE}(z)$, which are sometimes of physical interest from the phenomenological point of view.

Next, let us summarize in a pointwise manner some of the physics that we have been able to extract by applying our approach to the $\Lambda$CDM-like cosmological solution \eqref{eq:LCDM} and an almost $\Lambda$CDM-like phantom crossing cosmological solution \eqref{eq:almost_LCDM}
\begin{enumerate}
    \item {\bf $\Lambda$CDM-mimicking $f(R)$:}
    \begin{itemize}
    \item There has been some earlier works on reconstructing $f(R)$ gravity admitting as a solution the $\Lambda$CDM-like evolution \cite{Dunsby:2010wg,He:2012rf,Choudhury:2019zod}. All these works have reconstructed the underlying $f(R)$ implicitly in terms of hypergeometric functions, which is consistent with our reconstruction differential equation \eqref{eq:recon_LCDM_2}. However, unlike in the earlier works, here we dived a little deeper to uncover what it physically means to mimic a $\Lambda$CDM-like evolution in $f(R)$ gravity. In particular, we show that this does \emph{not} necessitate the effective curvature fluid in $f(R)$, defined in Eq.\eqref{eq:curv_fluid}, to behave like a cosmological constant. The effective curvature fluid can behave like a dynamical dark energy of the form \eqref{eq:DDE}, while still reproducing exactly a $\Lambda$CDM-like cosmic evolution.
    \item It is possible to realize a $\Lambda$CDM-like cosmic evolution in an $f(R)$ gravity that is free from ghost or tachyonic instability throughout its course of evolution, and hence theoretically viable. 
    \item Within the space of the possible $f(R)$ theories that admit a $\Lambda$CDM-like cosmic evolution as a solution, there is no tendency of GR to behave as a cosmological past or future attractor. In other words, a $\Lambda$CDM-like $f(R)$ cosmological solution does not have any generic tendency to asymptotically tend towards the General Relativistic $\Lambda$CDM solution in the past or the future. The implication is that, in the space of all possible $f(R)$ theories admitting a $\Lambda$CDM-like cosmic evolution as a solution, GR is not the most natural candidate.
    \item $\Lambda$CDM-like $f(R)$ cosmological solutions that do in fact tend to the General Relativistic $\Lambda$CDM solutions in the past ($q\lesssim1/2$), are unstable with respect to small homogeneous and isotropic perturbations within the solution space of their underlying driving $f(R)$. Hence, most likely, they can arise only from fine-tuned initial conditions.
    \end{itemize}
    \item {\bf $f(R)$ cosmologies reproducing an almost $\Lambda$CDM-like phantom crossing solution:} This is a classic example of the failure of the reconstruction method and the success of the theory space analysis.
    \begin{itemize}
    \item It is not possible to realize such a solution in a physically healthy $f(R)$ gravity. The underlying $f(R)$ gravity inevitably comes across tachyonic instability ($f''(R)<0$) along the course of cosmic evolution.
    \item If one considers two possible such almost $\Lambda$CDM-mimicking $f(R)$ cosmological solutions, whose underlying driving $f(R)$'s are close to each other in the past, the theories remain close to each other all the way until the present. This situation is in stark contrast with exact $\Lambda$CDM-mimicking $f(R)$ cosmologies (compare Figs.\ref{fig:LCDM_mimicking_m(r)s_forward} and \ref{fig:almost_LCDM_mimicking_m(r)s_forward}).
    \item Most, if not all, of the almost $\Lambda$CDM-mimicking $f(R)$ cosmological solutions are unstable with respect to small homogeneous and isotropic perturbations within the solution space of their underlying driving $f(R)$.
    \end{itemize}
\end{enumerate}

Some disclaimer about the stability analysis performed in our paper is in order. To assess the stability of a cosmological solution in the underlying theory, the technique found in \cite{Bamba:2013fha} is adopted here. The relevant cosmological solutions considered in our paper cannot be expressed as a single fixed point but rather as a whole phase trajectory in the solution space of the underlying $f(R)$. It is a mathematically hard problem in differential geometry to pose and address the issue of the stability of a trajectory. The stability analysis with respect to homogeneous and isotropic perturbations that we have performed here utilizes the Routh-Hurwitz stability criteria. Strictly speaking, the Routh-Hurwitz stability criteria, applied to our case, can provide only an instantaneous stability condition. The argument we present in support of our treatment is that a given trajectory corresponds to a particular solution of an autonomous system, which can be numerically obtained by setting initial conditions at some $N$-value. If we demand that the particular numerical solution (i.e. the given trajectory) is stable irrespective of at what $N$-value we set our initial conditions to obtain it, then the instantaneous Routh-Hurwitz criteria can be extended to \emph{all} $N$. Our results regarding the stability of the solutions is thus suggestive, and not necessarily conclusive. A more formal and mathematically rigorous framework to study the stability will be the Kosambi-Cartan analysis \cite{Boehmer:2010jqg,Harko_2016}, which is beyond the scope of the present work.

From the phenomenological point of view, an interesting question is, even if some $f(R)$ theory can mimic a given General Relativistic cosmological solution, how can one possibly obtain distinctive signatures between the two, so that one can later on try to test them against data? For this, one needs to go to the perturbation level. For example, distinctive signatures at the matter perturbation level between the General Relativistic $\Lambda$CDM model and $\Lambda$CDM-mimicking $f(R)$ models have recently been investigated \cite{MacDevette:2024wpg}. The same can be done for the GR and $f(R)$ models of almost $\Lambda$CDM phantom crossing cosmology. Typically, in such works, one tries to choose a viable background trajectory in the phase space $r$-$m$-$q$ (i.e., one that never encounters a ghost or a tachyonic instability) and calculates the evolution of matter perturbation along the trajectory \cite{MacDevette:2024wpg}. The theory space analysis that we have presented here can act as a guiding principle to choose the trajectories. For example, our theory space analysis for $\Lambda$CDM-mimicking $f(R)$ cosmology reveals that as long as one chooses a trajectory passing through a point like $(r_{\rm in},m_{\rm in},q_{\rm in})=\left(\frac{3q-3}{q-2}\vert_{q=0.49}-\epsilon,\epsilon,0.49\right)$, one can get a physically viable trajectory, while the trajectory passing through the point $(r_{\rm in},m_{\rm in},q_{\rm in})=\left(\frac{3q-3}{q-2}\vert_{q=0.49}+\epsilon,\epsilon,0.49\right)$ is not a physically viable one (see Fig.\ref{fig:LCDM_mimicking_m(r)s_forward}).

Lastly, let us spend a few words about the future potential of the approach presented in this work. An attentive reader may realize that one need not restrict oneself to only $f(R)$ gravity framework. The same approach can be applied equally well for other $f$-classes of modified gravity theories, e.g. $f(T),\,f(Q),\,f(G)$ (modified teleparallel, symmetric teleparallel and general teleparallel gravity). In particular, it would be really interesting to apply this to $f(Q)$ gravity, in light of the recent surge of interest in that sector. More realistic cosmological evolutions can also be imposed as a given solution by specifying a numerical fitting function of the form $j=j_0+j_1z+j_2z^2$, where $j_0,j_1,j_2$ can be constrained from combinations of datasets in a model-independent manner via a Gaussian process (see e.g. \cite[Sec 4.4]{Mukherjee:2020ytg}). Note that a numerical fitting function of the form $j=j_0+j_1z+j_2z^2$ can always be re-expressed in the form $j=j(q)$, since $q=q(z)$ is a monotonically increasing function and hence invertible.

\section*{Acknowledgements}

This research is supported by the Second Century Fund (C2F), Chulalongkorn University, Thailand. The authors acknowledge useful discussions with Christian Boehmer. P.B. is supported in part by National Research Council of Thailand (NRCT) and Chulalongkorn University under Grant N42A660500. 

\appendix

\section{More on the \emph{theory space}}\label{app:theory_space}

The 2-dimensional \emph{theory space} that we have introduced in this paper is the key idea that allows us to perform our analysis. In this appendix, we elaborate on the concept in some detail.

\subsection{Mathematical description}

The theory space is a 2-dimensional manifold created by the dimensionless coordinates $(r,m)\in\mathbb{R}^2$, where the coordinates $r\equiv\frac{d \ln f}{d \ln R}$ and $m\equiv\frac{d \ln f'}{d \ln R}$ quantify how the Lagrangian function and its first derivative behaves locally under logarithmic derivative with respect to $R$. Topologically, one can write
\begin{equation}
    \mathcal{T} = \lbrace (r,m)\in\mathbb{R}^2 \setminus \mathcal{S} \rbrace\,,
\end{equation}
where $\mathcal{S}$ is the singular submanifold where $f(R)$ or $f'(R)$ vanishes. The inherent assumption in our work is that such situations do not arise in the late-time context we are considering. In fact, $f'(R)=0$ will imply a divergence of the effective gravitational coupling, which jeopardizes the physical interpretation of the underlying situation. Such a situation is usually kept away from consideration when building cosmological models in $f(R)$ gravity.

In the late-time context which is the domain of application of our framework in this paper, one has $q$ monotonically varying from a value $\sim1/2$ during matter domination to a value $q\sim-0.55$ today. Then, one always has $R>0$ (see Eq.\eqref{eq:Ricci_CP}). Consequently, the region $m<0$ corresponds to either $\{f'(R)<0,f''(R)>0\}$ or $\{f'(R)>0,f''(R)<0\}$ in the theory space $r-m$. The conditions $f'(R)<0$ and $f''(R)<0$ are associated with ghost and tachyonic instability, respectively \cite{Sotiriou:2008rp,DeFelice:2010aj}. Therefore, in the late-time context, the lower half of the theory space is definitely unphysical
\begin{equation}
    \mathcal{T}_{\rm unphysical} = \lbrace (r,m)\in(\mathbb{R}\times\mathbb{R}^-)\setminus \mathcal{S} \rbrace\,.
\end{equation}
In the late-time context, any trajectory that crosses into the lower half of the theory space can be safely concluded to be an $f(R)$ cosmological solution plagued by either ghost or tachyonic instability. 

However, one needs to be cautious about the converse statement. A trajectory that always remains on the upper half of the theory space, $\lbrace (r,m)\in(\mathbb{R}\times\mathbb{R}^+)\setminus \mathcal{S} \rbrace$, need not necessarily be a cosmological solution free from any instability. Note that if a late-time $f(R)$ cosmological solution is plagued by both ghost and tachyonic instabilities ($R>0,\,f'(R)<0,\,f''(R)<0$), then we still have $m>0$ corresponding to this solution.

The notion of the theory space is introduced as an alternative local description of $f(R)$ theory. One can also describe the cosmology in the $f(R)$ framework in some kind of a 3-dimensional theory space $\lbrace (f,f',f'')\in\mathbb{R}^3\rbrace$. In a sense, this would have been a natural first choice, since the derivatives $\{\dot{f},\dot{f'},\dot{f''}\}$ clearly appear in the field equations \eqref{eq:f(R)_fieldeqs}. The theory space $\{r,m\}$ can be viewed as a coordinate transformation $\Phi:\mathbb{R}^3\to\mathbb{R}^2$ with $(r,m)=\Phi(f,f',f'')$. This coordinate transformation, which gives rise to a theory space with reduced dimensionality, is an equally valid candidate to describe cosmology in an $f(R)$ framework, because only the derivatives $\lbrace\frac{dr}{d(-q)},\frac{dm}{d(-q)}\rbrace$ appear in the field equations \eqref{eq:nonautonomous}. The crucial requirements of the success of the latter approach is the nonvanishing of the functions $f(R)$ and $f'(R)$, and the monotonically decreasing nature of the deceleration parameter $q$. The latter condition is definitely valid in the late-time context, whereas the former is taken as a preliminary requirement in late-time cosmological model building in $f(R)$.

We must, however, mention that the theory space is not a geometric Riemannian manifold in the sense of differential geometry, and no metric can be defined on this manifold. Let us mention that there exists a mathematical framework in the dynamical system analysis called the Kosambi-Cartan-Chern (KCC) theory. In the KCC approach, one describes the evolution of the dynamical system geometrically by considering the trajectories as geodesics in a Finsler space. For particular applications in the cosmological context the reader is referred to \cite{Boehmer:2010jqg,Harko_2016}. One can possibly do the same with our autonomous system \eqref{eq:autonomous}, but in this paper, we do not pursue in that direction. Rather, we have worked with the non-autonomous system \eqref{eq:nonautonomous}. Consequently, our theory space is essentially a non-Riemannian space endowed with a vector $\lbrace\frac{dr}{d(-q)},\frac{dm}{d(-q)}\rbrace$, not a Riemannian space endowed with a metric. The solution curves are the integral curves of the flow equations \eqref{eq:nonautonomous}, not the geodesics of a metric.

\subsection{Mapping a given $f(R)$ theory into the theory space}

In Section \ref{sec:theory_space}, we have mentioned that any given $f(R)$ theory maps into a curve or a family of curves into the theory space $m-r$, except for the special case of a monomial $f(R)$ theory, which maps into a single point. This kind of an $m-r$ parametrization of modified gravity theories first appeared in the well-known work by Amendola, Gannouji, Polarski, and Tsujikawa \cite{Amendola:2006we}, although the authors did not term it as \emph{theory space}. We list in the table \ref{tab:f(R)_to_m(r)} the $m-r$ curve corresponding to some theories as considered in \cite[Section V]{Amendola:2006we}. For details, the reader is referred to \cite{Amendola:2006we}. The reader is advised to take note of the fact that the definition of $r$ in our paper differs from that in \cite{Amendola:2006we} by a negative sign. 
\begin{table}[h]
    \centering
    \begin{tabular}{|c|c|p{4cm}|}
    \hline
        $f(R)$ & $m(r)$ & Comment 
        \\
        \hline
        $\frac{\alpha}{R^n}$ & \makecell{\\ $(r,m)=(-n,-n-1)$ \\} & A monomial theory mapped to a single point in the theory space
        \\
        \hline
        $R+\frac{\alpha}{R^n}$ & \makecell{\\ $m(r)=-n\left(1-\frac{1}{r}\right)$ \\} & A 2-parameter $f(R)$ theory mapped to a 1-parameter family of curves in the theory space
        \\
        \hline
        $R^p \exp(qR)$ & \makecell{\\ $m(r)=r-\frac{p}{r}$ \\} & A 2-parameter $f(R)$ theory mapped to a 1-parameter family of curves in the theory space
        \\
        \hline
        $R^p[\log(\alpha R)]^q$ & \makecell{\\ $m(r)=\frac{p^2-2pr+r(q+r-qr)}{-qr}$ \\} & A 3-parameter $f(R)$ theory mapped to a 2-parameter family of curves in the theory space.
        \\
        \hline
        $R^p \exp\left(\frac{q}{R}\right)$ & \makecell{\\ $m(r)=\frac{p-r(2-r)}{r}$ \\} & A 2-parameter $f(R)$ theory mapped to a 1-parameter family of curves in the theory space.
        \\
        \hline
        $f(R)=-2\Lambda+R+\alpha R^2$ & \makecell{\\ $m(r)=\frac{(r-1) \pm \sqrt{(r-1)^2-4\tilde{\alpha}r(2-r)}}{1 \pm \sqrt{(r-1)^2-4\tilde{\alpha}r(2-r)}}$, \\ \\ $\tilde{\alpha}=2\alpha\Lambda$} & A 2-parameter $f(R)$ theory mapped to a 1-parameter family of curves in the theory space.
        \\
        \hline
        $f(R)=R-\frac{\mu_1^4}{R}+\left(\frac{R}{\mu_2}\right)^2$ & \makecell{\\ $r(R)=\frac{2R^3 + \mu_2^2\left(\mu_1^4 + R^2\right)}{R^3 + \mu_2^2 \left(R^2 - \mu_1^4\right)}$ \\ \\ $m(R)=\frac{2\left(R^3 - \mu_1^4 \mu_2^2\right)}{2R^3 + \mu_2^2\left(\mu_1^4 + R^2\right)}$} & A 2-parameter $f(R)$ theory mapped to a 2-parameter family of curves in the theory space.
        \\
        \hline
    \end{tabular}
    \caption{The $m(r)$ curves in the theory space corresponding to different $f(R)$ forms, taken from \cite[Section V]{Amendola:2006we}. Notice that our definition of $r$ differs from that in \cite{Amendola:2006we} by a minus sign. Therefore, to tally with the $m=m(r)$ relations obtained in \cite{Amendola:2006we}, replace $r\to-r$.}
    \label{tab:f(R)_to_m(r)}
\end{table}

Except for the particular case $f(R)=R-\frac{\mu_1^4}{R}+\left(\frac{R}{\mu_2}\right)^2$, it is possible to invert the relation $r=r(R)$ to obtain $R=R(r)$, and hence obtain a unique function $m=m(r)$. For the particular theory $f(R)=R-\frac{\mu_1^4}{R}+\left(\frac{R}{\mu_2}\right)^2$, one gets 
\begin{equation}
    R^3\frac{2-r}{\mu_2^2} + R^2(1-r) + \mu_1^4(1+r) = 0\,,
\end{equation}
a relation which cannot be inverted uniquely to obtain an $R=R(r)$. Nonetheless, in this case (and similar cases where $r(R)$ is not invertible), one can still obtain the functions $\{r(R),m(R)\}$ for the given theory. A parametric plot then provides the corresponding family of $m(r)$ curves in the theory space.

In the paper \cite{Amendola:2006we}, the authors showed that given an $f(R)$ theory, one can obtain a corresponding family of curves in the $m-r$ plane. Our work takes the inverse approach. In our work, we do not know the underlying $f(R)$ theory beforehand, but we demand that it admits a given cosmological solution. What we have been able to show in this work is that this latter requirement enables us to \emph{reconstruct} the $m(r)$ curve of the underlying theory. Additionally, we could show that the quantities $r$ and $m$ can be used to express the $f(R)$ cosmological field equations in the form of a non-autonomous dynamical system \eqref{eq:nonautonomous}. This enabled us to indicate a sense of flow on these resulting $m(r)$ curves, which helped us understand how the deviation from GR ($f(R)=-2\Lambda+R$) is evolving along with the cosmic evolution.

\bibliographystyle{unsrt}
\bibliography{refs}

@article{Panpanich:2019fxq,
    author = "Panpanich, Sirachak and Burikham, Piyabut and Ponglertsakul, Supakchai and Tannukij, Lunchakorn",
    title = "{Resolving Hubble Tension with Quintom Dark Energy Model}",
    eprint = "1908.03324",
    archivePrefix = "arXiv",
    primaryClass = "gr-qc",
    doi = "10.1088/1674-1137/abc537",
    journal = "Chin. Phys. C",
    volume = "45",
    number = "1",
    pages = "015108",
    year = "2021"
}

@article{Sotiriou:2008rp,
    author = "Sotiriou, Thomas P. and Faraoni, Valerio",
    title = "{$f(R)$ theories of gravity}",
    eprint = "0805.1726",
    archivePrefix = "arXiv",
    primaryClass = "gr-qc",
    doi = "10.1103/RevModPhys.82.451",
    journal = "Rev. Mod. Phys.",
    volume = "82",
    pages = "451--497",
    year = "2010"
}

@article{DeFelice:2010aj,
    author = "De Felice, Antonio and Tsujikawa, Shinji",
    title = "{$f(R)$ theories}",
    eprint = "1002.4928",
    archivePrefix = "arXiv",
    primaryClass = "gr-qc",
    doi = "10.12942/lrr-2010-3",
    journal = "Living Rev. Rel.",
    volume = "13",
    pages = "3",
    year = "2010"
}

@article{Dunsby:2010wg,
    author = "Dunsby, Peter K. S. and Elizalde, Emilo and Goswami, Rituparno and Odintsov, Sergei and Gomez, Diego Saez",
    title = "{On the LCDM Universe in f(R) gravity}",
    eprint = "1005.2205",
    archivePrefix = "arXiv",
    primaryClass = "gr-qc",
    doi = "10.1103/PhysRevD.82.023519",
    journal = "Phys. Rev. D",
    volume = "82",
    pages = "023519",
    year = "2010"
}

@article{He:2012rf,
    author = "He, Jian-hua and Wang, Bin",
    title = "{Revisiting $f(R)$ gravity models that reproduce $\Lambda$CDM expansion}",
    eprint = "1208.1388",
    archivePrefix = "arXiv",
    primaryClass = "astro-ph.CO",
    doi = "10.1103/PhysRevD.87.023508",
    journal = "Phys. Rev. D",
    volume = "87",
    number = "2",
    pages = "023508",
    year = "2013"
}

@article{Choudhury:2019zod,
    author = "Choudhury, Shibendu Gupta and Dasgupta, Ananda and Banerjee, Narayan",
    title = "{Reconstruction of $f(R)$ gravity models for an accelerated universe using the Raychaudhuri equation}",
    eprint = "1903.04775",
    archivePrefix = "arXiv",
    primaryClass = "gr-qc",
    doi = "10.1093/mnras/stz731",
    journal = "Mon. Not. Roy. Astron. Soc.",
    volume = "485",
    number = "4",
    pages = "5693--5699",
    year = "2019"
}

@article{Carloni:2010ph,
    author = "Carloni, Sante and Goswami, Rituparno and Dunsby, Peter K. S.",
    title = "{A new approach to reconstruction methods in $f(R)$ gravity}",
    eprint = "1005.1840",
    archivePrefix = "arXiv",
    primaryClass = "gr-qc",
    doi = "10.1088/0264-9381/29/13/135012",
    journal = "Class. Quant. Grav.",
    volume = "29",
    pages = "135012",
    year = "2012"
}

@article{Nojiri:2010oco,
    author = "Nojiri, Shin'ichi and Odintsov, Sergei D. and Toporensky, Alexey and Tretyakov, Petr",
    title = "{Reconstruction and deceleration-acceleration transitions in modified gravity}",
    eprint = "0912.2488",
    archivePrefix = "arXiv",
    primaryClass = "hep-th",
    doi = "10.1007/s10714-010-0977-5",
    journal = "Gen. Rel. Grav.",
    volume = "42",
    pages = "1997--2008",
    year = "2010"
}

@article{Nojiri:2009kx,
    author = "Nojiri, Shin'ichi and Odintsov, Sergei D. and Saez-Gomez, Diego",
    title = "{Cosmological reconstruction of realistic modified F(R) gravities}",
    eprint = "0908.1269",
    archivePrefix = "arXiv",
    primaryClass = "hep-th",
    doi = "10.1016/j.physletb.2009.09.045",
    journal = "Phys. Lett. B",
    volume = "681",
    pages = "74--80",
    year = "2009"
}

@article{DESI:2025zgx,
    author = "Abdul Karim, M. and others",
    collaboration = "DESI",
    title = "{DESI DR2 Results II: Measurements of Baryon Acoustic Oscillations and Cosmological Constraints}",
    eprint = "2503.14738",
    archivePrefix = "arXiv",
    primaryClass = "astro-ph.CO",
    reportNumber = "FERMILAB-PUB-25-0169-PPD",
    month = "3",
    year = "2025"
}

@article{Rapetti:2006fv,
    author = "Rapetti, David and Allen, Steven W. and Amin, Mustafa A. and Blandford, Roger D.",
    title = "{A kinematical approach to dark energy studies}",
    eprint = "astro-ph/0605683",
    archivePrefix = "arXiv",
    reportNumber = "SLAC-PUB-11882",
    doi = "10.1111/j.1365-2966.2006.11419.x",
    journal = "Mon. Not. Roy. Astron. Soc.",
    volume = "375",
    pages = "1510--1520",
    year = "2007"
}

@article{Mukherjee:2024wix,
    author = "Mukherjee, Purba and Dainotti, Maria Giovanna and Dialektopoulos, Konstantinos F. and Levi Said, Jackson and Mifsud, Jurgen",
    title = "{Model-independent calibration of Gamma-Ray Bursts with neural networks}",
    eprint = "2411.03773",
    archivePrefix = "arXiv",
    primaryClass = "astro-ph.CO",
    month = "11",
    year = "2024"
}

@article{Dialektopoulos:2023dhb,
    author = "Dialektopoulos, Konstantinos F. and Mukherjee, Purba and Levi Said, Jackson and Mifsud, Jurgen",
    title = "{Neural network reconstruction of cosmology using the Pantheon compilation}",
    eprint = "2305.15499",
    archivePrefix = "arXiv",
    primaryClass = "gr-qc",
    doi = "10.1140/epjc/s10052-023-12124-3",
    journal = "Eur. Phys. J. C",
    volume = "83",
    number = "10",
    pages = "956",
    year = "2023"
}

@article{Visser:2003vq,
    author = "Visser, Matt",
    title = "{Jerk and the cosmological equation of state}",
    eprint = "gr-qc/0309109",
    archivePrefix = "arXiv",
    doi = "10.1088/0264-9381/21/11/006",
    journal = "Class. Quant. Grav.",
    volume = "21",
    pages = "2603--2616",
    year = "2004"
}

@article{Visser:2004bf,
    author = "Visser, Matt",
    editor = "McClelland, D. E. and Scott, S. M.",
    title = "{Cosmography: Cosmology without the Einstein equations}",
    eprint = "gr-qc/0411131",
    archivePrefix = "arXiv",
    doi = "10.1007/s10714-005-0134-8",
    journal = "Gen. Rel. Grav.",
    volume = "37",
    pages = "1541--1548",
    year = "2005"
}

@article{Dunsby:2015ers,
    author = "Dunsby, Peter K. S. and Luongo, Orlando",
    title = "{On the theory and applications of modern cosmography}",
    eprint = "1511.06532",
    archivePrefix = "arXiv",
    primaryClass = "gr-qc",
    doi = "10.1142/S0219887816300026",
    journal = "Int. J. Geom. Meth. Mod. Phys.",
    volume = "13",
    number = "03",
    pages = "1630002",
    year = "2016"
}

@article{Bolotin:2018xtq,
    author = "Bolotin, Yu. L. and Cherkaskiy, V. A. and Ivashtenko, O. Yu. and Konchatnyi, M. I. and Zazunov, L. G.",
    title = "{APPLIED COSMOGRAPHY: A Pedagogical Review}",
    eprint = "1812.02394",
    archivePrefix = "arXiv",
    primaryClass = "gr-qc",
    month = "12",
    year = "2018"
}

@article{Dunajski:2008tg,
    author = "Dunajski, Maciej and Gibbons, Gary",
    title = "{Cosmic Jerk, Snap and Beyond}",
    eprint = "0807.0207",
    archivePrefix = "arXiv",
    primaryClass = "gr-qc",
    reportNumber = "DAMTP-2008-58",
    doi = "10.1088/0264-9381/25/23/235012",
    journal = "Class. Quant. Grav.",
    volume = "25",
    pages = "235012",
    year = "2008"
}

@article{Sahni:2002fz,
    author = "Sahni, Varun and Saini, Tarun Deep and Starobinsky, Alexei A. and Alam, Ujjaini",
    title = "{Statefinder: A New geometrical diagnostic of dark energy}",
    eprint = "astro-ph/0201498",
    archivePrefix = "arXiv",
    doi = "10.1134/1.1574831",
    journal = "JETP Lett.",
    volume = "77",
    pages = "201--206",
    year = "2003"
}

@article{Alam:2003sc,
    author = "Alam, Ujjaini and Sahni, Varun and Saini, Tarun Deep and Starobinsky, A. A.",
    title = "{Exploring the expanding universe and dark energy using the Statefinder diagnostic}",
    eprint = "astro-ph/0303009",
    archivePrefix = "arXiv",
    doi = "10.1046/j.1365-8711.2003.06871.x",
    journal = "Mon. Not. Roy. Astron. Soc.",
    volume = "344",
    pages = "1057",
    year = "2003"
}

@article{Arroja:2010wy,
    author = "Arroja, Frederico and Sasaki, Misao",
    title = "{A note on the equivalence of a barotropic perfect fluid with a K-essence scalar field}",
    eprint = "1002.1376",
    archivePrefix = "arXiv",
    primaryClass = "astro-ph.CO",
    reportNumber = "YITP-10-6",
    doi = "10.1103/PhysRevD.81.107301",
    journal = "Phys. Rev. D",
    volume = "81",
    pages = "107301",
    year = "2010"
}

@article{Luongo:2014nld,
    author = "Luongo, Orlando and Quevedo, Hernando",
    title = "{A Unified Dark Energy Model from a Vanishing Speed of Sound with Emergent Cosmological Constant}",
    doi = "10.1142/S0218271814500126",
    journal = "Int. J. Mod. Phys. D",
    volume = "23",
    pages = "1450012",
    year = "2014"
}

@article{Chakraborty:2022evc,
    author = "Chakraborty, Saikat and Gregoris, Daniele and Mishra, B.",
    title = "{On the uniqueness of \ensuremath{\Lambda}CDM-like evolution for homogeneous and isotropic cosmology in General Relativity}",
    eprint = "2208.04596",
    archivePrefix = "arXiv",
    primaryClass = "gr-qc",
    doi = "10.1016/j.physletb.2023.137962",
    journal = "Phys. Lett. B",
    volume = "842",
    pages = "137962",
    year = "2023"
}

@article{Amendola:2006we,
    author = "Amendola, Luca and Gannouji, Radouane and Polarski, David and Tsujikawa, Shinji",
    title = "{Conditions for the cosmological viability of f(R) dark energy models}",
    eprint = "gr-qc/0612180",
    archivePrefix = "arXiv",
    doi = "10.1103/PhysRevD.75.083504",
    journal = "Phys. Rev. D",
    volume = "75",
    pages = "083504",
    year = "2007"
}

@article{Bamba:2013fha,
    author = "Bamba, Kazuharu and Makarenko, Andrey N. and Myagky, Alexandr N. and Nojiri, Shin'ichi and Odintsov, Sergei D.",
    title = "{Bounce cosmology from $F(R)$ gravity and $F(R)$ bigravity}",
    eprint = "1309.3748",
    archivePrefix = "arXiv",
    primaryClass = "hep-th",
    doi = "10.1088/1475-7516/2014/01/008",
    journal = "JCAP",
    volume = "01",
    pages = "008",
    year = "2014"
}

@article{Nesseris:2022hhc,
    author = "Nesseris, Savvas",
    title = "{The Effective Fluid Approach for Modified Gravity and Its Applications}",
    eprint = "2212.12768",
    archivePrefix = "arXiv",
    primaryClass = "astro-ph.CO",
    reportNumber = "IFT-UAM/CSIC-22-154",
    doi = "10.3390/universe9010013",
    journal = "Universe",
    volume = "9",
    number = "1",
    pages = "13",
    year = "2023"
}

@article{MacDevette:2024wpg,
    author = "MacDevette, Kelly and Worsley, Jess and Dunsby, Peter and Chakraborty, Saikat",
    title = "{A model-independent approach to the study of structure growth in f(R) gravity}",
    eprint = "2408.03998",
    archivePrefix = "arXiv",
    primaryClass = "gr-qc",
    doi = "10.1093/mnras/staf168",
    journal = "Mon. Not. Roy. Astron. Soc.",
    volume = "537",
    number = "3",
    pages = "2471--2495",
    year = "2025"
}

@article{Carloni:2007br,
    author = "Carloni, S. and Troisi, A. and Dunsby, P. K. S.",
    title = "{Some remarks on the dynamical systems approach to fourth order gravity}",
    eprint = "0706.0452",
    archivePrefix = "arXiv",
    primaryClass = "gr-qc",
    doi = "10.1007/s10714-008-0747-9",
    journal = "Gen. Rel. Grav.",
    volume = "41",
    pages = "1757--1776",
    year = "2009"
}

@article{Chakraborty:2021jku,
    author = "Chakraborty, Saikat and MacDevette, Kelly and Dunsby, Peter",
    title = "{A model independent approach to the study of $f(R)$ cosmologies with expansion histories close to $\Lambda$CDM}",
    eprint = "2103.02274",
    archivePrefix = "arXiv",
    primaryClass = "gr-qc",
    doi = "10.1103/PhysRevD.103.124040",
    journal = "Phys. Rev. D",
    volume = "103",
    number = "12",
    pages = "124040",
    year = "2021"
}

@article{Boehmer:2007tr,
    author = "Boehmer, Christian G. and Hollenstein, Lukas and Lobo, Francisco S. N.",
    title = "{Stability of the Einstein static universe in f(R) gravity}",
    eprint = "0706.1663",
    archivePrefix = "arXiv",
    primaryClass = "gr-qc",
    doi = "10.1103/PhysRevD.76.084005",
    journal = "Phys. Rev. D",
    volume = "76",
    pages = "084005",
    year = "2007"
}

@article{Seahra:2009ft,
    author = "Seahra, Sanjeev S. and Boehmer, Christian G.",
    title = "{Einstein static universes are unstable in generic f(R) models}",
    eprint = "0901.0892",
    archivePrefix = "arXiv",
    primaryClass = "gr-qc",
    doi = "10.1103/PhysRevD.79.064009",
    journal = "Phys. Rev. D",
    volume = "79",
    pages = "064009",
    year = "2009"
}

@article{Barrow:1983rx,
    author = "Barrow, John D. and Ottewill, A. C.",
    title = "{The Stability of General Relativistic Cosmological Theory}",
    reportNumber = "PRINT-83-0259 (SUSSEX)",
    doi = "10.1088/0305-4470/16/12/022",
    journal = "J. Phys. A",
    volume = "16",
    pages = "2757",
    year = "1983"
}

@article{delaCruz-Dombriz:2011oii,
    author = "de la Cruz-Dombriz, Alvaro and Saez-Gomez, Diego",
    title = "{On the stability of the cosmological solutions in $f(R,G)$ gravity}",
    eprint = "1112.4481",
    archivePrefix = "arXiv",
    primaryClass = "gr-qc",
    doi = "10.1088/0264-9381/29/24/245014",
    journal = "Class. Quant. Grav.",
    volume = "29",
    pages = "245014",
    year = "2012"
}

@article{Faraoni:2005vk,
    author = "Faraoni, Valerio",
    title = "{The Stability of modified gravity models}",
    eprint = "gr-qc/0511094",
    archivePrefix = "arXiv",
    doi = "10.1103/PhysRevD.72.124005",
    journal = "Phys. Rev. D",
    volume = "72",
    pages = "124005",
    year = "2005"
}

@article{Guzman:2024cwa,
    author = {Guzm{\'a}n, Mar{\'\i}a-Jos{\'e} and J{\"a}rv, Laur and Pati, Laxmipriya},
    title = "{Exploring the stability of f(Q) cosmology near general relativity limit with different connections}",
    eprint = "2406.11621",
    archivePrefix = "arXiv",
    primaryClass = "gr-qc",
    doi = "10.1103/PhysRevD.110.124013",
    journal = "Phys. Rev. D",
    volume = "110",
    number = "12",
    pages = "124013",
    year = "2024"
}

@article{Chakraborty:2025rvc,
    author = "Chakraborty, Saikat and Louw, Charlotte and Dunsby, Peter K. S. and MacDevette, Kelly and de la Cruz Dombriz, Alvaro",
    title = "{Dynamical dark energy in models close to $\Lambda$CDM}",
    eprint = "2508.09813",
    archivePrefix = "arXiv",
    primaryClass = "gr-qc",
    month = "8",
    year = "2025"
}

@article{Damour:1992kf,
    author = "Damour, Thibault and Nordtvedt, Kenneth",
    title = "{General relativity as a cosmological attractor of tensor scalar theories}",
    reportNumber = "IHES-P-92-94",
    doi = "10.1103/PhysRevLett.70.2217",
    journal = "Phys. Rev. Lett.",
    volume = "70",
    pages = "2217--2219",
    year = "1993"
}

@article{Damour:1993id,
    author = "Damour, T. and Nordtvedt, K.",
    title = "{Tensor - scalar cosmological models and their relaxation toward general relativity}",
    reportNumber = "IHES-P-93-16",
    doi = "10.1103/PhysRevD.48.3436",
    journal = "Phys. Rev. D",
    volume = "48",
    pages = "3436--3450",
    year = "1993"
}

@article{Mimoso:1998dn,
    author = "Mimoso, J. P. and Nunes, A. M.",
    title = "{General relativity as a cosmological attractor of scalar tensor gravity theories}",
    doi = "10.1016/S0375-9601(98)00724-5",
    journal = "Phys. Lett. A",
    volume = "248",
    pages = "325--331",
    year = "1998"
}

@article{Mimoso:1999ai,
    author = "Mimoso, Jose P. and Nunes, Ana",
    title = "{General relativity as an attractor to scalar tensor gravity theories}",
    doi = "10.1023/A:1002025505686",
    journal = "Astrophys. Space Sci.",
    volume = "261",
    pages = "327--330",
    year = "1999"
}

@article{Saal:2012zb,
    author = "Saal, Margus and Jarv, Laur and Kuusk, Piret",
    editor = "Zhou, Yu-Feng",
    title = "{Scalar-tensor cosmological models converging to general relativity: Potential dominated and matter dominated cases}",
    doi = "10.1088/1742-6596/384/1/012029",
    journal = "J. Phys. Conf. Ser.",
    volume = "384",
    pages = "012029",
    year = "2012"
}

@article{Jarv:2015odu,
    author = "Jarv, Laur and Toporensky, Alexey",
    title = "{General relativity as an attractor for scalar-torsion cosmology}",
    eprint = "1511.03933",
    archivePrefix = "arXiv",
    primaryClass = "gr-qc",
    doi = "10.1103/PhysRevD.93.024051",
    journal = "Phys. Rev. D",
    volume = "93",
    number = "2",
    pages = "024051",
    year = "2016"
}

@article{Dinda:2025svh,
    author = "Dinda, Bikash R. and Maartens, Roy and Saito, Shun and Clarkson, Chris",
    title = "{Improved null tests of {\ensuremath{\Lambda}}CDM and FLRW in light of DESI DR2}",
    eprint = "2504.09681",
    archivePrefix = "arXiv",
    primaryClass = "astro-ph.CO",
    doi = "10.1088/1475-7516/2025/08/018",
    journal = "JCAP",
    volume = "08",
    pages = "018",
    year = "2025"
}

@article{deSouza:2007zpn,
    author = "de Souza, Jose C. C. and Faraoni, Valerio",
    title = "{The Phase space view of f(R) gravity}",
    eprint = "0706.1223",
    archivePrefix = "arXiv",
    primaryClass = "gr-qc",
    doi = "10.1088/0264-9381/24/14/006",
    journal = "Class. Quant. Grav.",
    volume = "24",
    pages = "3637--3648",
    year = "2007"
}

@article{Rodrigues:2025tfg,
    author = "Rodrigues, Gabriel and de Souza, Rayff and Alcaniz, Jailson",
    title = "{Cosmography with DESI DR2 and SN data}",
    eprint = "2506.22373",
    archivePrefix = "arXiv",
    primaryClass = "astro-ph.CO",
    month = "6",
    year = "2025"
}

@article{Nojiri:2010wj,
    author = "Nojiri, Shin'ichi and Odintsov, Sergei D.",
    title = "{Unified cosmic history in modified gravity: from F(R) theory to Lorentz non-invariant models}",
    eprint = "1011.0544",
    archivePrefix = "arXiv",
    primaryClass = "gr-qc",
    doi = "10.1016/j.physrep.2011.04.001",
    journal = "Phys. Rept.",
    volume = "505",
    pages = "59--144",
    year = "2011"
}

@article{Boehmer:2010jqg,
    author = "Boehmer, Christian G. and Harko, T. and Sabau, S. V.",
    title = "{Jacobi stability analysis of dynamical systems: Applications in gravitation and cosmology}",
    eprint = "1010.5464",
    archivePrefix = "arXiv",
    primaryClass = "math-ph",
    doi = "10.4310/ATMP.2012.v16.n4.a2",
    journal = "Adv. Theor. Math. Phys.",
    volume = "16",
    number = "4",
    pages = "1145--1196",
    year = "2012"
}

@article{Harko_2016,
   title={Kosambi–Cartan–Chern (KCC) theory for higher-order dynamical systems},
   volume={13},
   ISSN={1793-6977},
   url={http://dx.doi.org/10.1142/S0219887816500146},
   DOI={10.1142/s0219887816500146},
   number={02},
   journal={International Journal of Geometric Methods in Modern Physics},
   publisher={World Scientific Pub Co Pte Lt},
   author={Harko, Tiberiu and Pantaragphong, Praiboon and Sabau, Sorin V.},
   year={2016},
   month=jan, pages={1650014} 
}

@article{Mukherjee:2020ytg,
    author = "Mukherjee, Purba and Banerjee, Narayan",
    title = "{Non-parametric reconstruction of the cosmological $jerk$ parameter}",
    eprint = "2007.10124",
    archivePrefix = "arXiv",
    primaryClass = "astro-ph.CO",
    doi = "10.1140/epjc/s10052-021-08830-5",
    journal = "Eur. Phys. J. C",
    volume = "81",
    number = "1",
    pages = "36",
    year = "2021"
}

@article{Faraoni:2004dn,
    author = "Faraoni, Valerio",
    title = "{De Sitter attractors in generalized gravity}",
    eprint = "gr-qc/0407021",
    archivePrefix = "arXiv",
    doi = "10.1103/PhysRevD.70.044037",
    journal = "Phys. Rev. D",
    volume = "70",
    pages = "044037",
    year = "2004"
}

@article{Faraoni:2005ie,
    author = "Faraoni, Valerio",
    title = "{Modified gravity and the stability of de Sitter space}",
    eprint = "gr-qc/0509008",
    archivePrefix = "arXiv",
    doi = "10.1103/PhysRevD.72.061501",
    journal = "Phys. Rev. D",
    volume = "72",
    pages = "061501",
    year = "2005"
}

@article{Boehmer:2009fey,
    author = "Boehmer, Christian G. and Lobo, Francisco S. N.",
    title = "{Stability of the Einstein static universe in modified Gauss-Bonnet gravity}",
    eprint = "0902.2982",
    archivePrefix = "arXiv",
    primaryClass = "gr-qc",
    doi = "10.1103/PhysRevD.79.067504",
    journal = "Phys. Rev. D",
    volume = "79",
    pages = "067504",
    year = "2009"
}

@article{Pozdeeva:2019agu,
    author = "Pozdeeva, Ekaterina O. and Sami, Mohammad and Toporensky, Alexey V. and Vernov, Sergey Yu.",
    title = "{Stability analysis of de Sitter solutions in models with the Gauss-Bonnet term}",
    eprint = "1905.05085",
    archivePrefix = "arXiv",
    primaryClass = "gr-qc",
    doi = "10.1103/PhysRevD.100.083527",
    journal = "Phys. Rev. D",
    volume = "100",
    number = "8",
    pages = "083527",
    year = "2019"
}

@article{Barrow:2006xb,
    author = "Barrow, John D. and Hervik, Sigbjorn",
    title = "{On the evolution of universes in quadratic theories of gravity}",
    eprint = "gr-qc/0610013",
    archivePrefix = "arXiv",
    doi = "10.1103/PhysRevD.74.124017",
    journal = "Phys. Rev. D",
    volume = "74",
    pages = "124017",
    year = "2006"
}

@article{Toporensky:2006kc,
    author = "Toporensky, A. V. and Tretyakov, P. V.",
    title = "{De Sitter stability in quadratic gravity}",
    eprint = "gr-qc/0611068",
    archivePrefix = "arXiv",
    doi = "10.1142/S0218271807010572",
    journal = "Int. J. Mod. Phys. D",
    volume = "16",
    pages = "1075--1086",
    year = "2007"
}

@article{Toporensky:2016kss,
    author = {Toporensky, A. and M{\"u}ller, D.},
    title = "{On stability of the Kasner solution in quadratic gravity}",
    eprint = "1603.02851",
    archivePrefix = "arXiv",
    primaryClass = "gr-qc",
    doi = "10.1007/s10714-016-2172-9",
    journal = "Gen. Rel. Grav.",
    volume = "49",
    number = "1",
    pages = "8",
    year = "2017"
}

@article{Bahamonde:2017ize,
    author = {Bahamonde, Sebastian and B{\"o}hmer, Christian G. and Carloni, Sante and Copeland, Edmund J. and Fang, Wei and Tamanini, Nicola},
    title = "{Dynamical systems applied to cosmology: dark energy and modified gravity}",
    eprint = "1712.03107",
    archivePrefix = "arXiv",
    primaryClass = "gr-qc",
    doi = "10.1016/j.physrep.2018.09.001",
    journal = "Phys. Rept.",
    volume = "775-777",
    pages = "1--122",
    year = "2018"
}

@article{Myrzakulov:2010gt,
    author = "Myrzakulov, Ratbay and Saez-Gomez, Diego and Tureanu, Anca",
    title = "{On the $\Lambda$CDM Universe in $f(G)$ gravity}",
    eprint = "1009.0902",
    archivePrefix = "arXiv",
    primaryClass = "gr-qc",
    doi = "10.1007/s10714-011-1149-y",
    journal = "Gen. Rel. Grav.",
    volume = "43",
    pages = "1671--1684",
    year = "2011"
}

@article{Elizalde:2010jx,
    author = "Elizalde, E. and Myrzakulov, R. and Obukhov, V. V. and Saez-Gomez, D.",
    title = "{LambdaCDM epoch reconstruction from F(R,G) and modified Gauss-Bonnet gravities}",
    eprint = "1001.3636",
    archivePrefix = "arXiv",
    primaryClass = "gr-qc",
    doi = "10.1088/0264-9381/27/9/095007",
    journal = "Class. Quant. Grav.",
    volume = "27",
    pages = "095007",
    year = "2010"
}

@article{Ortiz-Banos:2021jgg,
    author = "Ortiz-Ba{\~n}os, Mar{\'\i}a and Bouhmadi-L{\'o}pez, Mariam and Lazkoz, Ruth and Salzano, Vincenzo",
    title = "{${\Lambda}$CDM suitably embedded in f(R) with a non-minimal coupling to matter}",
    eprint = "2103.01982",
    archivePrefix = "arXiv",
    primaryClass = "gr-qc",
    doi = "10.1140/epjc/s10052-021-09004-z",
    journal = "Eur. Phys. J. C",
    volume = "81",
    number = "3",
    pages = "237",
    year = "2021"
}

@article{Chakraborty:2025qlv,
    author = "Chakraborty, Saikat and Dutta, Jibitesh and Gregoris, Daniele and Karwan, Khamphee and Khyllep, Wompherdeiki",
    title = "{Reproducing {\ensuremath{\Lambda}}CDM-like solutions in f(Q) gravity: a~comprehensive study across all connection branches}",
    eprint = "2501.15159",
    archivePrefix = "arXiv",
    primaryClass = "gr-qc",
    doi = "10.1088/1475-7516/2025/05/098",
    journal = "JCAP",
    volume = "05",
    pages = "098",
    year = "2025"
}

@article{Kavya:2024bpj,
    author = "Kavya, N. S. and Venkatesha, V.",
    title = "{Embedding the {\ensuremath{\Lambda}}CDM framework in non-minimal f(Q) gravity with matter-coupling}",
    doi = "10.1016/j.physletb.2024.138927",
    journal = "Phys. Lett. B",
    volume = "856",
    pages = "138927",
    year = "2024"
}

@book{Papantonopoulos2015,
  title     = {Modifications of Einstein's Theory of Gravity at Large Distances},
  editor    = {Eleftherios Papantonopoulos},
  series    = {Lecture Notes in Physics},
  volume    = {892},
  year      = {2015},
  publisher = {Springer},
  address   = {Cham},
  isbn      = {978-3-319-10069-2},
  doi       = {10.1007/978-3-319-10070-8},
  url       = {https://link.springer.com/book/10.1007/978-3-319-10070-8}
}

@book{CANTATA:2021asi,
    author = "Akrami, Yashar and others",
    editor = "Saridakis, Emmanuel N. and Lazkoz, Ruth and Salzano, Vincenzo and Vargas Moniz, Paulo and Capozziello, Salvatore and Beltr{\'a}n Jim{\'e}nez, Jose and De Laurentis, Mariafelicia and Olmo, Gonzalo J.",
    collaboration = "CANTATA",
    title = "{Modified Gravity and Cosmology. An Update by the CANTATA Network}",
    eprint = "2105.12582",
    archivePrefix = "arXiv",
    primaryClass = "gr-qc",
    doi = "10.1007/978-3-030-83715-0",
    isbn = "978-3-030-83714-3, 978-3-030-83717-4, 978-3-030-83715-0",
    publisher = "Springer",
    year = "2021"
}

@article{Creminelli:2017sry,
    author = "Creminelli, Paolo and Vernizzi, Filippo",
    title = "{Dark Energy after GW170817 and GRB170817A}",
    eprint = "1710.05877",
    archivePrefix = "arXiv",
    primaryClass = "astro-ph.CO",
    doi = "10.1103/PhysRevLett.119.251302",
    journal = "Phys. Rev. Lett.",
    volume = "119",
    number = "25",
    pages = "251302",
    year = "2017"
}

@article{Ezquiaga:2017ekz,
    author = "Ezquiaga, Jose Mar{\'\i}a and Zumalac{\'a}rregui, Miguel",
    title = "{Dark Energy After GW170817: Dead Ends and the Road Ahead}",
    eprint = "1710.05901",
    archivePrefix = "arXiv",
    primaryClass = "astro-ph.CO",
    reportNumber = "IFT-UAM-CSIC-17-096, NORDITA-2017-109",
    doi = "10.1103/PhysRevLett.119.251304",
    journal = "Phys. Rev. Lett.",
    volume = "119",
    number = "25",
    pages = "251304",
    year = "2017"
}

@article{Baker:2017hug,
    author = "Baker, T. and Bellini, E. and Ferreira, P. G. and Lagos, M. and Noller, J. and Sawicki, I.",
    title = "{Strong constraints on cosmological gravity from GW170817 and GRB 170817A}",
    eprint = "1710.06394",
    archivePrefix = "arXiv",
    primaryClass = "astro-ph.CO",
    doi = "10.1103/PhysRevLett.119.251301",
    journal = "Phys. Rev. Lett.",
    volume = "119",
    number = "25",
    pages = "251301",
    year = "2017"
}

@article{Copeland:2018yuh,
    author = "Copeland, Edmund J. and Kopp, Michael and Padilla, Antonio and Saffin, Paul M. and Skordis, Constantinos",
    title = "{Dark energy after GW170817 revisited}",
    eprint = "1810.08239",
    archivePrefix = "arXiv",
    primaryClass = "gr-qc",
    doi = "10.1103/PhysRevLett.122.061301",
    journal = "Phys. Rev. Lett.",
    volume = "122",
    number = "6",
    pages = "061301",
    year = "2019"
}

@article{Kuusk:2008ak,
    author = "Kuusk, Piret and Jarv, Laur and Saal, Margus",
    editor = "Bezerra, V. B. and Mostepanenko, V. M. and Romero, Carlos",
    title = "{Scalar-tensor cosmologies: General relativity as a fixed point of the Jordan frame scalar field}",
    eprint = "0810.5038",
    archivePrefix = "arXiv",
    primaryClass = "gr-qc",
    doi = "10.1142/S0217751X09045133",
    journal = "Int. J. Mod. Phys. A",
    volume = "24",
    pages = "1631--1638",
    year = "2009"
}

@article{Jarv:2008eb,
    author = "Jarv, Laur and Kuusk, Piret and Saal, Margus",
    title = "{Scalar-tensor cosmologies: Fixed points of the Jordan frame scalar field}",
    eprint = "0807.2159",
    archivePrefix = "arXiv",
    primaryClass = "gr-qc",
    doi = "10.1103/PhysRevD.78.083530",
    journal = "Phys. Rev. D",
    volume = "78",
    pages = "083530",
    year = "2008"
}

@article{Jarv:2010xm,
    author = "Jarv, Laur and Kuusk, Piret and Saal, Margus",
    title = "{Scalar-tensor cosmologies with a potential in the general relativity limit: time evolution}",
    eprint = "1006.1246",
    archivePrefix = "arXiv",
    primaryClass = "gr-qc",
    doi = "10.1016/j.physletb.2010.09.029",
    journal = "Phys. Lett. B",
    volume = "694",
    pages = "1--5",
    year = "2011"
}

@article{Jarv:2011sm,
    author = "Jarv, Laur and Kuusk, Piret and Saal, Margus",
    title = "{Scalar-tensor cosmologies with dust matter in the general relativity limit}",
    eprint = "1112.5308",
    archivePrefix = "arXiv",
    primaryClass = "gr-qc",
    doi = "10.1103/PhysRevD.85.064013",
    journal = "Phys. Rev. D",
    volume = "85",
    pages = "064013",
    year = "2012"
}

@article{Roy:2025cxk,
    author = "Roy, Nandan and Chakrabarti, Soumya",
    title = "{Is Phantom Barrier Crossing Inevitable? A Cosmographic Analysis}",
    eprint = "2508.13740",
    archivePrefix = "arXiv",
    primaryClass = "astro-ph.CO",
    month = "8",
    year = "2025"
}

@article{Plaza:2025gcv,
    author = "Plaza, Francisco and Kraiselburd, Lucila",
    title = "{Testing $f(R)$-gravity models with DESI DR2 2025-BAO and other cosmological data}",
    eprint = "2504.05432",
    archivePrefix = "arXiv",
    primaryClass = "gr-qc",
    doi = "10.1103/gtrg-56fj",
    journal = "Phys. Rev. D",
    volume = "112",
    number = "2",
    pages = "023554",
    year = "2025"
}

@article{Odintsov:2024woi,
    author = "Odintsov, Sergei D. and S{\'a}ez-Chill{\'o}n G{\'o}mez, Diego and Sharov, German S.",
    title = "{Modified gravity/dynamical dark energy vs $\Lambda $CDM: is the game over?}",
    eprint = "2412.09409",
    archivePrefix = "arXiv",
    primaryClass = "gr-qc",
    doi = "10.1140/epjc/s10052-025-14013-3",
    journal = "Eur. Phys. J. C",
    volume = "85",
    number = "3",
    pages = "298",
    year = "2025"
}

@article{Odintsov:2025jfq,
    author = "Odintsov, S. D. and Oikonomou, V. K. and Sharov, G. S.",
    title = "{Dynamical Dark Energy from $F(R)$ Gravity Models Unifying Inflation with Dark Energy: Confronting the Latest Observational Data}",
    eprint = "2506.02245",
    archivePrefix = "arXiv",
    primaryClass = "gr-qc",
    month = "6",
    year = "2025"
}

@article{Cattoen:2007sk,
    author = "Cattoen, Celine and Visser, Matt",
    title = "{The Hubble series: Convergence properties and redshift variables}",
    eprint = "0710.1887",
    archivePrefix = "arXiv",
    primaryClass = "gr-qc",
    doi = "10.1088/0264-9381/24/23/018",
    journal = "Class. Quant. Grav.",
    volume = "24",
    pages = "5985--5998",
    year = "2007"
}

@article{Gruber:2013wua,
    author = "Gruber, Christine and Luongo, Orlando",
    title = "{Cosmographic analysis of the equation of state of the universe through Pad{\'e} approximations}",
    eprint = "1309.3215",
    archivePrefix = "arXiv",
    primaryClass = "gr-qc",
    doi = "10.1103/PhysRevD.89.103506",
    journal = "Phys. Rev. D",
    volume = "89",
    number = "10",
    pages = "103506",
    year = "2014"
}

@article{Aviles:2014rma,
    author = "Aviles, Alejandro and Bravetti, Alessandro and Capozziello, Salvatore and Luongo, Orlando",
    title = "{Precision cosmology with Pad{\'e} rational approximations: Theoretical predictions versus observational limits}",
    eprint = "1405.6935",
    archivePrefix = "arXiv",
    primaryClass = "gr-qc",
    doi = "10.1103/PhysRevD.90.043531",
    journal = "Phys. Rev. D",
    volume = "90",
    number = "4",
    pages = "043531",
    year = "2014"
}

\end{document}